\documentclass[]{aastex631}
\draft
\usepackage{natbib}
\usepackage{graphicx}
\usepackage{url}
\usepackage{xcolor}
\newcommand{\cmd}{\,cm$^{-2}$}   
\newcommand{\kms}{\,km\,s$^{-1}$}

\newcommand{\myr}{\,$M_{\odot}\,{\rm yr}^{-1}$}

\newcommand{\hII}{H\,{\scriptsize II}}

\newcommand{\ovi}{O\,{\scriptsize VI}}  

\newcommand{\ro}{\,$R_{\odot}$}

\newcommand{\lo}{\,$L_{\odot}$}
\shorttitle{Neutral wind in the orbital plane during outbursts 
              of symbiotic stars}
\shortauthors{Skopal}
\begin{document} 

\title{The emergence of a neutral wind region in the orbital 
       plane of symbiotic binaries \\
       during their outbursts}
\author[0000-0002-8312-3326]{Augustin Skopal}
\affiliation{Astronomical Institute, Slovak Academy of Sciences, \\
             059\,60 Tatransk\'a Lomnica, Slovakia}

\begin{abstract}
Accretion of mass onto a white dwarf (WD) in a binary system 
can lead to stellar explosions. If a WD accretes from stellar 
wind of a distant evolved giant in a symbiotic binary, it can 
undergo occasional outbursts in which it brightens by several 
magnitudes, produces a low- and high-velocity mass-outflow, and, 
in some cases, ejects bipolar jets. 
In this paper, we complement the current picture of these outbursts 
by the transient emergence of a neutral region in the orbital 
plane of symbiotic binaries consisting of wind from the giant. 
We prove its presence by determining H$^0$ column densities 
($N_{\rm H}$) in the direction of the WD and at any orbital phase 
of the binary by modeling the continuum depression around 
the Ly$\alpha$ line caused by Rayleigh scattering on 
atomic hydrogen for all suitable objects, i.e., eclipsing 
symbiotic binaries, for which a well-defined ultraviolet 
spectrum from an outburst is available. 
The $N_{\rm H}$ values follow a common course along the orbit 
with a minimum and maximum of a few times $10^{22}$ and 
$10^{24}$\,cm$^{-2}$ around the superior and inferior 
conjunction of the giant, respectively. 
Its asymmetry implies an asymmetric density distribution of 
the wind from the giant in the orbital plane with respect to 
the binary axis. The neutral wind is observable in the orbital 
plane due to the formation of a dense disk-like structure 
around the WD during outbursts, which blocks ionizing radiation 
from the central burning WD in the orbital plane. 
\end{abstract}


\section{Introduction}
\label{s:intro}
Currently, symbiotic stars (SySts) are understood as the widest 
interacting binaries comprising a cool giant as the donor and 
a compact star, mostly a white dwarf (WD), as the accretor. 
Their orbital periods are typically as long as a few years 
\citep[S-type systems containing a normal giant, 
see][]{2000A&AS..146..407B,2013AcA....63..405G}, but for 
systems containing a Mira variable 
\citep[D-type systems, see][]{
1975MNRAS.171..171W,
1988AJ.....95.1817K,
1990Ap&SS.174..321S} 
they can be significantly longer, but mostly they are unknown 
\citep[e.g.,][]{
2002A&A...395..117S,
2006ApJ...637L..49M,
2013ApJ...770...28H}. 
SySts are detached binaries \cite[][]{1999A&AS..137..473M}; 
their activity is thus triggered via the wind mass transfer. 
The accretion process is responsible for the WD's temperatures 
of $1-2\times10^5$\,K and luminosities of 
$10^{1}-10^{4}$\lo\ \citep[][]{
1991A&A...248..458M,
2005A&A...440..995S}. 

According to the light variations in the optical, we distinguish 
between the quiescent and active phases of SySts. 
%
During quiescent phases, a SySt releases its energy 
at a constant rate. 
%
The hot and luminous accreting WD ionizes the neighboring part 
of the wind from the giant giving rise to the nebular emission 
\citep[e.g.,][]{1966SvA....10..331B}, while its portion 
around the cool giant remains neutral. 
The boundary between the ionized and neutral regions is given 
by the balance between the flux of ionizing photons from the hot 
component and the flux of neutral particles from the cool giant, 
the shape of which determines the ionization structure of SySts 
during the quiescent phase 
\citep[see e.g., Fig.~6 of][]{1984ApJ...284..202S}. 
%
%
This configuration causes the characteristic feature in the light 
curves of SySts during their quiescent phases -- a periodic 
wave-like variation as a function of the orbital phase 
\citep[see light curves presented by, e.g.,][]{
1992IzKry..84...49B,
2000AstL...26..520D,
2008JAVSO..36....9S}. 
%
%
On the other hand, active phases of SySts result from 
unstable nuclear burning on the WD surface that gives rise to 
transient outbursts indicated by brightening of a few magnitudes 
in the optical. 
The most characteristic outbursts for SySts are of `Z And-type' 
that result from an increase in the accretion rate above that 
sustaining the stable burning
\citep[][]{
2017A&A...604A..48S,
2020A&A...636A..77S}. 
They show 1--3\,mag (multiple) brightening(s) evolving on the 
time-scale of a few months to years/decades
\citep[e.g.,][]{
2005A&A...440..239B,
2008MNRAS.385..445L,
2019CoSka..49...19S} 
with signatures of a mass outflow 
\citep[e.g.,][]{
1995ApJ...442..366F,
2000AstL...26..162E,
2006A&A...457.1003S,
2011PASP..123.1062M}. 
The periodic wave-like variation in the light curve disappears 
and, in the case of eclipsing systems, deep narrow minima 
caused by eclipses of the hot component by the red giant, appear 
in the light curve \citep[e.g.,][and Appendix~\ref{app:B} 
for our targets]{1976IBVS.1169....1B} -- one of the most prominent 
indications of the change in the ionization structure of the hot 
component during active phases (see below). 
%

During outbursts, the stellar radiation from the hot component 
and the nebular radiation markedly change. 
For systems with a high orbital inclination, the hot component 
temperature significantly declines, often showing spectral 
features of an A--F type star in the optical
\citep[][]{
1992AJ....103..579M,
1996AJ....112.1659M,
1992AJ....104..262M,
2002A&A...387..139Q,
2005A&A...440..239B}. 
Such a warm WD pseudophotosphere with an effective radius of 
several solar radii \citep[][]{2005A&A...440..995S} becomes 
a significant source of optical radiation. Therefore, its 
eclipse by the red giant will cause narrow minima in the light 
curve \citep[see e.g., Fig.~4 of][]{2008JAVSO..36....9S}.
\cite{1997ARep...41..802M}, while studying the active phase 
of YY~Her, pointed out that its hot component cannot always 
be spherically symmetric, as they indicated its occasionally 
low temperature ($\sim$1.5$\times 10^4 $\,K) in the simultaneous 
presence of a hot subdwarf radiation ($\sim$10$^5$\,K) in 
the system.
Similarly, during the active phase of AS~338, 
\cite{2000AstL...26..162E} indicated a low hot component 
temperature ($<$1.5$\times 10^4$\,K) in the optical with 
the simultaneous presence of an extended \hII\ region 
indicating the presence of a high-temperature region 
($\sim$10$^5$\,K) in the `pseudosphere'. In this respect, 
the authors pointed out the similarity with other 
classical SySts. 
Another interesting effect, considering the aim of this work, 
was the appearance of a substantial amount of neutral hydrogen 
on the line of sight to the hot component of YY~Her after its 
short-duration outburst in December 1981 
\citep[][]{1997ARep...41..802M}. 
Also, \cite{2002A&A...387..139Q} found that the nebula in 
AR~Pav is probably bounded on all sides by significant 
amount of neutral material at least near the orbital plane. 
In contrast, for non-eclipsing systems, seen $\sim$\,pole-on, 
the temperature increases to around 2$\times 10^5$\,K 
\citep[e.g.,][]{2016MNRAS.462.4435T,2020A&A...636A..77S}. 
In both cases, eclipsing and non-eclipsing, the nebular continuum 
increases by a factor of $\sim$10 relative to quiescent phases 
\citep[][]{2005A&A...440..995S,2017A&A...604A..48S}. 
The nature of these dramatic changes between the quiescence and 
the activity was clarified by disentangling the ultraviolet (UV) 
to near-infrared spectra of S-type symbiotic stars 
\citep[][]{2005A&A...440..995S}. This revealed that all eclipsing 
systems develop a two-temperature type of the hot component 
spectrum during active phases: 
The cool spectrum is produced by a $1-3\times 10^{4}$\,K warm 
pseudophotosphere, while the hot one is represented by highly 
ionized emission lines and a strong nebular continuum. 
The former is not capable of producing the measured nebular 
emission, which thus signals the presence of a strong ionizing 
source in the system -- the essential condition for its 
interpretation: 
The hot component has a disk-like structure, whose flared outer 
rim (which is the warm WD's pseudophotosphere) occults 
the central ionizing source in the line of sight, while 
the nebula above/below the disk is ionized 
(see Fig.~27 of \cite{2005A&A...440..995S} 
and Fig.~6 of \cite{2012A&A...548A..21C}). 
On the other hand, non-eclipsing systems, seen $\sim$\,pole-on, 
show an increase in the WD temperature ($\sim 2\times 10^{5}$\,K) 
and the nebular emission from the very beginning of the outbursts 
\citep[e.g.,][]{2017A&A...604A..48S}. 
%

%
The key phenomenon, the emergence of the disk-like structure 
around the WD during active phases, allows us to measure the 
column density of neutral hydrogen between the observer and 
the warm WD's pseudophotosphere ($N_{\rm H}^{\rm obs}$) for 
{\em eclipsing} SySts at any orbital phase of the binary 
\citep[see Figs. 8, 10, 19 and 21 of][]{2005A&A...440..995S}. 
This is because the flared disk blocks the ionizing photons 
from the central hot WD in the orbital plane. Consequently, 
the lack of ionizing photons allows the giant's wind 
to remain neutral in the orbital plane. 

Accordingly, measuring $N_{\rm H}^{\rm obs}$ from Rayleigh 
scattering on H atoms around the Ly$\alpha$ line 
\citep[e.g.,][]{1989A&A...219..271I} along the whole orbit, 
and for all suitable objects (Sect.~\ref{ss:targets}) represents 
the main aim of this paper. 
In this way, we revealed the transient emergence of a neutral 
wind region in the orbital plane, and its asymmetry with 
respect to the binary axis (Sect.~\ref{s:results}). 
In Sect.~\ref{s:discuss} we discuss how our results aid us in 
understanding the geometric and ionization structure of SySts, 
while in Sect.~\ref{s:concl} we summarize our findings and 
propose tasks for future investigation. 

\section{Observations and methods}
\label{s:obs}
\subsection{Selection of targets}
\label{ss:targets}
According to the ionization structure of hot components during 
active phases (see above), the number of objects suitable for 
the objective of this paper is restricted only to the eclipsing 
systems, for which a well-defined far-UV spectrum from the 
active phase is available. This allows us to model the effect 
of Rayleigh scattering around the Ly$\alpha$ line. 
Only in such the case, it is possible to expect comparable 
$N_{\rm H}^{\rm obs}$ for different objects at any orbital phase, 
because their values depend on the orbital inclination. 
For systems that may no longer be eclipsing but have a relatively 
high orbital inclination, up to values given by the vertical 
extension of the WD’s pseudophotosphere, lower 
$N_{\rm H}^{\rm obs}$ values can be expected. 
For even lower inclinations, when the line of sight passes only 
through the ionized region, no neutral hydrogen is measurable. 
Moreover, the wind density dispersion with the orbital 
inclination can be accentuated by focusing of the giant's 
wind to the orbital plane \citep[see][]{2016A&A...588A..83S} 
that results in its dilution around the giant's poles 
\citep[see][]{2021A&A...646A.116S}. 
%

Accordingly, we first inspected the historical light curves of 
SySts available from the literature, and, on the basis 
of our own experience with photometric monitoring of these objects 
for more than 30 years \citep[][]{2019CoSka..49...19S} we 
found that BF~Cyg, CH~Cyg, CI~Cyg, YY~Her, BX~Mon, AR~Pav, 
AX~Per, FN~Sgr, PU~Vul, AS~338 and AS~296 show eclipses during 
outbursts. 
Second, searching the archive of the {\it International 
Ultraviolet Explorer} (\textsl{IUE}) we found that the spectra 
of Z~And, CD-43$^{\circ}$14304, TX~CVn, BF~Cyg, CH~Cyg, CI~Cyg, 
YY~Her, AR~Pav, AX~Per, PU~Vul, AS~338 and AS~296 show 
signatures of the Rayleigh scattering attenuation around 
the Ly$\alpha$ line during active phases that signals a large 
amount of H atoms on the line of sight. 
The intersection of both sets of objects then should represent 
those that best suit the goals of this study. 
However, some systems from these groups cannot be included 
in the final set of our targets. 
They are discussed in Appendix~\ref{app:targets}. 

As a result, targets for which comparable $N_{\rm H}^{\rm obs}$ 
at any orbital phase of the binary can be expected are the 
eclipsing systems: BF~Cyg, CI~Cyg, YY~Her, AR~Pav, AX~Per and 
PU~Vul. Log of their \textsl{IUE} spectra is introduced in 
Table~\ref{tab:nh}, and Appendix~\ref{app:param} summarizes 
their basic characteristics. 
%
%
%
\begin{table*}
\caption{Log of the low-resolution \textsl{IUE} spectra of our 
targets and corresponding column densities $N_{\rm H}^{\rm obs}$ 
(see text and Fig.~\ref{fig:nhfi}).} 
\label{tab:nh}
\begin{center}
\begin{tabular}{cccccc}
\hline
\hline
\noalign{\smallskip}
Object & Spectrum & Date (UT)$^{\dagger}$ & Julian Date$^{\dagger}$
       & $\varphi^{\star}$ & $N_{\rm H}^{\rm obs}$ \\
       &          &yyyy-mm-dd& JD~2\,44... &          & (cm$^{-2}$) \\
\noalign{\smallskip}
\hline
\noalign{\smallskip}
BF~Cyg$^a$ & SWP31540+LWP11376 & 1987-08-11 & 7019.35 & 0.145 & 
           $(2.5\pm 1.0)\times 10^{23}$\\
           & SWP32341+LWP12109 & 1987-11-16 & 7116.00 & 0.273 &
           (9.0 +2.4/-1.9)$\times 10^{22}$\\
           & SWP33113+LWP12886 & 1988-03-19 & 7239.88 & 0.436 &
           $(3.5\pm 1.1)\times 10^{22}$\\
           & SWP39163+LWP18251 & 1990-06-30 & 8072.58 & 0.536 &
           $(4.8\pm 1.5)\times 10^{22}$\\
           & SWP39188+LWP18302 & 1990-07-06 & 8079.35 & 0.545 &
           $(6.0\pm 1.8)\times 10^{22}$\\
           & SWP40055+LWP19153(4) & 1990-11-05 & 8201.25 & 0.706 &
           (4.0 +1.2/-1.5)$\times 10^{22}$\\
\noalign{\smallskip}
\hline
\noalign{\smallskip}
CI~Cyg$^b$ 
& SWP03816+LWR03396 & 1979-01-05 & 3879.35 & 0.392 & 
  $(7.0 \pm 2.7)\times 10^{22}$\\
& SWP05485+LWR04757 & 1979-06-11 & 4035.74 & 0.576 & 
  $(3.0\pm 1.3)\times 10^{22}$\\ 
& SWP05672+LWR04916 & 1979-06-29 & 4054.03 & 0.597 & 
  $(3.5\pm 1.4)\times 10^{22}$\\
& SWP07818+LWR06831 & 1980-01-30 & 4269.40 & 0.850 & 
  $(1.6\pm 0.7)\times 10^{23}$\\
& SWP08757          & 1980-04-14 & 4344.17 & 0.937 & 
  (2.5 +0.9/-0.8)$\times 10^{24}$\\
& SWP09664+LWR08408 & 1980-08-01 & 4453.21 & 0.065 & 
  $(1.5\pm 0.7)\times 10^{24}$\\
& SWP09830+LWR08542 & 1980-08-18 & 4469.96 & 0.085 & 
  (1.0 +0.6/-0.4)$\times 10^{24}$\\
& SWP09942+LWR08651 & 1980-08-28 & 4480.44 & 0.097 & 
  (6.0 +1.6/-2.6)$\times 10^{23}$\\
& SWP10602+LWR09303 & 1980-11-14 & 4558.37 & 0.188 & 
  $(2.5\pm 0.8)\times 10^{23}$\\
& SWP11003+LWR09671 & 1981-01-08 & 4613.44 & 0.253 & 
  (1.0 +0.5/-0.3)$\times 10^{23}$\\
& SWP14755+LWR11318 & 1981-08-14 & 4830.93 & 0.508 & 
  $(6.0\pm 2.6)\times 10^{22}$\\
& SWP15713+LWR12127 & 1981-12-11 & 4950.19 & 0.648 & 
  $(3.0\pm 1.3)\times 10^{22}$\\
\noalign{\smallskip}
\hline
\noalign{\smallskip}
YY~Her$^c$
& SWP15652+LWR12080 & 1981-12-04 & 4943.46 & 0.265 & 
  (1.4 +0.4/-0.6)$\times 10^{23}$\\
\noalign{\smallskip}
\hline
\noalign{\smallskip}
AR~Pav$^d$
 & SWP17070+LWR13346 & 1982-05-30 & 5119.91 &  0.008 &  eclipse \\
 & SWP10415+LWR09095 & 1980-10-19 & 4532.05 &  0.035 &  eclipse \\
 & SWP10497+LWR09184 & 1980-10-29 & 4541.85 &  0.051 &  eclipse \\
 & SWP10510+LWR09199 & 1980-10-30 & 4543.45 &  0.054 &  
   $(3.0\pm 1.3)\times 10^{24}$ \\
 & SWP10520+LWR09207 & 1980-11-01 & 4544.90 &  0.056 &  
   $(3.0\pm 1.3)\times 10^{24}$ \\
 & SWP10524+LWR09212 & 1980-11-02 & 4545.74 &  0.058 &  
   $(2.5\pm 1.3)\times 10^{24}$ \\
 & SWP10527+LWR09217 & 1980-11-03 & 4546.70 &  0.059 &  
   $(2.5\pm 1.3)\times 10^{24}$ \\
 & SWP17331+LWR13580 & 1982-07-01 & 5152.09 &  0.061 &  
   (2.5 +0.8/-1.3)$\times 10^{24}$ \\
 & SWP22434+LWP02903 & 1984-03-07 & 5766.70 &  0.078 &  
   $(2.0\pm 0.7)\times 10^{24}$ \\
 & SWP45099+LWP23467 & 1992-07-08 & 8812.01 &  0.116 &  
   (5.0 +1.5/-2.5)$\times 10^{23}$ \\
 & SWP22707+LWP03139 & 1984-04-10 & 5800.76 &  0.134 &  
   $(4.0\pm 1.4)\times 10^{23}$ \\
 & SWP05828+LWR05077 & 1979-07-17 & 4071.74 &  0.274 &  
   (7.5 +1.8/-3.3)$\times 10^{22}$ \\
 & SWP13956+LWR10570 & 1981-05-10 & 4735.37 &  0.371 &  
   $(5.0\pm 2.0)\times 10^{22}$ \\
 & SWP01561          & 1978-05-16 & 3645.37 &  0.568 &  
   (5.0 +3.5/-2.5)$\times 10^{22}$ \\
 & SWP02236+LWR02023 & 1978-08-08 & 3728.89 &  0.706 &  
   $(3.0\pm 1.3)\times 10^{22}$ \\
 & SWP03310+LWR02917 & 1978-11-11 & 3824.39 &  0.864 &  
   (2.0 +1.7/-0.7)$\times 10^{23}$ \\
 & SWP09646+LWR08393 & 1980-07-30 & 4451.29 &  0.901 &  
   (1.9 +0.7/-0.8)$\times 10^{24}$ \\
 & SWP16711+LWR12974 & 1982-04-07 & 5067.04 &  0.920 &  
   (1.0 +0.6/-0.4)$\times 10^{24}$ \\
 & SWP16857+LWR13104 & 1982-04-29 & 5088.99 &  0.956 &  eclipse \\
 & SWP16938+LWR13208 & 1982-05-09 & 5098.88 &  0.973 &  eclipse \\
 & SWP16949+LWR13235 & 1982-05-13 & 5102.93 &  0.980 &  eclipse \\
\noalign{\smallskip}
\hline
\noalign{\smallskip}
AX~Per$^e$
 & SWP03755+LWR03332 & 1978-12-31 & 3873.65 & 0.598 & 
   (5.0 +2.5/-1.5)$\times 10^{22}$ \\
 & SWP03814+LWR03332 & 1979-01-05 & 3879.19 & 0.606 & 
   (4.5 +2.5/-1.5)$\times 10^{22}$ \\
\noalign{\smallskip}
\hline 
\end{tabular}
\end{center}
\end{table*}
\addtocounter{table}{-1}
\begin{table*}[!ht]
\begin{center}
\caption{continued}
\begin{tabular}{cccccc}
\hline
\noalign{\smallskip}
PU~Vul$^f$
 & SWP33833+LWP13531 & 1988-06-29 & 7342.30 & 0.570 & 
   $(5.6\pm 1.8)\times 10^{22}$ \\
 & SWP33930+LWP13669 & 1988-07-16 & 7359.45 & 0.574 & 
   $(5.8\pm 1.9)\times 10^{22}$ \\
 & SWP34406+LWP14173 & 1988-10-03 & 7437.82 & 0.590 & 
   $(3.9\pm 1.5)\times 10^{22}$ \\ 
 & SWP35960+LWP15328 & 1989-04-08 & 7624.82 & 0.628 & 
   $(3.0\pm 0.9)\times 10^{22}$ \\
 & SWP36301+LWP15549 & 1989-05-19 & 7666.20 & 0.636 & 
   $(2.9\pm 1.2)\times 10^{22}$ \\
 & SWP37185+LWP16416 & 1989-09-25 & 7794.53 & 0.663 & 
   $(3.8\pm 1.4)\times 10^{22}$ \\
\noalign{\smallskip}
\hline
\noalign{\smallskip}
AS~338$^g$
 & SWP37290+LWP16522 & 1989-10-09 & 7809.16 & 0.670 & 
   $< 5\times 10^{23}$ \\
\noalign{\smallskip}
\hline
\noalign{\smallskip}
AS~296$^h$
 & SWP34080+LWP13843 & 1988-08-13 & 7387.15 & 0.321 & 
   (1.0 +2.1/-0.6)$\times 10^{23}$ \\
 & SWP34725+LWP14446 & 1988-11-11 & 7477.07 & 0.458 & 
   $< 3\times 10^{23}$ \\
 & SWP38353+LWP17523 & 1990-03-13 & 7963.90 & 0.197 & 
   $(3.0\pm 2.3)\times 10^{23}$ \\
\noalign{\smallskip}
\hline
\hline
\end{tabular}
\end{center}
{\bf Notes:}\\
$^{\dagger}$ the dates correspond to the start of the observation 
             with the SWP camera, \\
$^{\star}$ orbital phase as given by the ephemerides below: \\
$^a$ JD$_{spec. conj.}$ = 2\,445\,395.1 + 757.2$\times$E 
     \citep[][]{2001AJ....121.2219F}, 
  E$_{B-V}$ = 0.35, $d=3.4$\,kpc \citep[][]{1991A&A...248..458M}, \\
$^b$ JD$_{ecl.}$ = 2\,441\,838.8 + 852.98$\times$E 
     \citep[][]{2012AN....333..242S}, 
  E$_{B-V}$ = 0.35 \citep[][]{2005A&A...440..995S}, 
  $d=1.6$\,kpc \citep[][]{1993ApJ...410..260S}, \\
$^c$ JD$_{ecl.}$ = 2\,440\,637 + 592.8$\times$E 
     \citep[][]{1998A&A...338..599S}, 
  $d=6.3$\,kpc \citep[][]{2005A&A...440..995S}, 
  E$_{B-V}$ = 0.20 \citep[][]{1997ARep...41..802M}, \\
$^d$ JD$_{ecl.}$ = 2\,411\,266.1 + 604.45$\times$E 
  \citep[][]{2005A&A...440..995S}, E$_{B-V}$ = 0.26, 
  $d=4.9$\,kpc \citep[][]{2001A&A...366..972S}, \\
$^e$ JD$_{ecl.}$ = 2\,447\,551.26 + 680.83$\times$E 
  \citep[][]{2011A&A...536A..27S}, 
  E$_{B-V}$ = 0.27 \citep[][]{1984ApJ...279..252K}, 
  $d=1.7$\,kpc \citep[][]{2001A&A...367..199S}, \\
$^f$ JD$_{ecl.}$ = 2\,444\,550 + 4897$\times$E 
  \citep[][]{2012BaltA..21..150S}, E$_{B-V}$ = 0.30, 
  $d=4.7$\,kpc \citep[][]{2012ApJ...750....5K}, \\
$^g$ JD$_{ecl.}$ = 2\,446\,650 + 434.1$\times$E 
  \citep[][]{2007BaltA..16...55S}, E$_{B-V}$ = 0.50, 
  $d=6$\,kpc \citep[][]{2000AstL...26..162E}, \\
$^h$ JD$_{ecl.}$ = 2\,448\,492 + 658$\times$E 
  \citep[][]{1995AJ....109.1740M}, E$_{B-V}$ = 1.0, 
  $d=1$\,kpc \citep[][]{1992AJ....104..262M}. 
\end{table*}
\subsection{Measuring $N_{\rm H}^{\rm obs}$ from Rayleigh 
            scattering}
\label{ss:ray}
Rayleigh scattering represents the process, where the atom 
is excited by an incident photon to the intermediate state 
and is immediately stabilized by a transition to the same 
bound state, re-emitting a photon of the original energy 
\citep[e.g.,][]{1989A&A...211L..27N}. For hydrogen, 
the effect is best observable around the Ly$\alpha$ line as 
a continuum depression with zero residual intensity and 
extended wings. 
The strength of Rayleigh scattering is determined by the number 
of scatterers on the path between the emitting source and 
the observer, i.e., on the value of $N_{\rm H}^{\rm obs}$, and 
its profile is given by the cross-section, 
$\sigma_{\rm Ray}(\lambda)$, which yields the optical depth, 
$\tau_{\rm Ray} = \sigma_{\rm Ray}(\lambda) N_{\rm H}^{\rm obs}$. 
Therefore, fitting the far-UV continuum of an object attenuated 
by Rayleigh scattering, we can obtain the corresponding value of 
$N_{\rm H}^{\rm obs}$. 
It is important to emphasize that the Rayleigh scattering effect 
is dominant in modeling the continuum depression around the 
Ly$\alpha$ line\footnote{It is meant relative to other line 
broadening effects in the neutral part of the giant's wind.} 
during quiescent phases 
\citep[e.g.,][]{1989A&A...219..271I,
                1991A&A...249..173V,
                1999A&A...349..169D,
                2005ESASP.560..343C}, 
and also in our case, during active phases, because the width 
of possible `atmospheric' Ly$\alpha$ absorption profile 
from/above the warm WD pseudophotosphere can be neglected 
(see Appendix~\ref{app:lya}). 
%

According to the configuration of SySts, we assume that the UV 
continuum of eclipsing systems during outbursts consists of 
the stellar continuum from the warm WD's pseudophotosphere and 
the nebular continuum from the ionized circumbinary matter 
(see Sect.~\ref{s:intro}). 
We compare the former with the blackbody radiation at 
a temperature $T_{\rm BB}$, and approximate the latter by 
contributions from free-bound and free-free transitions in 
hydrogen plasma radiating at the electron temperature $T_{\rm e}$. 
Then, using Eqs.~(5) and (11) of \cite{2005A&A...440..995S}, 
the reddening-free UV continuum of a SySt, $F(\lambda)$, 
measured at the Earth, can be written in the form, 
%
%
\begin{equation}
 F(\lambda) =   
   \theta_{\rm WD}^2 \pi B_{\lambda}(T_{\rm BB})\,
    e^{-\sigma_{\rm Ray}(\lambda)\,N_{\rm H}^{\rm obs}} 
   +\, k_{\rm N} \times \varepsilon_{\lambda}({\rm H},T_{\rm e}), 
\label{eq:uvsed}
\end{equation}
where 
$\theta_{\rm WD} = R_{\rm WD}^{\rm eff}/d$ is the angular radius 
of the warm WD's pseudophotosphere, given by its effective radius 
$R_{\rm WD}^{\rm eff}$ (i.e., the radius of a sphere with the 
same luminosity) and the distance $d$. 
The Rayleigh cross-section for scattering by hydrogen in its ground 
state can be expressed as \citep[see][]{1989A&A...211L..27N},
%
%
\begin{equation}
 \sigma_{\rm Ray}(\lambda) = \sigma_{\rm e}
    \Big[\sum_k\frac{f_{1k}}{(\lambda/\lambda_{1k})^2 - 1}\Big]^2,
\end{equation}
where $\sigma_{\rm e} = 6.65\times 10^{-25}$\,cm$^2$ 
is the Thomson cross-section, $f_{1k}$ are the oscillator 
strengths of the hydrogen Lyman series and $\lambda_{1k}$ are 
the corresponding wavelengths. In calculating 
$\sigma_{\rm Ray}(\lambda)$ we used all bound states tabulated 
by \cite{1966atp..book.....W}, i.e., $k$ = 2 to 40. 
The second term on the right is the nebular continuum given by 
its volume emission coefficient, 
$\varepsilon_{\lambda}({\rm H},T_{\rm e})$, scaled with 
the factor $k_{\rm N} = EM/4\pi d^2$, where $EM$ is the emission 
measure of the nebula. 
The variables determining the model continuum are, $\theta_{\rm WD}$, 
$T_{\rm BB}$, $N_{\rm H}^{\rm obs}$, $k_{\rm N}$ and $T_{\rm e}$. 
They are obtained by comparing Eq.~(\ref{eq:uvsed}) with 
the continuum fluxes selected from the dereddened spectra of 
our targets. In selecting suitable flux points of the true 
continuum, we proceeded as follows. 

First, we considered the influence of the line blanketing effect, 
which represents an additional source of absorption in the 
neutral wind, mostly by Fe\,{\scriptsize II} transitions 
\citep[see][]{1993ApJ...416..355S}. 
This so-called iron curtain can strongly depress the level of 
the continuum at 1500 to 1800\,\AA\ and 2300 to 2800\,\AA\ 
(see e.g., BF~Cyg, PU~Vul, AR~Pav in Appendix~\ref{app:A} 
and \ref{app:B}). 
To eliminate this effect we used a template of three 
representative low-resolution \textsl{IUE} spectra with a low, 
modest and strong influence of the continuum by the iron 
curtain \citep[see Fig.~1 of][]{2005A&A...440..995S}. 
With the aid of these observations and theoretical calculations 
of, e.g., \cite{1993ApJ...416..355S} and 
\cite{1994ApJ...426..294H}, we selected spectral regions with 
the lowest effect of the intervening absorption. 
%
Second, we took into account the often very different errors of 
the measured fluxes for different spectra, which are also 
wavelength dependent. For example, in the short-wavelength 
part of the spectrum, especially below 1200\,\AA, the standard 
deviation of the flux measurement can be compared to the flux 
itself \footnote{However, there are spectra that are quite well 
exposed even below 1200\,\AA. In these cases the model matches 
also the short-wavelength wing of the continuum depression 
(e.g., PU~Vul and BF~Cyg, see Appendix~\ref{app:A}).}. 
Fortunately, the use of only the fluxes of the long-wavelength 
wing of the Rayleigh attenuated continuum is sufficient to 
determine the values of $N_{\rm H}^{\rm obs}$ 
\citep[e.g.,][]{1991A&A...249..173V,1995ApJ...442..366F}. 
Furthermore, the very extended absorption profile for 
$N_{\rm H}^{\rm obs} \gtrsim 2\times 10^{22}$\cmd\ causes 
a measurable Rayleigh scattering effect even for $\lambda > 1300$\,\AA\ 
(see Fig.~\ref{fig:appE} in Appendix~\ref{app:lya}), where the 
spectrum is in most cases sufficiently well determined to allow 
us to at least estimate the corresponding value of the H$^0$ 
column density. 
However, for some strongly underexposed spectra, it was possible 
to estimate only the upper limit of $N_{\rm H}^{\rm obs}$ 
or its wide range. 
Therefore, according to the quality of the used spectra, we 
proceeded to determine the $N_{\rm H}^{\rm obs}$ values 
as follows: 

%
(i) For well-exposed spectra, according to the signal-to-noise 
ratio and for simplicity, we set the errors to 7 -- 10\%\ for 
all selected continuum fluxes. In these cases, 
we calculated a grid of models (\ref{eq:uvsed}) for reasonable 
ranges of the fitting parameters and selected that corresponding 
to a minimum of the reduced $\chi^{2}$ 
function\footnote{The corresponding software and application 
example are available at 
\url{https://doi.org/10.5281/zenodo.7695015}}. 

(ii) In cases where the spectrum is underexposed, mostly 
in the far-UV region and/or heavily affected by the iron curtain 
absorptions, we compared the spectrum to a set of plausible 
synthetic models, 
($\lambda, B_{\lambda}(T_{\rm BB})exp[-\sigma_{\rm Ray}(\lambda)
N_{\rm H}^{\rm obs}], \varepsilon_{\lambda}({\rm H},T_{\rm e}$), 
and selected the best match by eye. 
 
(iii) Finally, for severely underexposed spectra of AS~338 and 
AS~296, we estimated only the upper limits or wide ranges of 
H$^0$ column densities (see Appendix~\ref{app:targets} and 
examples in the bottom row of Fig.~\ref{fig:sediue}). 
Some example models for our targets are shown in 
Fig.~\ref{fig:sediue} (Appendix~\ref{app:A}). 
Examples of the selected continuum fluxes for better- (PU~Vul) 
and poorly-exposed (AR~Pav) far-UV spectra with the 
corresponding models are illustrated in Fig.~\ref{fig:ebv2} 
(Appendix~\ref{app:ebv}). 
%

The most important parameter for the purpose of this paper is 
the column density $N_{\rm H}^{\rm obs}$ (Table~\ref{tab:nh}), 
and, in particular, its dependence on the orbital phase 
(Fig.~\ref{fig:nhfi}). 
Its uncertainty is estimated at $\sim$20 to $\sim$60\,\%, 
depending on the signal-to-noise ratio and the influence of other 
emission features around the Ly$\alpha$ transition. 
For illustration, the models corresponding to the upper and 
lower value of $N_{\rm H}^{\rm obs}$ are depicted by dotted 
lines in Figs.~\ref{fig:sediue} and \ref{fig:sedqa}. 
In addition, we also investigated the possible dependence of 
the $N_{\rm H}^{\rm obs}$ determination on interstellar 
reddening. Our analysis suggests that uncertainties in the 
$E_{\rm B-V}$ color excess may cause additional errors in the 
$N_{\rm H}^{\rm obs}$ determination of about 10\%\ (see 
Appendix~\ref{app:ebv}). Table~\ref{tab:nh} and Fig.~\ref{fig:nhfi} 
indicate the total $N_{\rm H}^{\rm obs}$ errors. 
For completeness, Table~\ref{tab:apar} (Appendix~\ref{app:par}) 
presents the values of 
other parameters determining the UV continuum. Their uncertainties 
would be estimated at $\sim$10 to $\sim$50\% of their best values 
\citep[see Sect.~3.3 of][]{2005A&A...440..995S}. 

\section{Results}
\label{s:results}
\subsection{Indication of the neutral wind region in 
             the orbital plane during outbursts}
\label{ss:res1}
\begin{figure*}
\begin{center}
\resizebox{16cm}{!}
          {\includegraphics[angle=-90]{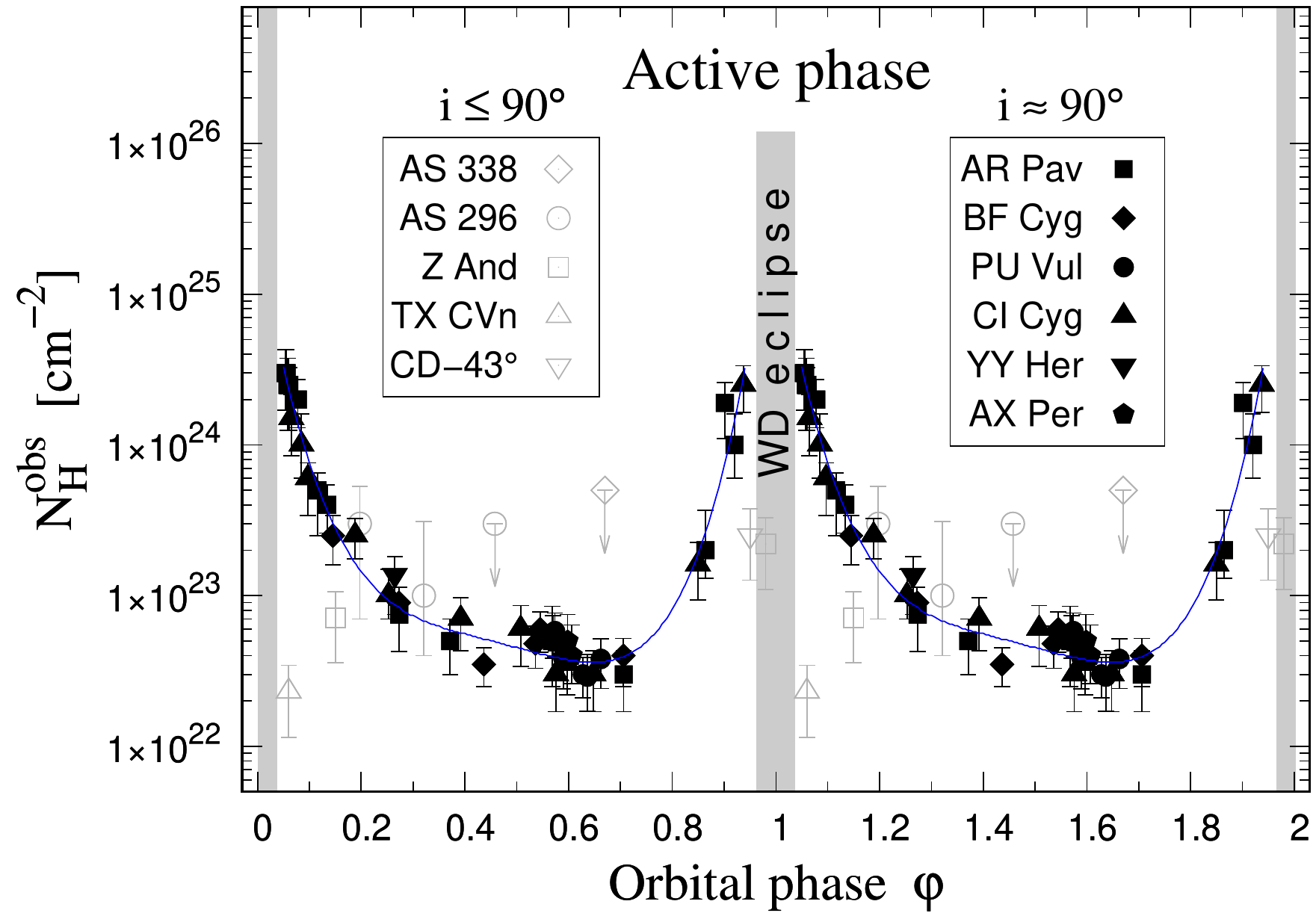}}
\end{center}
\caption{
Column densities of atomic hydrogen, $N_{\rm H}^{\rm obs}$, 
between the observer and the WD, measured for our targets 
(Sect.~\ref{ss:targets}) during active phases as a 
function of the orbital phase $\varphi$ ($\varphi = 0$ 
corresponds to the inferior conjunction of the giant). 
For better visualization, the values are plotted over 
two orbital cycles. 
Eclipsing objects (black solid symbols) follow a common 
course. 
The blue line indicates their fit with a 4th-degree polynomial. 
Open gray symbols denote values for objects discussed 
in Appendix~\ref{app:targets}, i.e., for those with poorly defined 
UV spectra (AS~338, AS~296) or not eclipsing, but with a high 
orbital inclination (Z~And, TX~CVn, CD-43$^{\circ}$14304 
(CD-43$^{\circ}$ in the legend)). 
Values for the latter were taken from \cite{2005A&A...440..995S}. 
All data measured in this work are summarised in 
Table~\ref{tab:nh}, and examples of the corresponding 
continuum models are shown in Appendix~\ref{app:A}. 
          }
\label{fig:nhfi}
\end{figure*}

The main result of this study is depicted in Fig.~\ref{fig:nhfi}: 
The $N_{\rm H}^{\rm obs}$ values for our targets 
show a strong dependence on the orbital 
phase, with a difference of about two orders of magnitude 
between the values around the inferior and superior conjunction 
of the giant. 
High values of $N_{\rm H}^{\rm obs}$ at any position of the 
binary reflects the presence of a neutral region in the orbital 
plane, because our targets are seen $\sim$edge-on. 
Its emergence is transient, being connected solely with 
active phases (Sect.~\ref{ss:res2}, Appendix~\ref{app:B}). 
The values for eclipsing systems (i.e., BF~Cyg, CI~Cyg, YY~Her, 
AR~Pav, AX~Per and PU~Vul) follow a common course that suggests 
similar properties of the neutral zone for these objects. 
In contrast, during quiescent phases of eclipsing systems, 
the neutral hydrogen zone is limited only to the vicinity of 
the inferior conjunction of the giant, where it represents 
the neutral wind from the giant 
\citep[e.g.,][]{
1991A&A...249..173V,
1999A&A...349..169D,
2017A&A...602A..71S}. 
Therefore, a comparison of H$^0$ column densities from quiescent 
and active phases at these positions (see Fig.~\ref{fig:qa}, 
panel b and e) suggests that our $N_{\rm H}^{\rm obs}$ values 
also represent the densities of the stellar wind from the giant. 

The case of measuring the H$^0$ column densities in the direction 
to the WD, while the source of neutral hydrogen is associated 
with the red giant, will give a large difference between their 
values measured around the inferior and the superior conjunction 
of the giant, which is in agreement with our result 
(see Fig.~\ref{fig:nhfi}). 
Thus, the high-amplitude orbital-dependent variation of 
$N_{\rm H}^{\rm obs}$ values confirms that we measure H$^0$ 
column densities of the neutral wind from the red 
giant\footnote{If the broad Ly$\alpha$ line profile was formed 
in the WD atmosphere, there would be no reason to measure the 
dependence of the $N_{\rm H}^{\rm obs}$ values on the orbital 
phase (see the last paragraph of Appendix~\ref{app:lya}).}. 
%
%
%
%
\subsection{The cause of the neutral region during outbursts} 
\label{ss:res2}
The emergence of the neutral wind region in the orbital plane 
is conditioned by the formation of an optically thick disk-like 
structure around the WD in the orbital plane during outbursts 
(see Sect.~\ref{s:intro}). 
Consequently, the disk blocks the ionizing radiation from the 
central burning WD within its vertical extension from the 
orbital plane, which makes the wind of the giant neutral in the 
orbital plane, thus making it detectable by eclipsing systems in 
any orbital phase. 
During active phases, the wind from the giant represents 
a cool absorbing gas veiling the WD's pseudophotosphere in 
the orbital plane. 
Therefore, the $N_{\rm H}^{\rm obs}$ values of this medium show 
a clear dependence on the orbital phase, with a high amplitude 
(Fig.~\ref{fig:nhfi}, Sect.~\ref{ss:res1}). 

The presence of the neutral giant's wind in the orbital plane is 
primarily indicated by the deep continuum depression in the 
spectrum around the Ly$\alpha$ line due to Rayleigh scattering on 
H$^0$ atoms, and by broad absorption bands due to iron curtain 
(see Fig.~\ref{fig:sediue})\footnote{The prominence of the iron 
curtain in the spectrum varies from object to object, depending 
mainly on the temperature of the WD's pseudophotosphere.}, 
the features that are absent during quiescent phases around the 
whole orbit. For our targets, we demonstrate the corresponding 
change of the UV spectrum between the quiescent and active phase 
in Fig.~\ref{fig:sedqa} (Appendix~\ref{app:B}). 

\subsection{Wind asymmetry}
\label{ss:res3}
\begin{figure*}
\begin{center}
\resizebox{\hsize}{!}
          {\includegraphics[angle=-90]{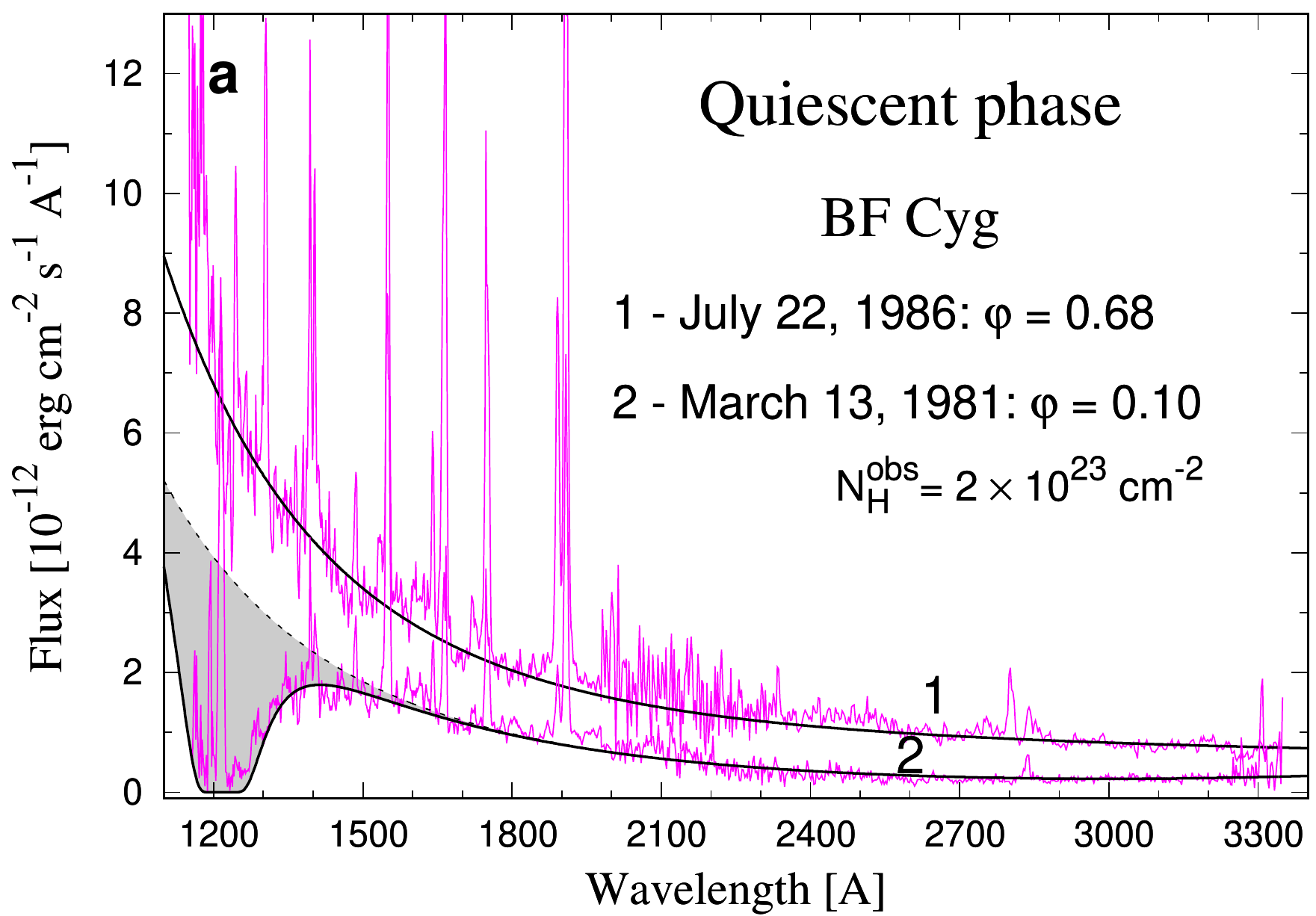}
           \includegraphics[angle=-90]{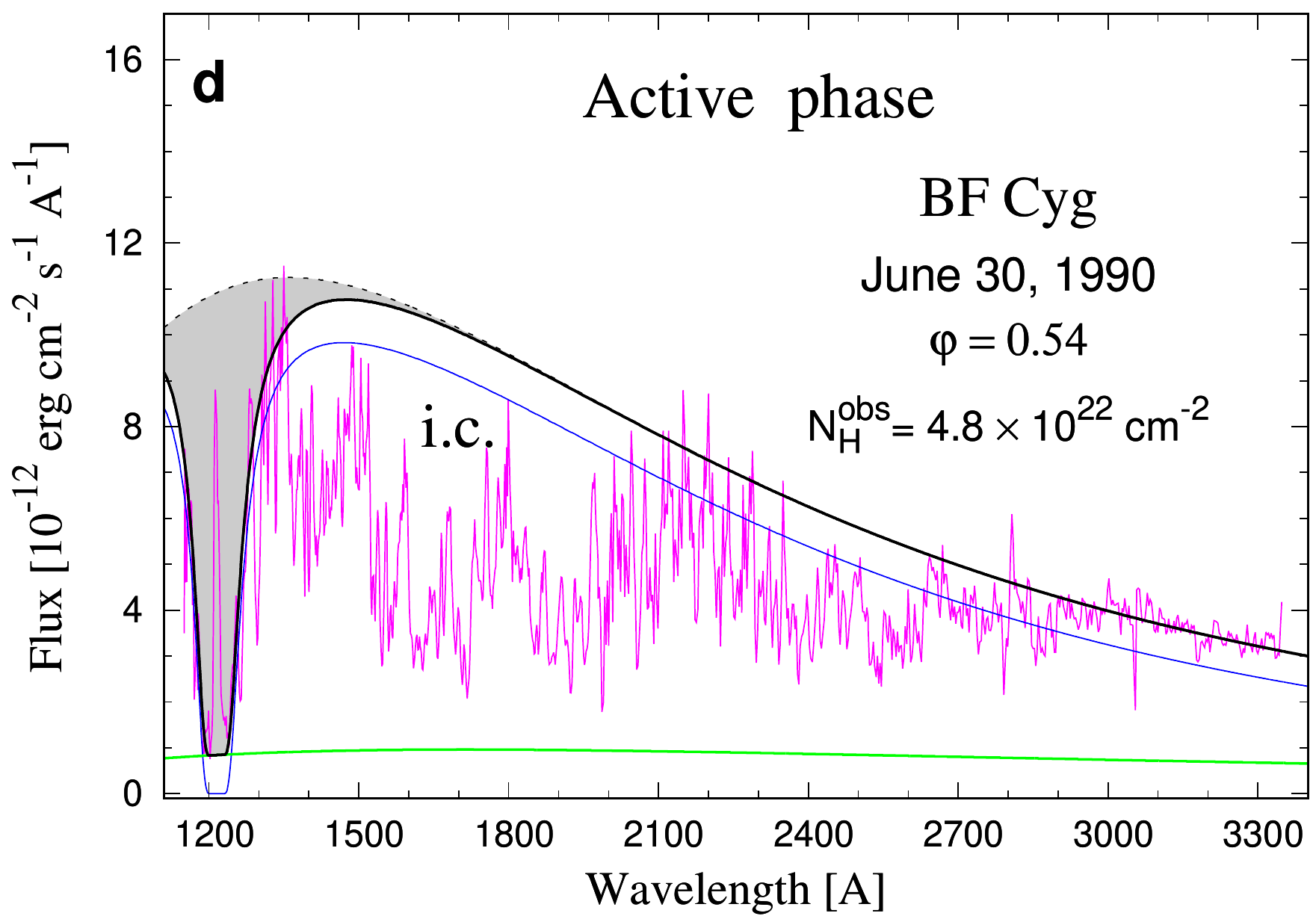}}
\vspace{2mm}
\resizebox{\hsize}{!}
          {\includegraphics[angle=-90]{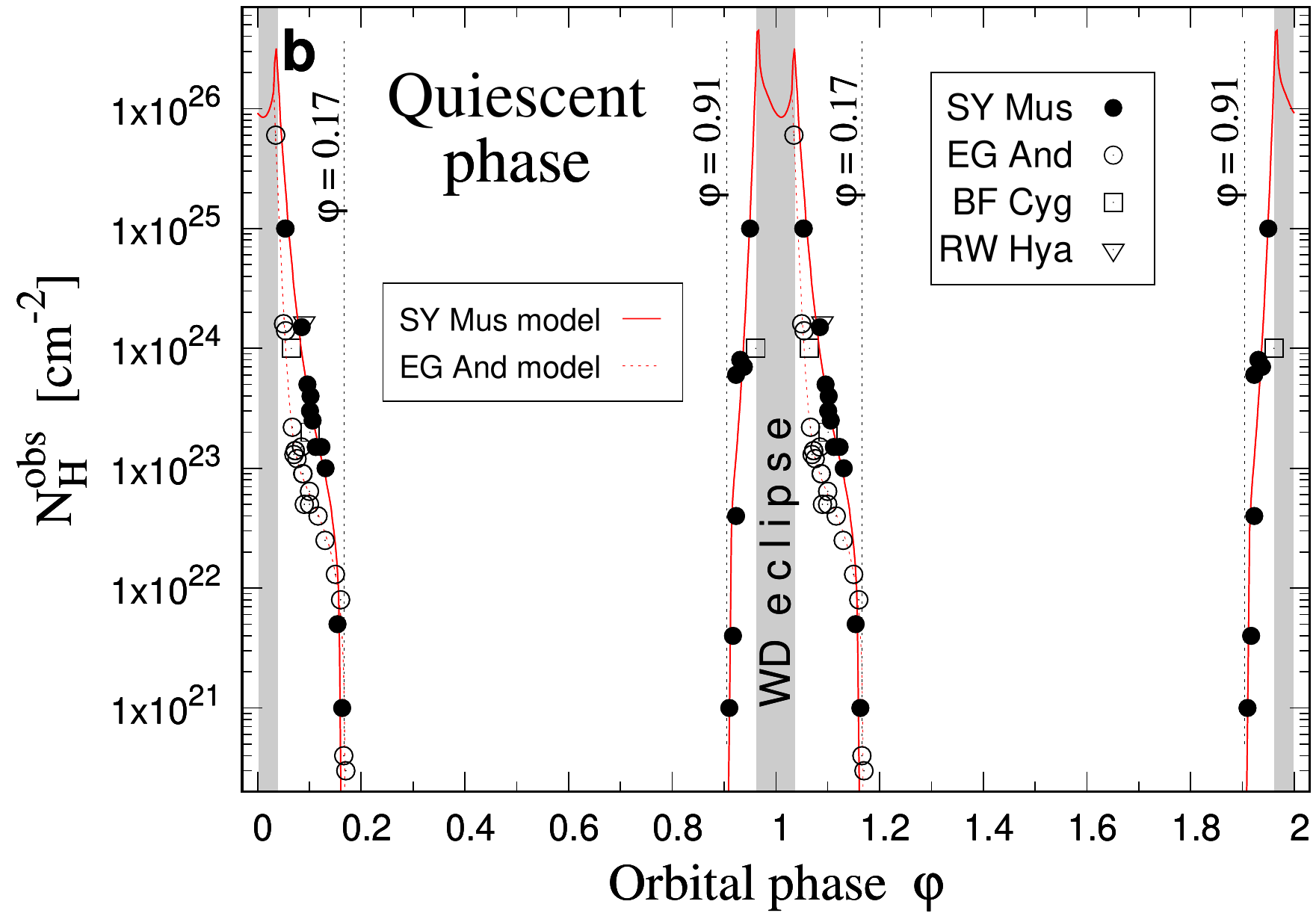}
           \includegraphics[angle=-90]{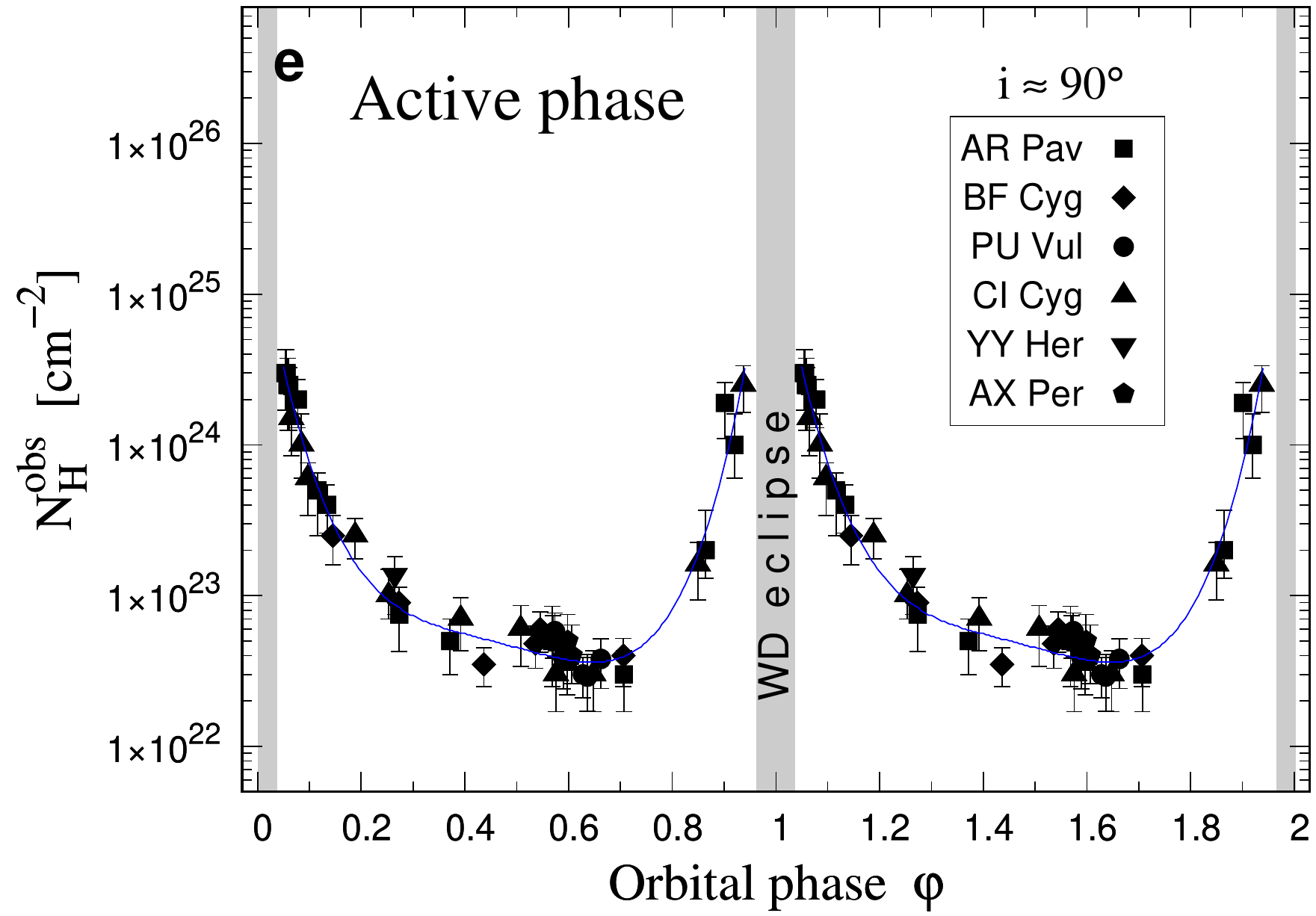}}
\vspace*{2mm}
\resizebox{\hsize}{!}
          {\includegraphics[angle=-90]{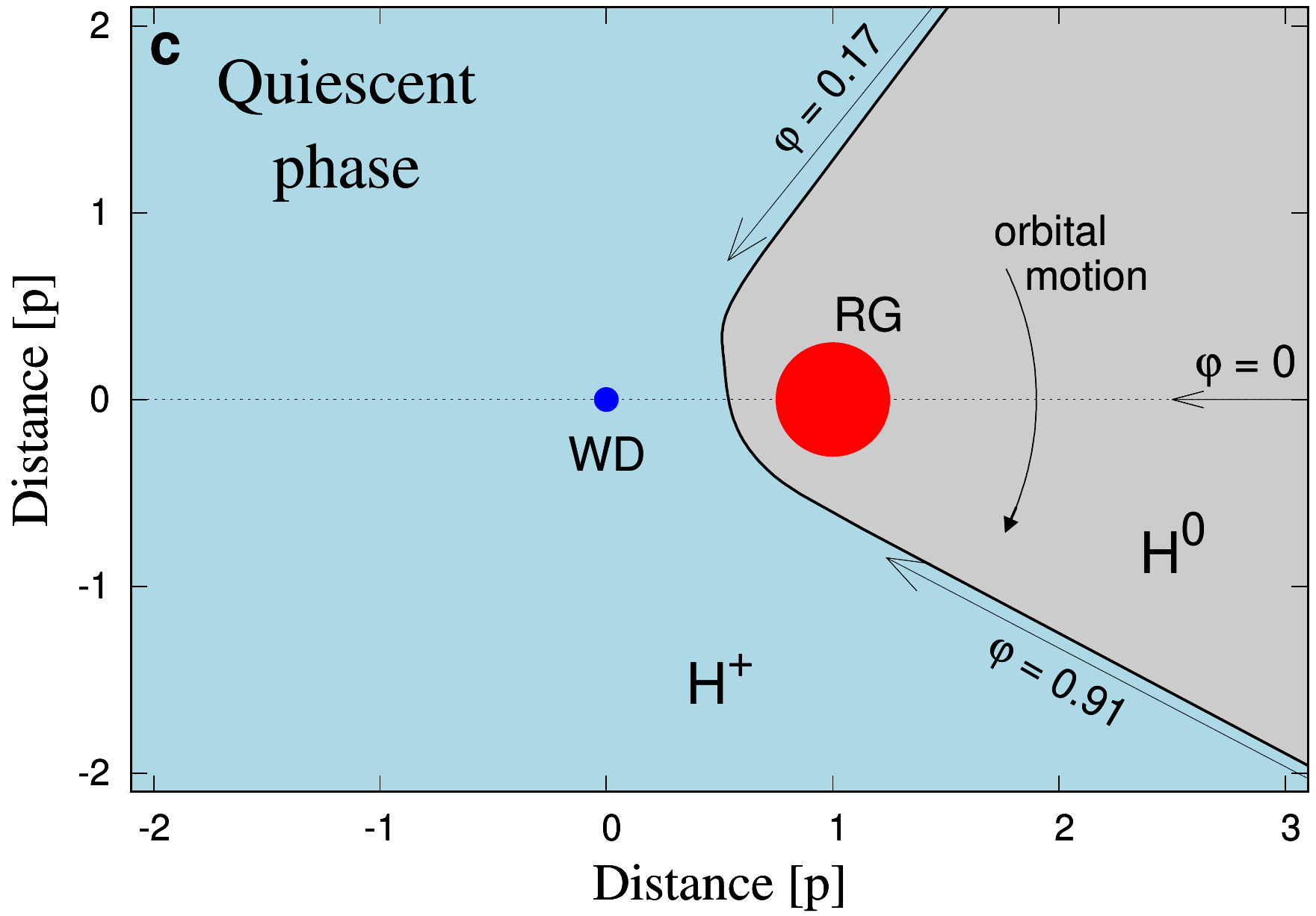}
           \includegraphics[angle=-90]{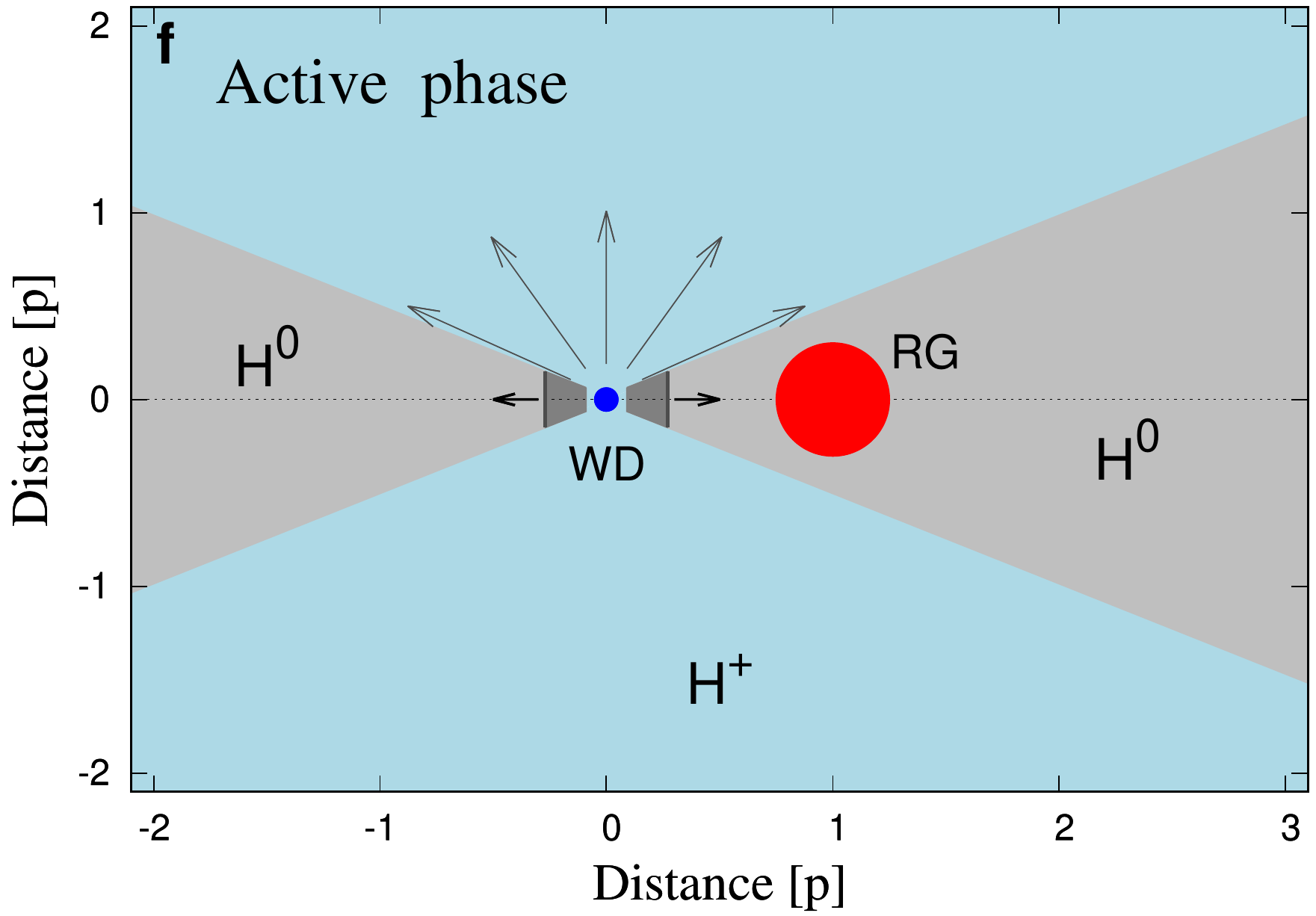}}
\end{center}
\caption{The symbiotic phenomenon in a nutshell. Characteristic 
UV spectra (top panels), hydrogen column densities 
$N_{\rm H}^{\rm obs}$ (middle panels), and sketch of the 
corresponding ionization structure (bottom panels) for eclipsing 
SySts (see keys) during quiescent and active phases are shown in 
the left and right columns, respectively. 
During the quiescent phase, the ionization structure is given by 
the location of $N_{\rm H}^{\rm obs}$ values only in the vicinity 
of the inferior conjunction of the giant 
\citep[pole-on view; see][]{2017A&A...602A..71S}, while 
during the active phase a biconical ionization structure is 
constrained by the H$^0$ wind region in the orbital plane 
(edge-on view; Sect.~\ref{ss:iona}). The two-velocity type of 
mass outflow during outbursts is denoted by arrows of different 
sizes (Sect.~\ref{ss:iona}). 
In both cases, the asymmetric profile of $N_{\rm H}^{\rm obs}(\varphi)$ 
(panels {\bf b} and {\bf e}) imprints the asymmetric wind from 
the giant. 
Spectra from the quiescent phase (panel {\bf a}) are 
introduced in Table~\ref{tab:qpar} (Appendix~\ref{app:par}). 
Description of SED models as in Appendix~\ref{app:A}, and 
column densities in panel {\bf e} as in Fig.~\ref{fig:nhfi}. 
The data and models in panel {\bf b} are from the literature 
(see Sect.~\ref{ss:ionq}), i.c. denotes the continuum 
depression caused by the iron curtain (see Sect.~\ref{ss:ray}), 
and p is in units of the binary components separation. 
Other denotations are obvious. 
          }
\label{fig:qa}
\end{figure*}

Figure~\ref{fig:nhfi} shows the asymmetric distribution 
of $N_{\rm H}^{\rm obs}$ values along the orbit with a minimum 
of $\sim 4\times 10^{22}$\,cm$^{-2}$ between 
$\varphi\sim$0.6 and $\sim$0.7 and a maximum of a few 
times $10^{24}$\,cm$^{-2}$ just before and after eclipses 
($\varphi \sim0.95$ and $\sim0.03$). 
The asymmetry is also evidenced by the steeper values of 
$N_{\rm H}^{\rm obs}$ before the eclipse ($0.85 < \varphi < 0.95$) 
than after the eclipse ($\varphi > 0.05$), and 
a small local maximum around $\varphi = 0.5$. 
Such the asymmetric distribution of $N_{\rm H}^{\rm obs}$ along 
the orbit implies the asymmetric density distribution of the 
giant's wind in the near-orbital-plane region with respect to 
the binary axis. 
This property of the wind was previously indicated also for 
quiescent eclipsing SySts SY~Mus and EG~And by the same type 
of the asymmetric distribution of H$^0$ column densities, 
although they are measurable only around the inferior 
conjunction of their giants in quiescent phases 
\citep[][ here Fig.~\ref{fig:qa}b, Sect.~\ref{ss:ionq}]{
1999A&A...349..169D,2016A&A...588A..83S}. 
Recently, this sort of asymmetric shaping of the wind in 
the orbital plane was confirmed by modeling the UV light curves 
of SY~Mus \citep[][]{2017A&A...602A..71S}, and by the H$\alpha$ 
line profile variation along the orbit of EG~And 
\citep[][]{2021A&A...646A.116S}. 
Therefore, the orbital-dependent asymmetry of 
$N_{\rm H}^{\rm obs}$ values measured for more objects and during 
the active phases (Fig.~\ref{fig:nhfi}) suggests that the 
asymmetric distribution of the giant's wind in the orbital 
plane with respect to the binary axis may possibly be a common 
property of winds from giants in SySts. 

Its origin could in principle be similar to that simulated 
and measured for detached high-mass X-ray binaries consisting 
of a hot O/B (super)giant and a neutron star or a black hole, 
where a similar type of orbital-related asymmetry in 
N$_{\rm H}$ is attributed to tidal streams and accretion wakes, 
distorted by the orbital motion 
\citep[e.g.,][]{1990ApJ...356..591B,
                1991ApJ...371..684B,
                2012A&A...547A..20M,
                2021A&A...652A..95K,
                2023arXiv230210953R}. 
%
Modeling of our $N_{\rm H}^{\rm obs}(\varphi)$ values in 
this way, but for parameters of S-type SySts, should clarify 
the cause of their asymmetric distribution throughout 
the orbit. 
%
%
\section{Discussion}
\label{s:discuss}
Here we describe how our findings complement the overall picture 
of the symbiotic phenomenon. For this purpose, we present the 
characteristic UV spectra of SySts, the measured H$^0$ column 
densities as a function of the orbital phase, and discuss how 
they relate to the ionization structure of SySts during both, 
the quiescent and active phases. 
Figure~\ref{fig:qa} summarizes our view. 

\subsection{Quiescent phase}
\label{ss:ionq}
During quiescent phases, the far-UV spectrum is dominated by 
the Rayleigh-Jeans tail of a very hot source of the stellar 
radiation ($T_{\rm BB}\gtrsim 10^5$\,K), while the near-UV is 
usually dominated by the nebular continuum 
\citep[e.g.,][]{1991A&A...248..458M,2005A&A...440..995S}. 
The effect of Rayleigh scattering on hydrogen atoms, and overall 
decrease of the continuum level is measured only for eclipsing 
systems in the vicinity of the inferior conjunction of the giant 
as a result of the passage of light from the WD through the 
absorbing medium of the neutral wind from the giant 
\citep[e.g.,][]{1989A&A...219..271I,1994ApJ...426..294H}.
%
%
Here, an example of BF~Cyg is depicted in Fig.~\ref{fig:qa}a. 
%
Figure~\ref{fig:qa}b then shows all $N_{\rm H}^{\rm obs}$ 
values of eclipsing SySts determined during their 
quiescent phases: 
SY~Mus and EG~And 
\citep[][]{1999A&A...349..169D,2016A&A...588A..83S}, 
BF~Cyg 
\citep[][]{1996AJ....111.1329P}, 
and RW~Hya 
\citep[][]{1999A&A...349..169D}. 
%
Accordingly, the observed distribution of H$^0$ column densities 
along the orbit suggests that the neutral zone is located around 
the giant and beyond it in the direction away from the WD, and is 
distributed asymmetrically with respect to the binary axis, while 
the rest of the binary environment around the WD is ionized. 
Example of SY~Mus that corresponds to the 
$N_{\rm H}^{\rm obs}(\varphi)$ model depicted in 
Fig.~\ref{fig:qa}b \citep[red line,][]{2016A&A...588A..83S} 
is shown in Fig.~\ref{fig:qa}c 
\citep[adapted according to][]{2017A&A...602A..71S}. 
A simplified calculation of 
the ionization H$^0$/H$^{+}$ boundary during quiescent phase 
is introduced by 
\cite{1984ApJ...284..202S} and \cite{1987A&A...182...51N}. 

\subsection{Active phase}
\label{ss:iona}
During active phases, the UV spectrum of eclipsing SySts can 
be fitted by a low-temperature stellar-type of radiation 
($T_{\rm BB}\approx 1-3\times 10^4$\,K) emitted by the warm 
WD's pseudophotosphere often dominating the far-UV, and 
a strong nebular continuum pronounced in the near-UV 
(see examples in Appendix~\ref{app:A} and \ref{app:B}). 
The former is strongly attenuated around the Ly$\alpha$ line 
by Rayleigh scattering on H$^0$ atoms at any orbital phase, 
and shows signatures of the iron curtain absorptions, while 
the latter increases compared to the quiescent phase 
(Fig.~\ref{fig:qa}d, and Appendix~\ref{app:A} and \ref{app:B}). 
The corresponding high values of $N_{\rm H}^{\rm obs}$ along 
the orbit (Fig.~\ref{fig:nhfi}) indicate the presence of 
the neutral region in the orbital plane 
(Sect.~\ref{s:results})\footnote{During outbursts of 
non-eclipsing systems, the neutral wind region in the orbital 
plane is indicated by significant broadening and high fluxes 
of the Raman-scattered \ovi\,6825\,\AA\ line relative 
to the quiescent phase. This is because the new neutral hydrogen 
in the orbital plane increases the Raman scattering 
efficiency \citep[see][]{2020A&A...636A..77S}.}. 
%
Such the neutral zone determines a biconical shaping of the 
ionized region distributed above/below it with the tops at 
the burning WD (see Fig.~\ref{fig:qa}f). 
In this case, the nebular radiation is produced by the ionized 
wind from the WD \citep[][]{2006A&A...457.1003S}. 
Such an ionization structure is responsible for observing 
two different types of spectra depending on the orbital 
inclination. 
This suggests a classification of outbursts into two types -- 
the warm- and hot-type 
\citep[see also][]{2005A&A...440..995S,2020A&A...636A..77S}: 

(i) Outbursts in systems with a high orbital inclination 
(i.e., seen $\sim$\,edge-on) show the two-temperature type of 
the hot component spectrum (see Sect.~\ref{s:intro}). These 
outbursts are classified as the `warm-type', because the stellar 
component of radiation is emitted by the warm ($1-3\times 10^4$\,K) 
WD's pseudophotosphere. 
During the warm-type of outbursts, the two-velocity type of mass 
outflow from the hot component is indicated. 
A rather slow outflow is indicated by the absorption component 
in P-Cygni line profiles created in the expanding optically 
thick wind from the WD in the orbital plane, while a fast 
outflow is indicated by the broad emission wings produced by 
the fast optically thin ionized wind from the WD over 
the remainder of the star 
\citep[see Fig.~\ref{fig:qa}f, and][ in detail]{
2006A&A...453..279S}. 
All our targets show signatures of the warm-type of outbursts. 

(ii) Outbursts in systems with a low orbital inclination 
(i.e., seen $\sim$\,pole-on) show the spectrum characterized 
by the immediate occurrence of strong nebular radiation that 
dominates the optical. These outbursts are classified as 
`hot-type', because the stellar component of radiation is 
emitted by the hot ($\sim2\times 10^5$\,K) WD's pseudophotosphere, 
the contribution of which is negligible in the optical 
\citep[e.g.,][]{2020A&A...636A..77S}. 
The strong nebular radiation is emitted by the enhanced 
ionized wind from the WD \citep[][]{2006A&A...457.1003S}. 
Here, the well-observed SySts, AG~Dra, AG~Peg, LT~Del and 
V426~Sge show features of the hot-type of outbursts 
\citep[e.g.,][]{2008A&A...481..725G,2016MNRAS.456.2558L,
2017A&A...604A..48S,
2019AstL...45..217I,
2020A&A...636A..77S}. 
\section{Conclusion and future work}
\label{s:concl}
We have found that a deep and wide absorption is formed around 
the Ly$\alpha$ line during outbursts of all eclipsing symbiotic 
binaries around the whole orbit, for which there are relevant 
observations (Sect.~\ref{ss:targets}). Modeling this feature by 
Rayleigh scattering on atomic hydrogen we determined 
the corresponding H$^0$ column densities in the direction 
of the WD (Fig.~\ref{fig:nhfi}). 
The high values of $N_{\rm H}^{\rm obs}$ measured at any orbital 
phase indicate the presence of a neutral region in the orbital 
plane. Their large-amplitude orbital-dependent variation indicates 
that this region consists of the wind from the giant 
(Sect.~\ref{ss:res1}). 
It is observable due to the emergence of a dense 
disk-like structure around the WD during outbursts 
\citep[][]{2005A&A...440..995S,2012A&A...548A..21C} 
that blocks the ionizing photons from the central hot WD 
in the orbital plane (Sect.~\ref{ss:res2}). 
The asymmetric course of $N_{\rm H}^{\rm obs}(\varphi)$ values 
implies an asymmetric density distribution of the wind from 
the giant in the orbital plane with respect to the binary 
axis (Sect.~\ref{ss:res3}). 
Finally, the emergence of the neutral near-orbital-plane region 
changes significantly the ionization structure of the symbiotic 
binary during active phases: The ionized region is distributed 
above/below it, having the tops at the burning WD 
(Sect.~\ref{ss:iona}, Fig.~\ref{fig:qa}f). 
%

The unique distribution of neutral hydrogen in the orbital plane 
of symbiotic binaries as indicated for different objects during 
different stages of their activity (Fig.~\ref{fig:nhfi}, 
Appendix~\ref{app:B}) suggests common properties of stellar 
winds from their giants. 
Our findings can aid us in further theoretical modeling and can 
be conducive to the explanation of more violent classical nova 
outbursts. For example: 
\begin{enumerate}
\item
Modeling $N_{\rm H}^{\rm obs}$ values along the whole orbit for 
more objects during active phases should confirm and generalize 
the substantial focusing of the wind from the giant in S-type 
SySts towards the orbital plane as found for quiescent SySts 
SY~Mus and EG~And by \cite{2016A&A...588A..83S,2021A&A...646A.116S}. 
\item
The $N_{\rm H}^{\rm obs}$ values measured around the orbit will 
serve as a benchmark for testing the theoretical modeling of the wind 
morphology of wide interacting binaries containing an evolved giant 
\citep[e.g.,][]{2020MNRAS.493.2606B,
                2020A&A...637A..91E,
                2022ApJ...931..142L}. 
In particular, modeling the morphology of a massive slow wind 
blowing from red giants in S-type SySts with terminal velocity 
of a few times 10\kms\ \citep[e.g.,][]{1999A&A...349..169D}, at 
rates of a few times ($10^{-7} - 10^{-6}$)\myr\ 
\citep[e.g.,][]{1993ApJ...410..260S,2016A&A...588A..83S}, which 
is disrupted by the accreting WD, in a similar way to what has 
been done for massive X-ray binaries (see references in 
Sect.~\ref{ss:res3}), should lead to a better understanding of 
the mass-transfer problem in symbiotic binaries. 
\item
The emergence of a slowly expanding disk-like structure 
around the exploding WD 
appears to be common also for more violent outbursts of 
classical novae, where it is indicated by direct radio imaging 
\citep[][]{2014Natur.514..339C} and is constrained by modeling 
the energy distribution in the nova spectrum 
\cite[][]{2019ApJ...878...28S}. 
During outbursts of SySts, the flared disk blocks the ionizing 
radiation from the center resulting in the creation of the neutral 
near-orbital-plane region throughout the binary. During the 
classical nova explosions, the disk represents a vital element 
to explain the $\gamma$-ray emission measured near the optical 
maximum as a result of shocks developing between the first slow 
and secondary fast outflow \citep[e.g.,][]{2017NatAs...1..697L}. 
Also, the disk structure facilitates the creation of dust and 
its persistence during extreme conditions of the super-soft 
X-ray phase \citep[e.g.,][]{2019ApJ...878...28S}. 
\end{enumerate}
%
%
\begin{acknowledgments}
I thank my colleague J\'{a}n Budaj for having calculated for 
me the Ly$\alpha$ profiles discussed in Appendix~\ref{app:lya}. 
This work was supported by a grant of the Slovak Academy 
of Sciences, VEGA No. 2/0030/21, and by the Slovak Research 
and Development Agency under contract No. APVV-20-0148. 
This research is based on observations made with the 
{\it International Ultraviolet Explorer}, obtained from 
the MAST data archive at the Space Telescope Science Institute, 
which is operated by the Association of Universities for 
Research in Astronomy, Inc., under NASA contract NAS 5-26555. 
\end{acknowledgments}
\facilities{IUE(SWP, LWP), AAVSO}

The data presented in this paper were obtained from the Mikulski 
Archive for Space Telescopes (MAST) at the Space Telescope Science 
Institute. The specific observations analyzed can be accessed via 
\dataset[10.17909/0mkj-f579]{https://doi.org/10.17909/0mkj-f579}. 
%

\appendix
%
%
\section{Indication of neutral hydrogen in the orbital plane of 
         eclipsing symbiotic stars during their active phases}
\label{app:A}
\begin{figure*}
\begin{center}
%
\resizebox{\hsize}{!}
          {\includegraphics[angle=-90]{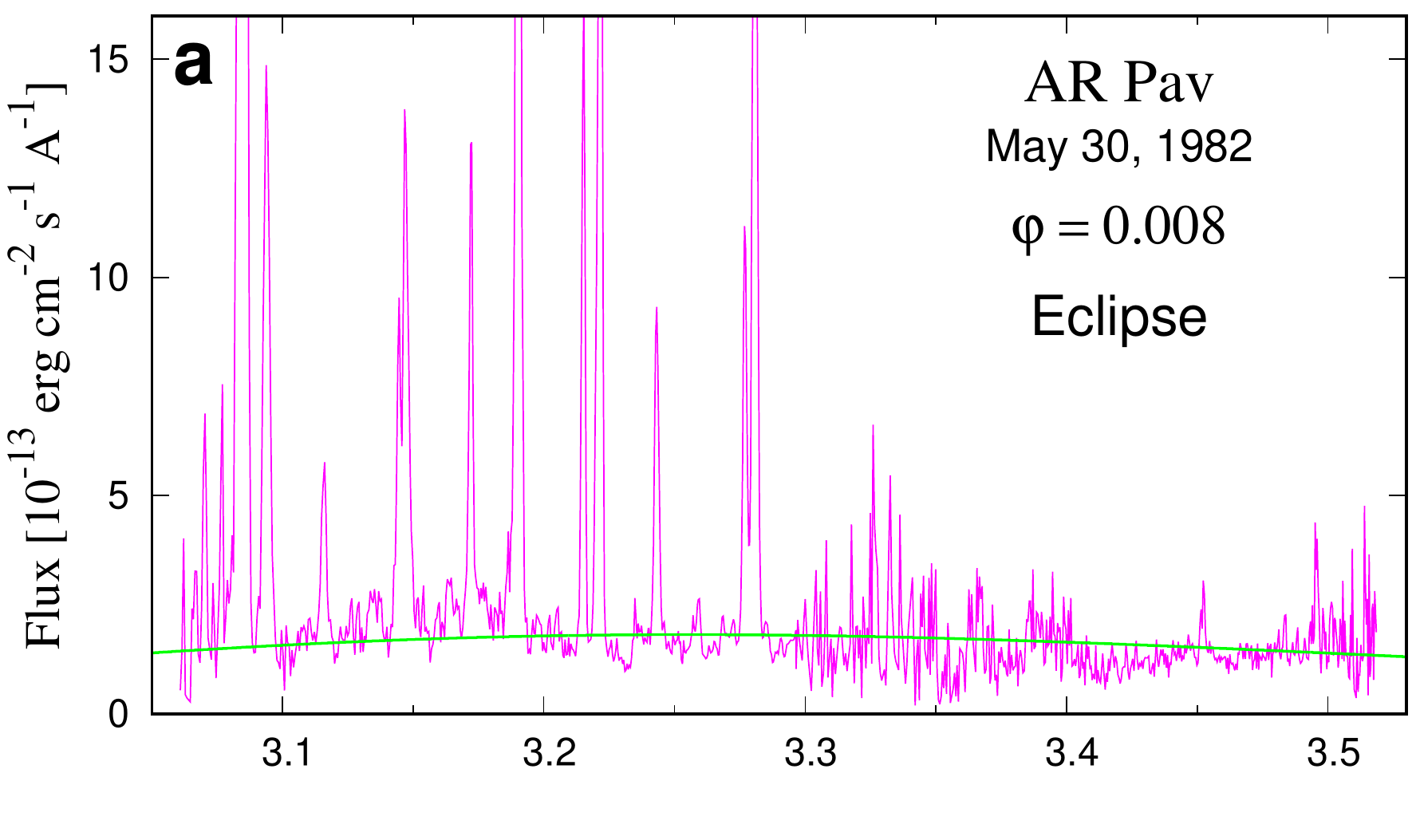} 
           \includegraphics[angle=-90]{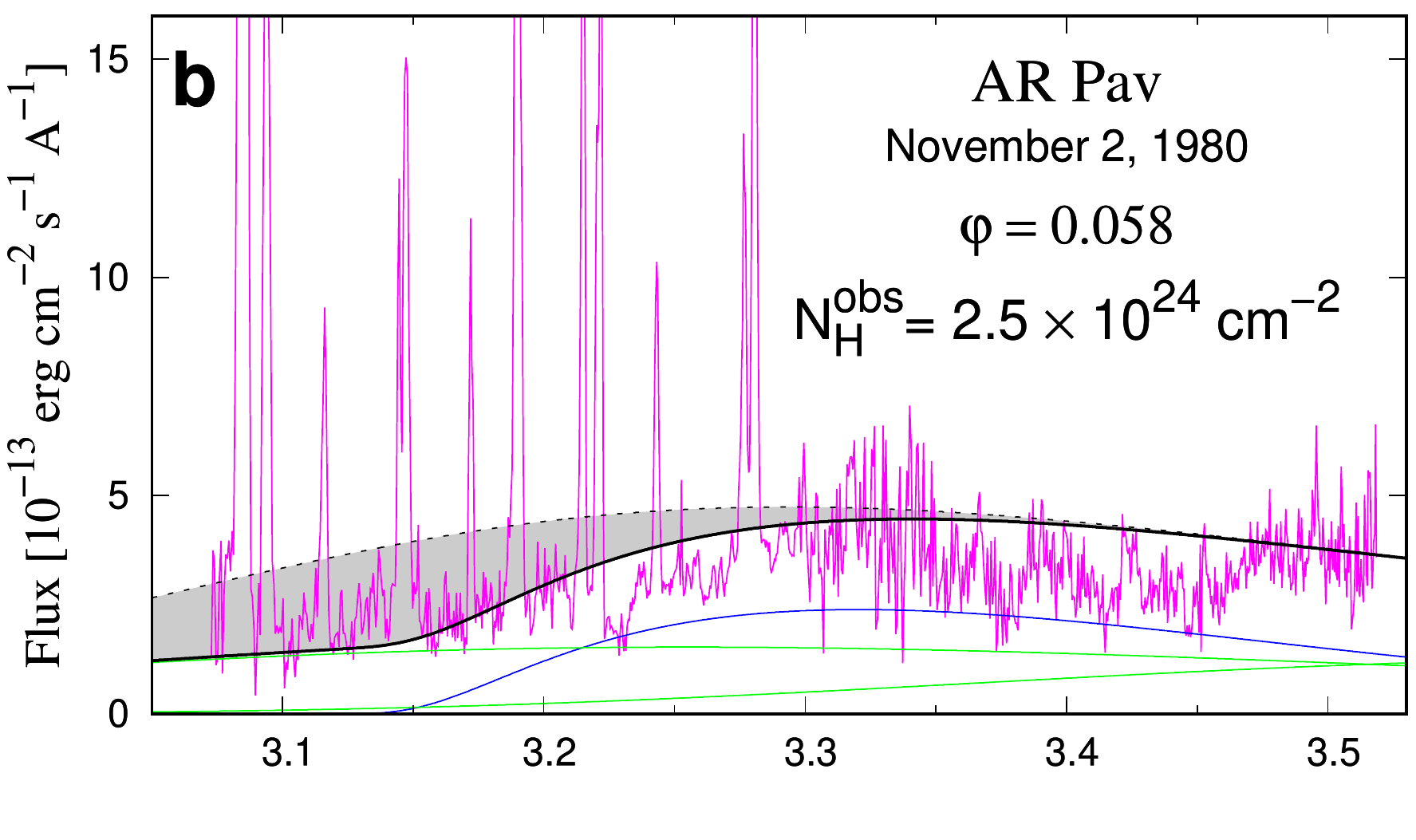} 
\includegraphics[angle=-90]{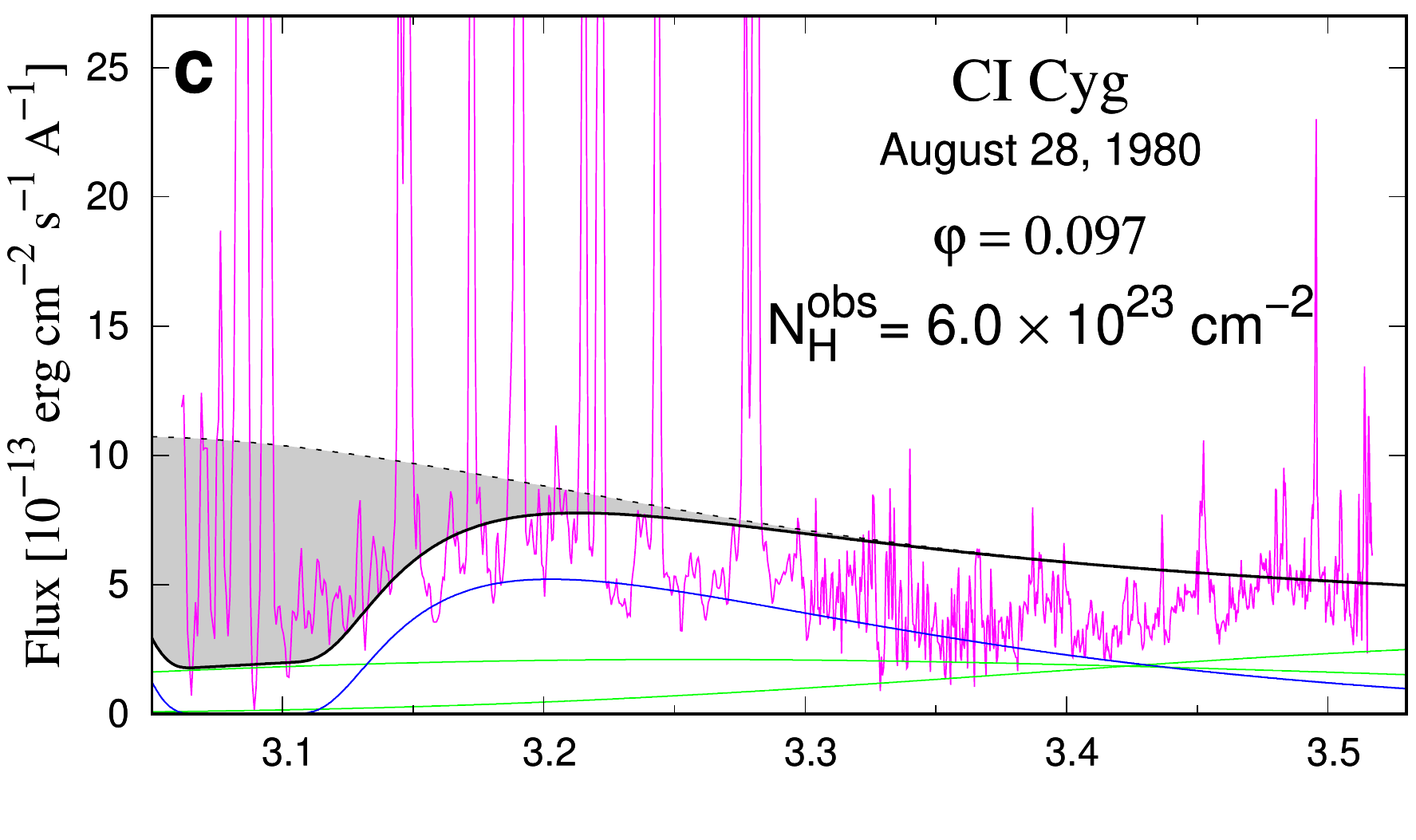}}\vspace*{-1mm} 
\resizebox{\hsize}{!}
          {\includegraphics[angle=-90]{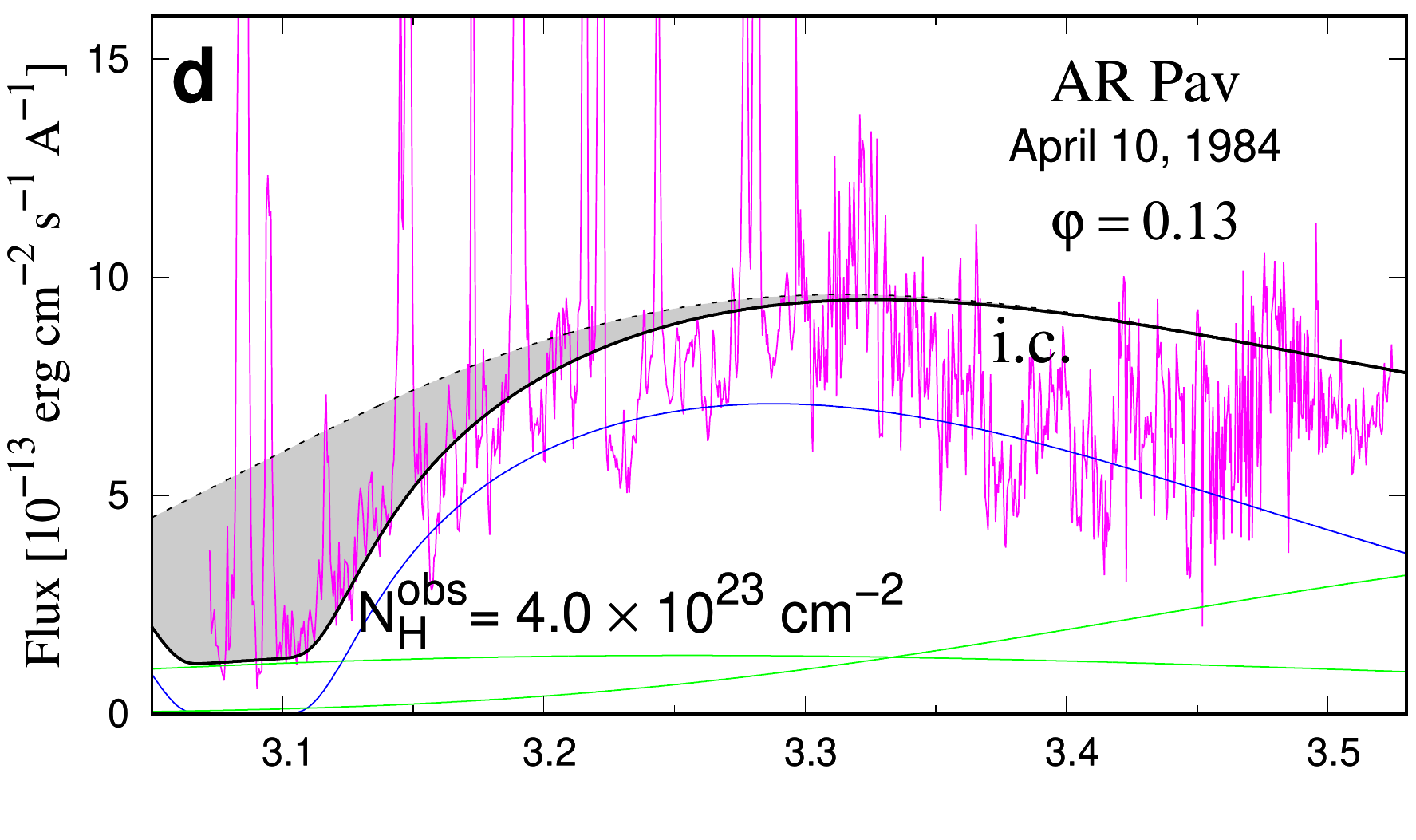} 
           \includegraphics[angle=-90]{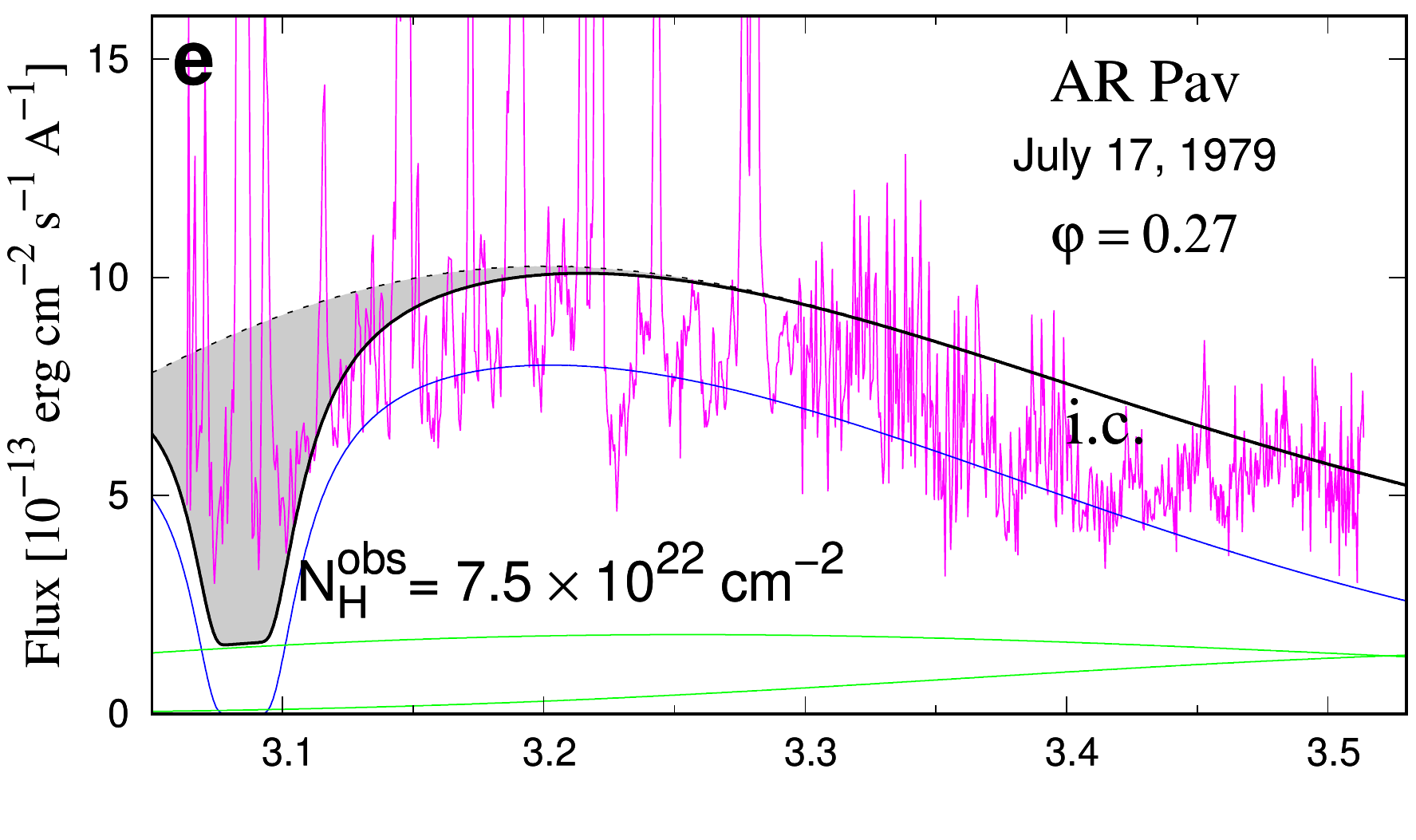} 
\includegraphics[angle=-90]{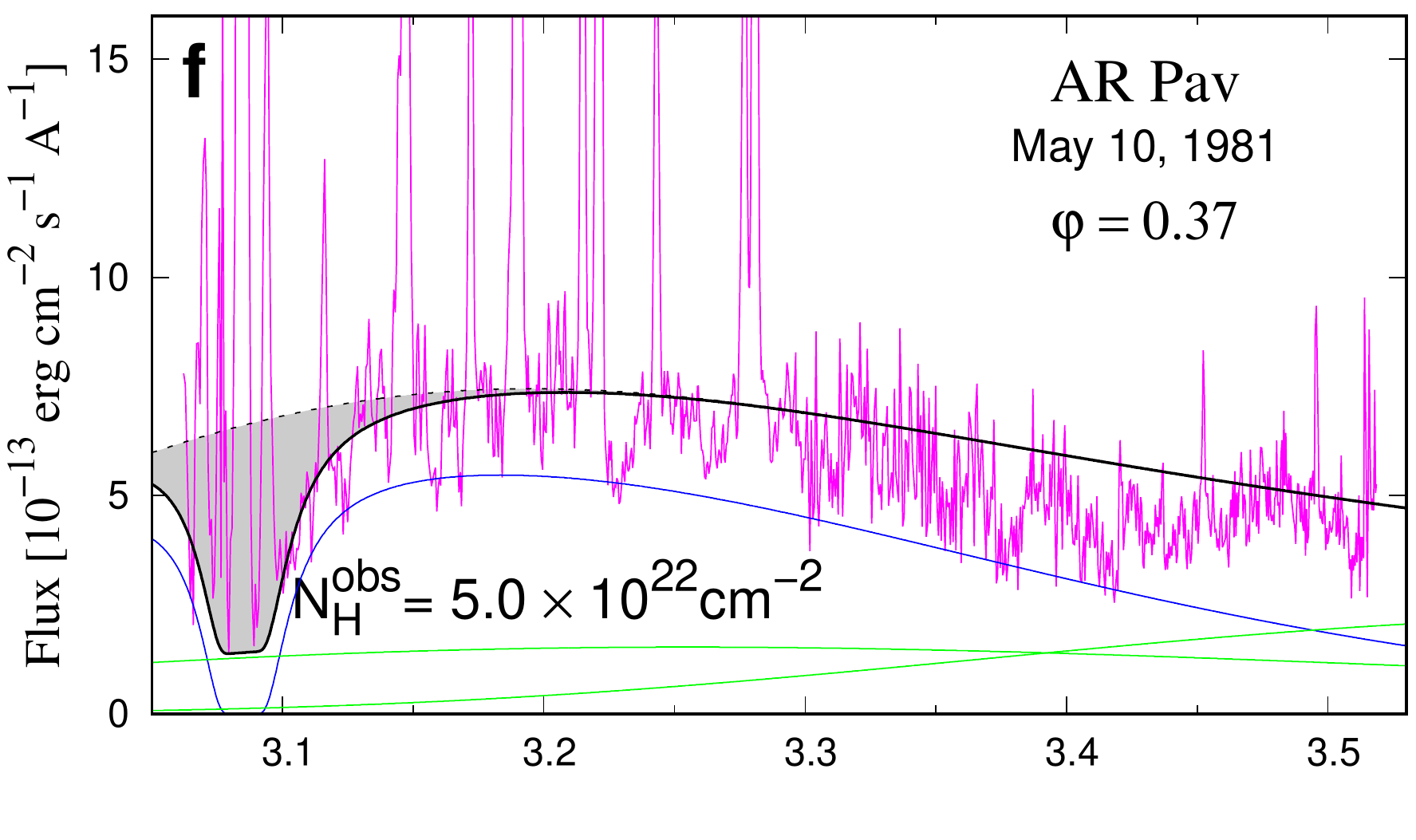}}\vspace*{-1mm} 
\resizebox{\hsize}{!}
          {\includegraphics[angle=-90]{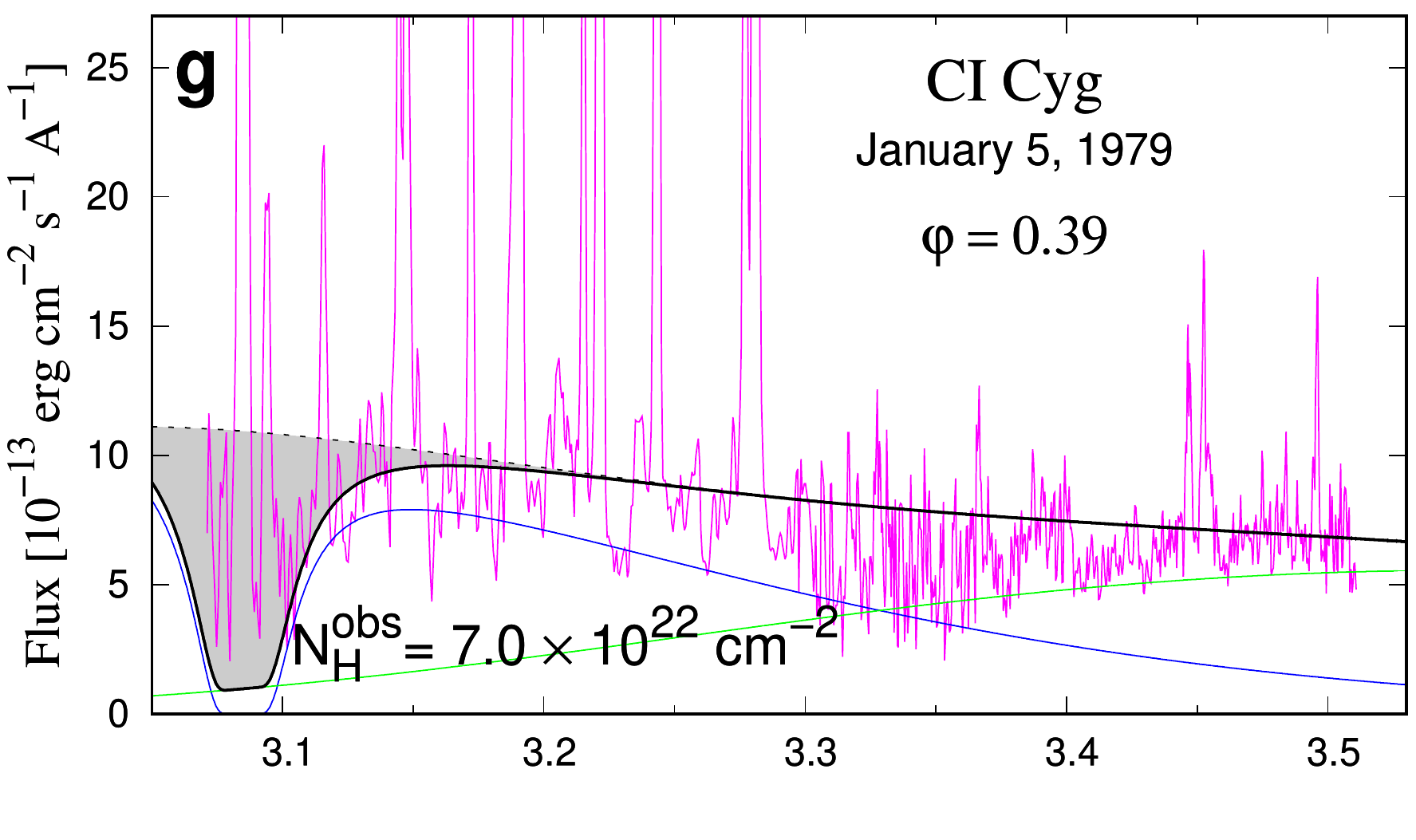} 
           \includegraphics[angle=-90]{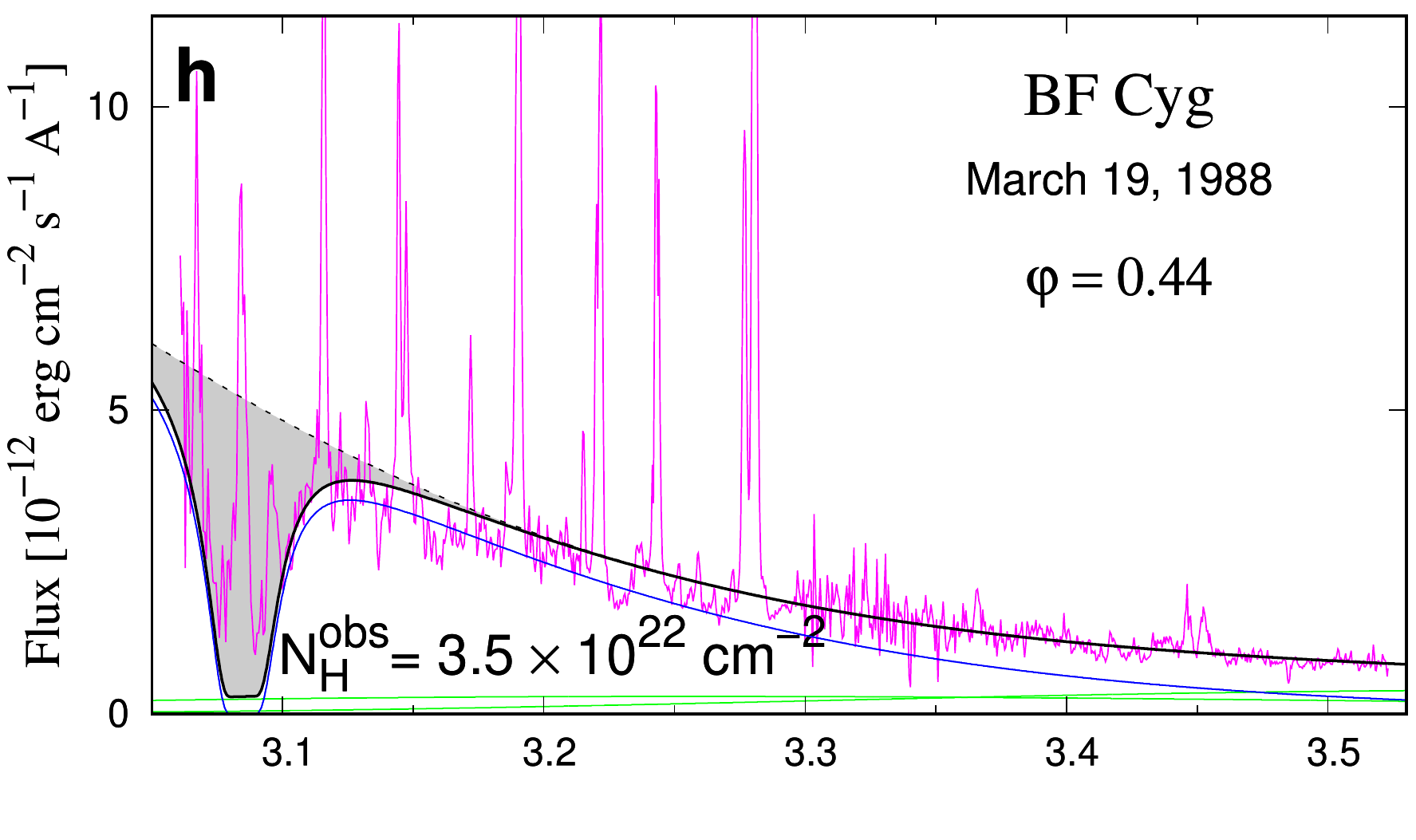} 
\includegraphics[angle=-90]{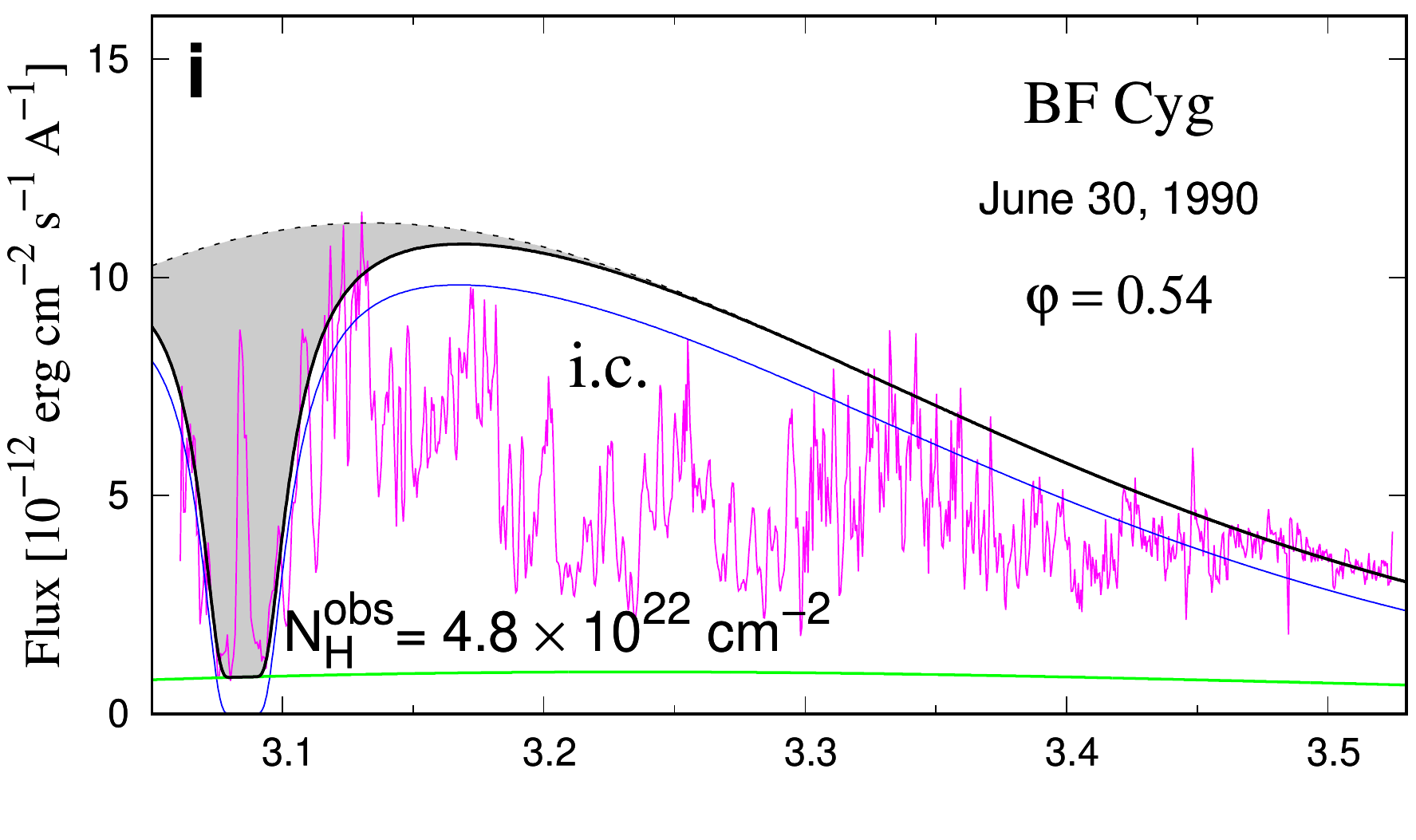}}\vspace*{-1mm} 
\resizebox{\hsize}{!}
          {\includegraphics[angle=-90]{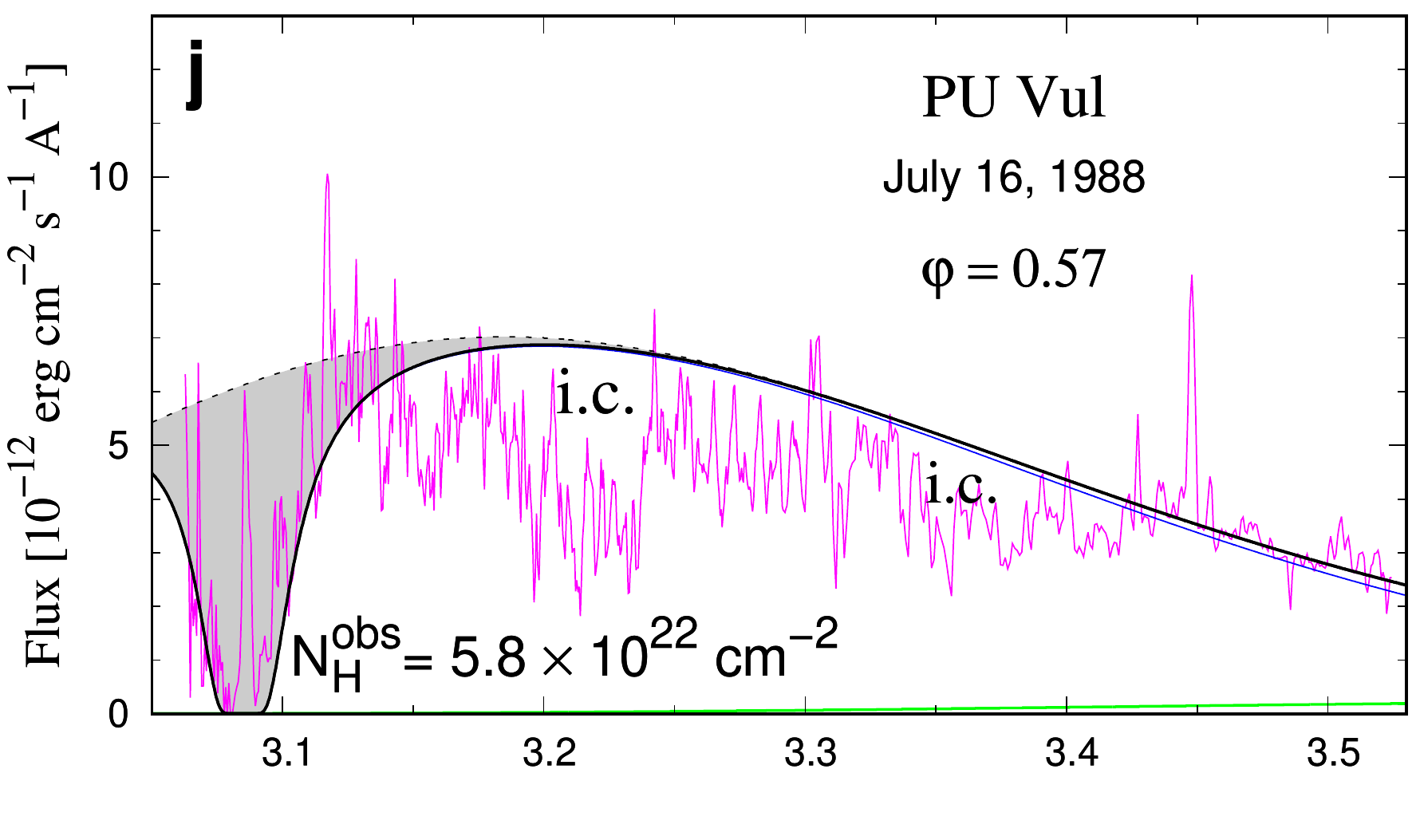} 
           \includegraphics[angle=-90]{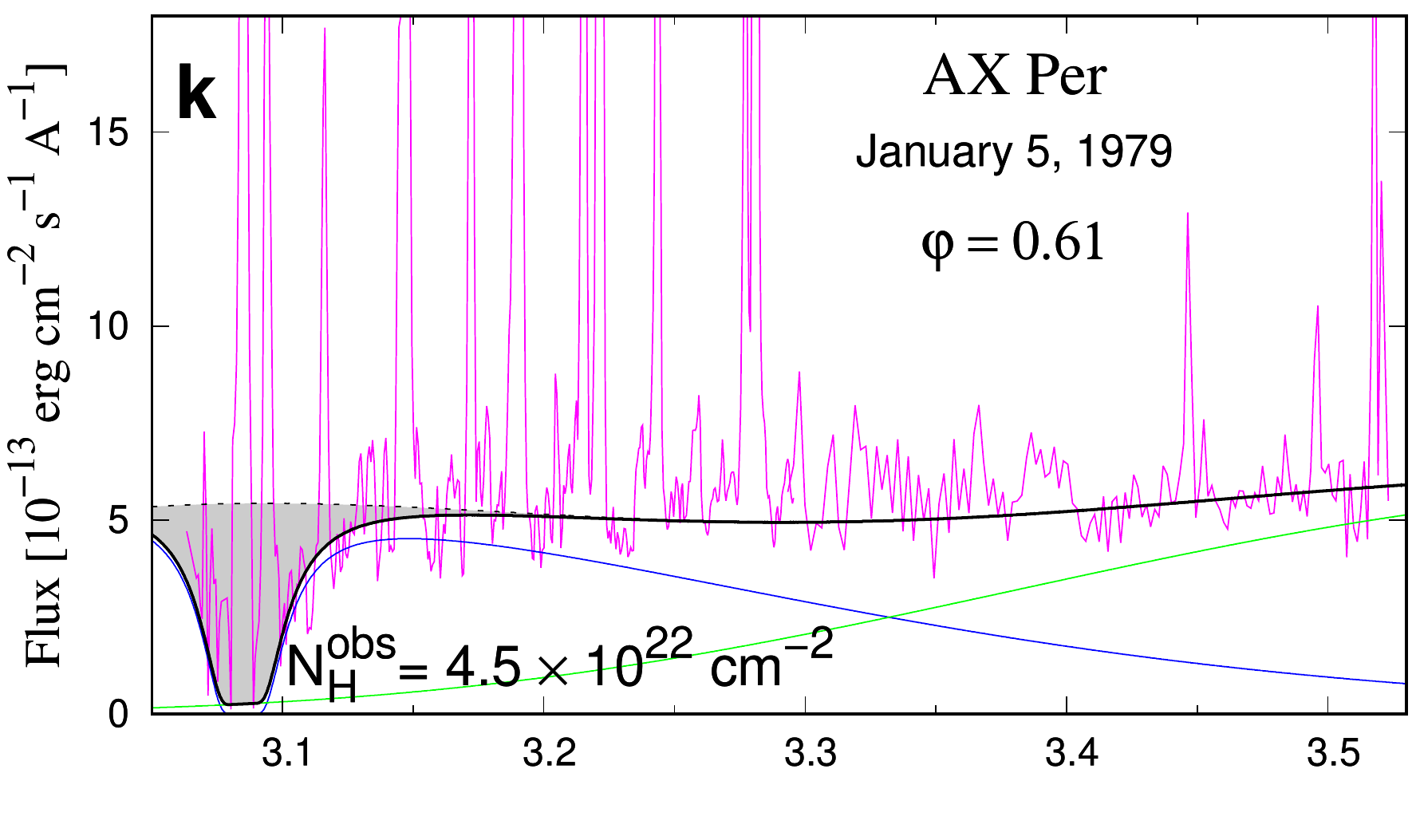} 
\includegraphics[angle=-90]{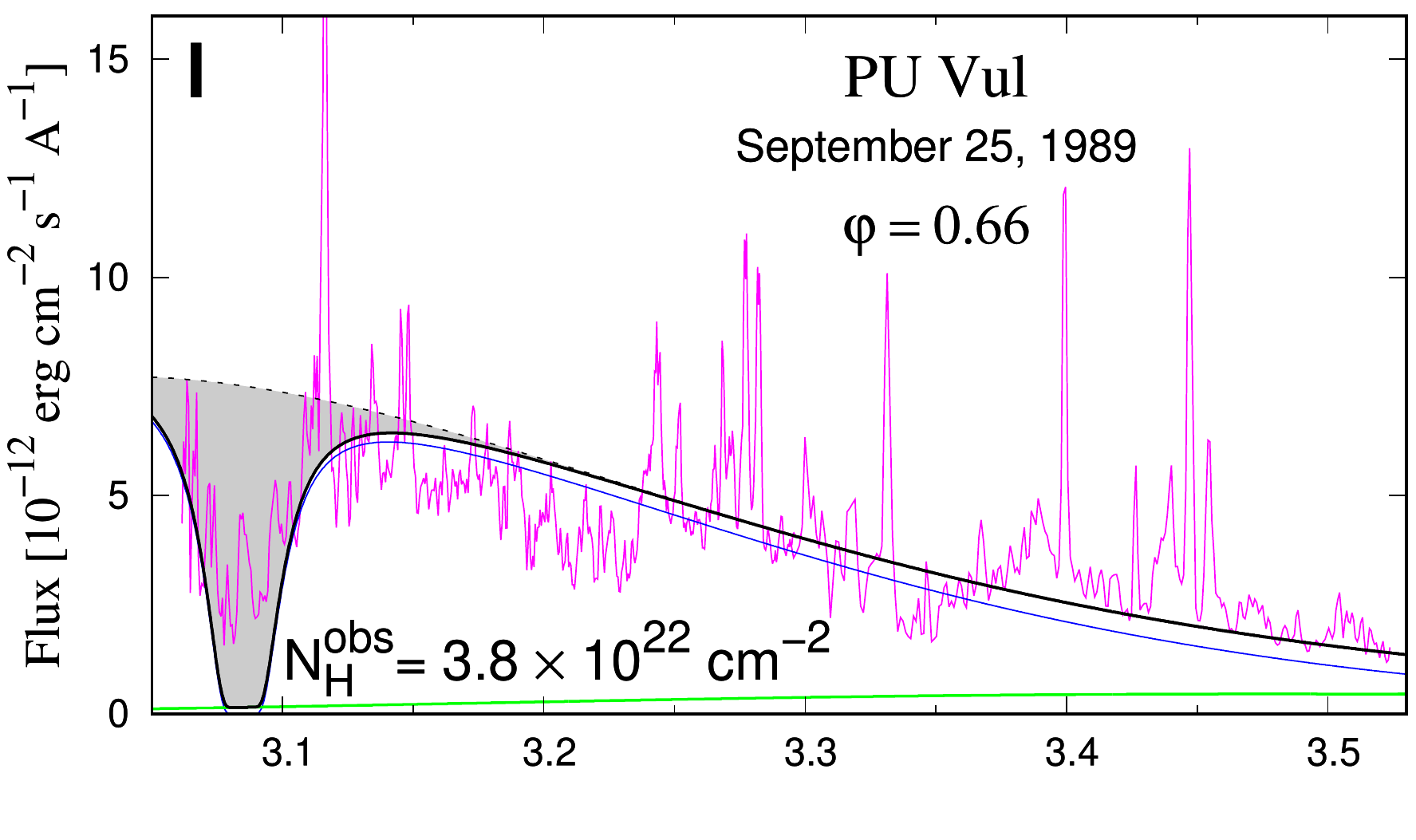}}\vspace*{-1mm} 
\resizebox{\hsize}{!}
          {\includegraphics[angle=-90]{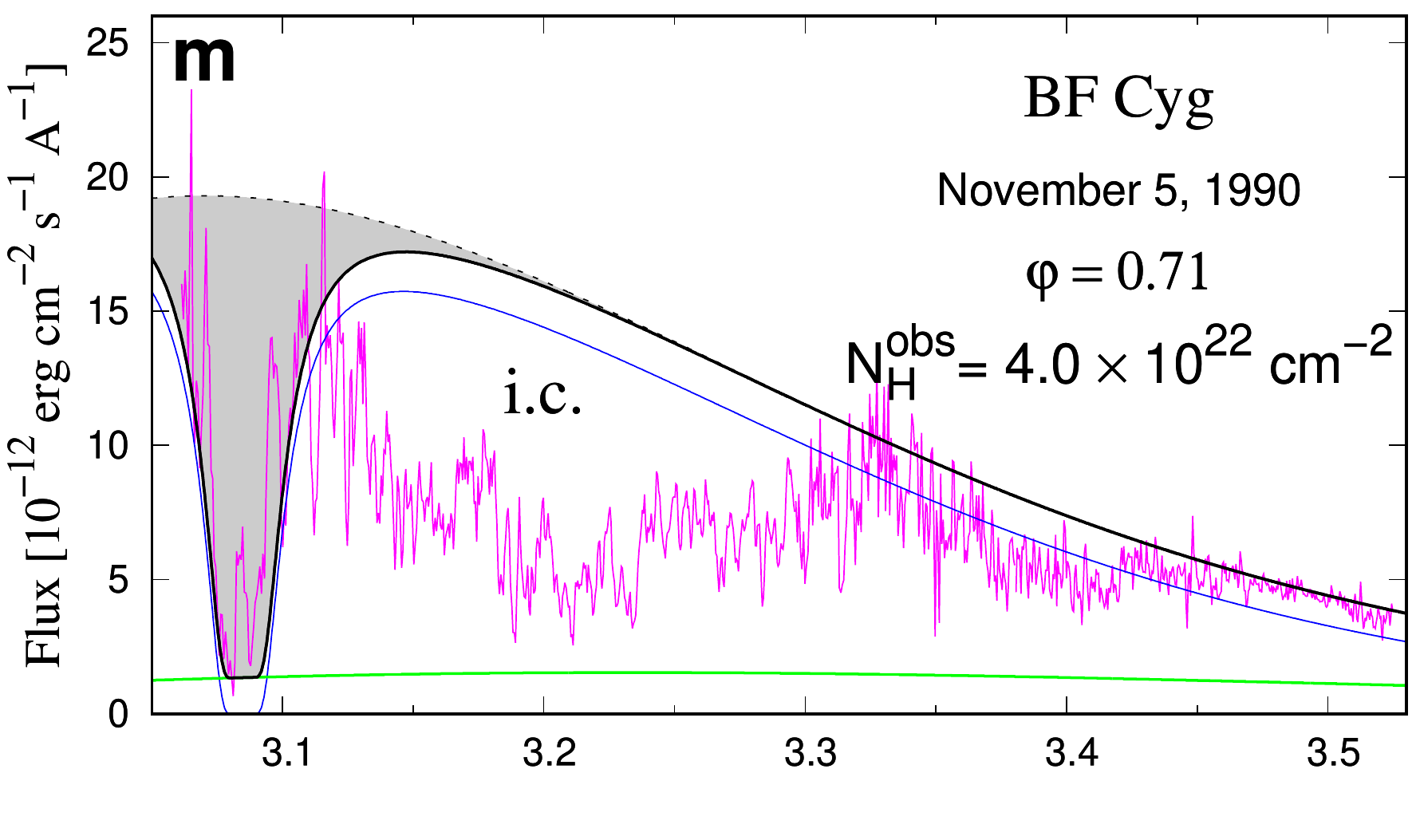} 
           \includegraphics[angle=-90]{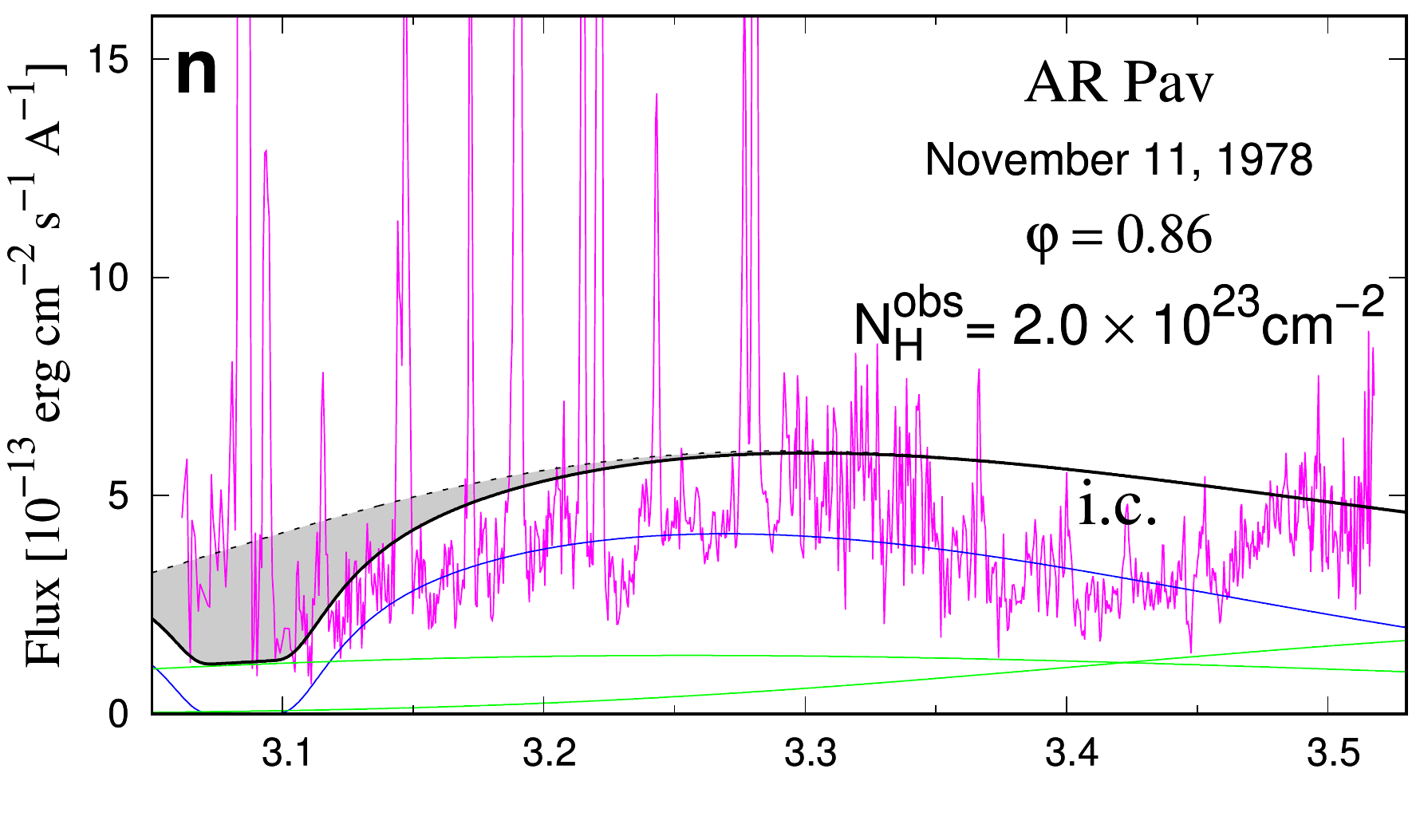} 
\includegraphics[angle=-90]{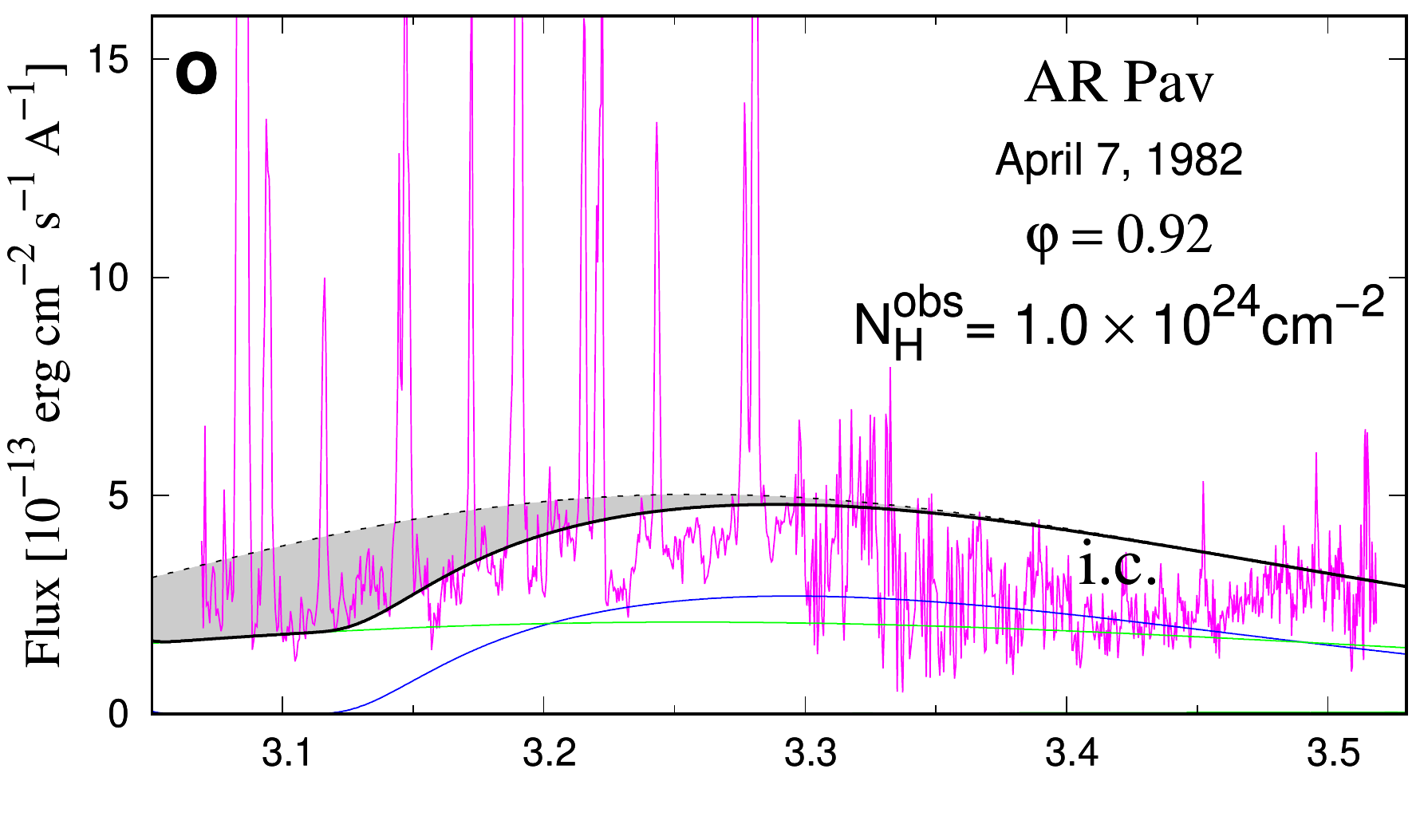}}\vspace*{-1mm} 
\resizebox{\hsize}{!}
          {\includegraphics[angle=-90]{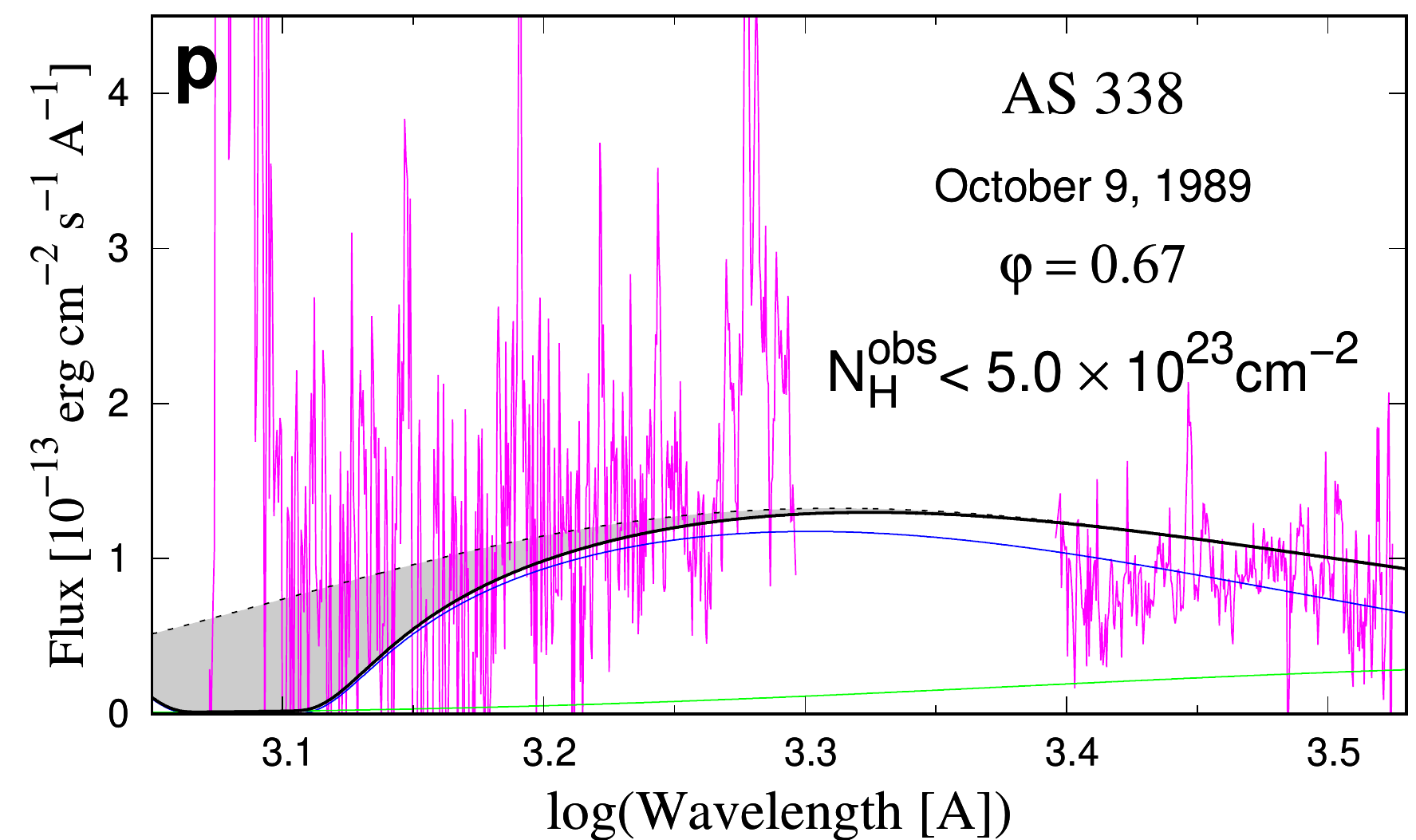} 
           \includegraphics[angle=-90]{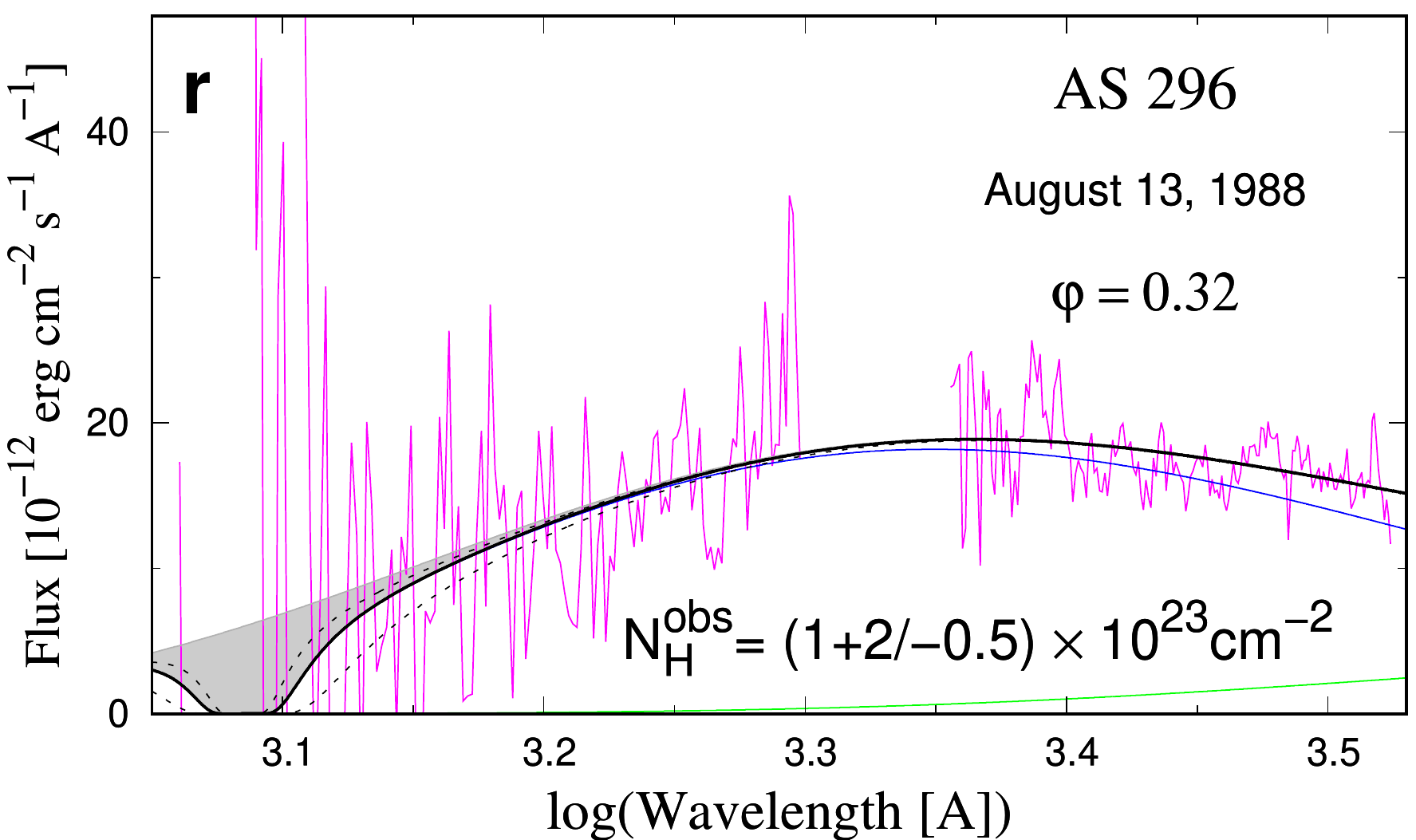} 
\includegraphics[angle=-90]{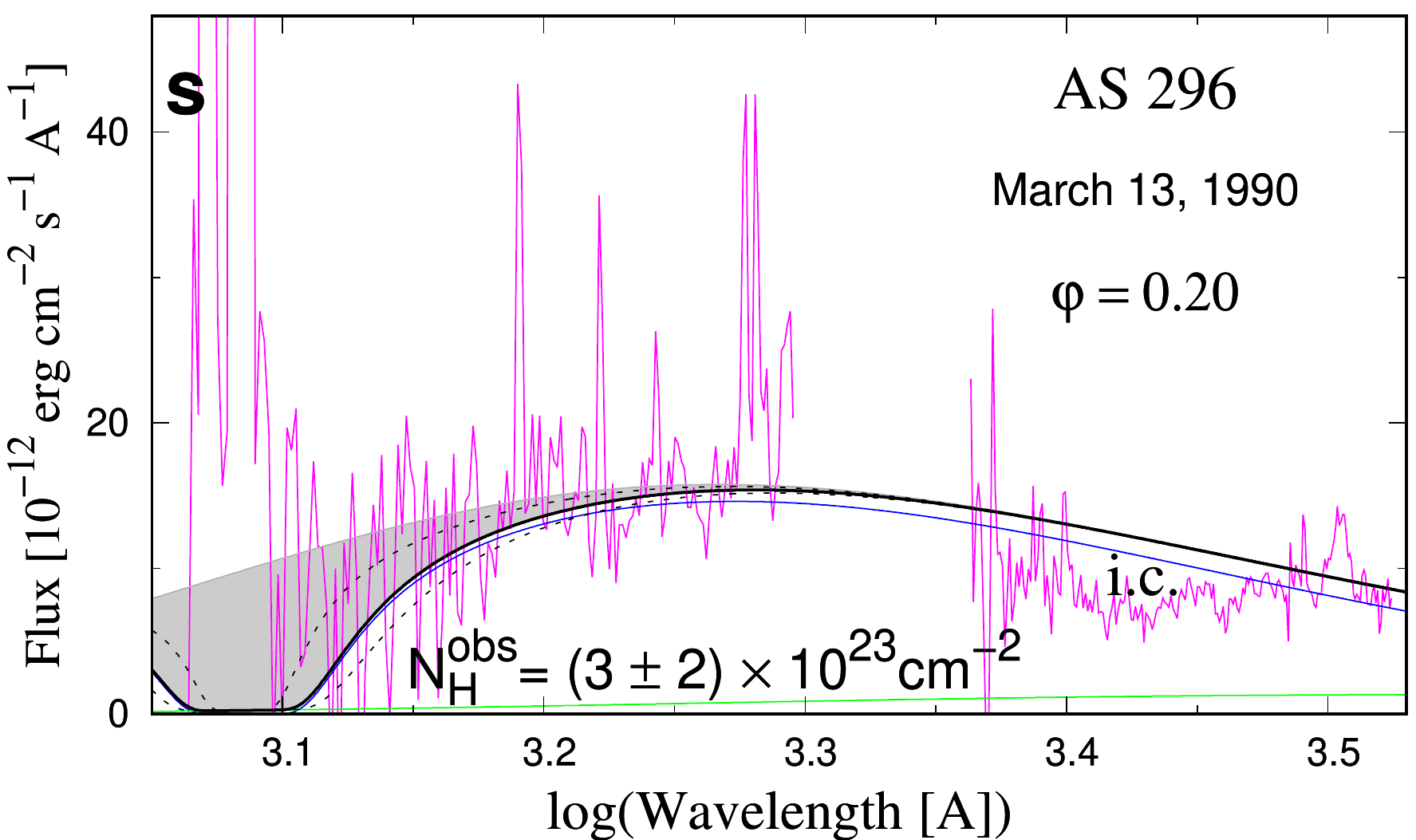}}\vspace*{-1mm} 
\end{center}
\caption{
Examples of \textsl{IUE} spectra (in magenta) of our targets 
(Sect.~\ref{ss:targets}) during active phases observed at 
different orbital phases $\varphi$ and their SED model 
(black line, Eq.~(\ref{eq:uvsed})). The model consists of 
the nebular (green lines) and the stellar continuum from 
the hot component (blue line; Sect.~\ref{ss:ray}). 
A strong depression of the continuum around the Ly$\alpha$ 
line (gray area) is caused by Rayleigh scattering of the WD 
radiation on hydrogen atoms with column densities, 
$N_{\rm H}^{\rm obs}$ (Table~\ref{tab:nh}). 
The bottom row shows examples of AS~338 and AS~296, whose spectra 
are underexposed (see Appendix~\ref{app:targets}). 
Models for very different $N_{\rm H}^{\rm obs}$ estimated for 
AS~296 are plotted with dotted lines. 
Additional depression of the continuum is caused by the iron 
curtain absorptions (Sect.~\ref{ss:ray}), here denoted by `i.c.'. 
          }
\label{fig:sediue}
\end{figure*}

Figure~\ref{fig:sediue} of this appendix shows examples of SED 
models in the ultraviolet for the investigated objects at 
different orbital phases (see Sect.~\ref{ss:ray}). 
%
%
\clearpage
%
%
\section{Spectral energy distribution during quiescent 
         and active phases of our targets}
\label{app:B}
\begin{figure*}
\begin{center}
\resizebox{\hsize}{!}
          {\includegraphics[angle=-90]{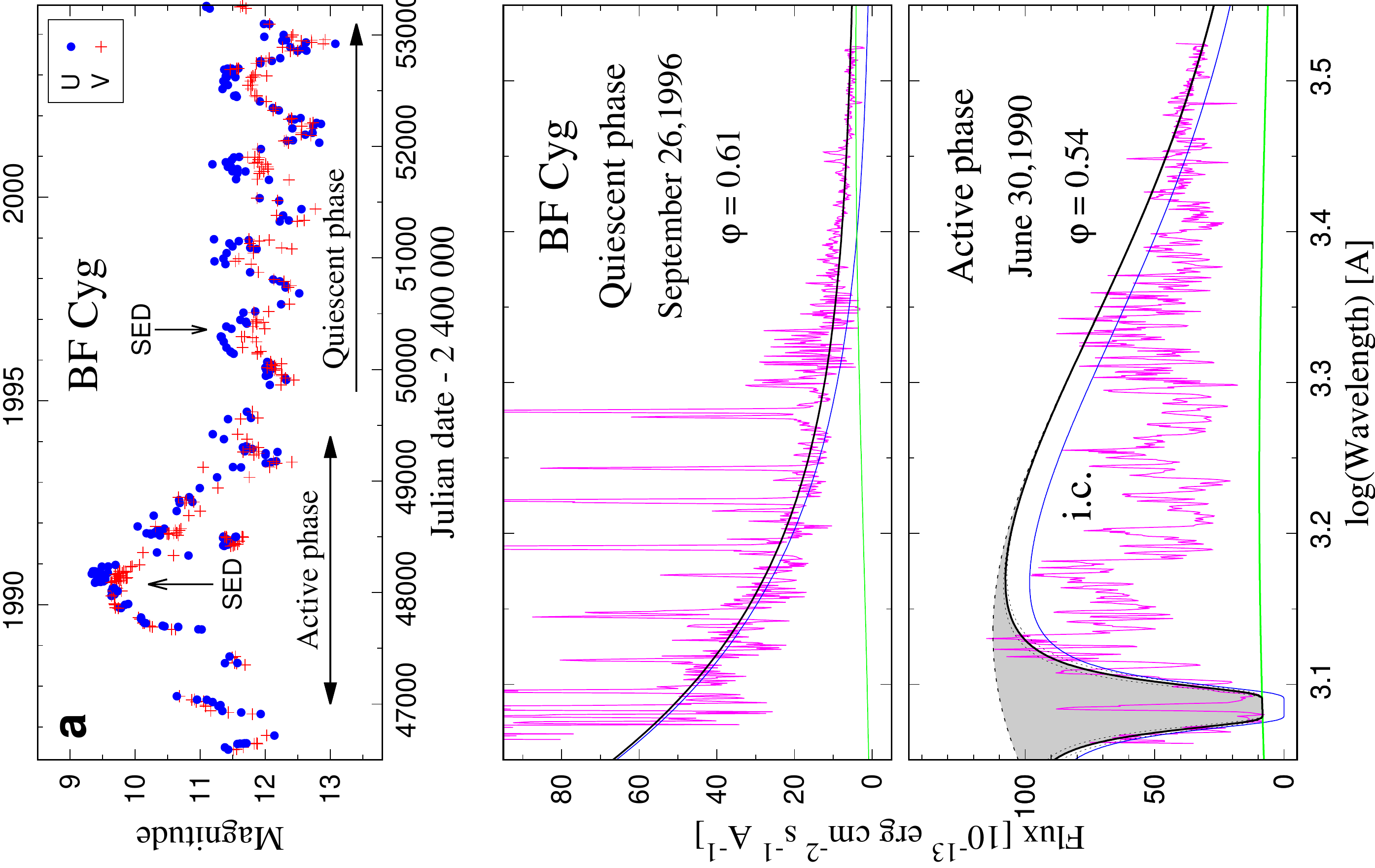}
           \includegraphics[angle=-90]{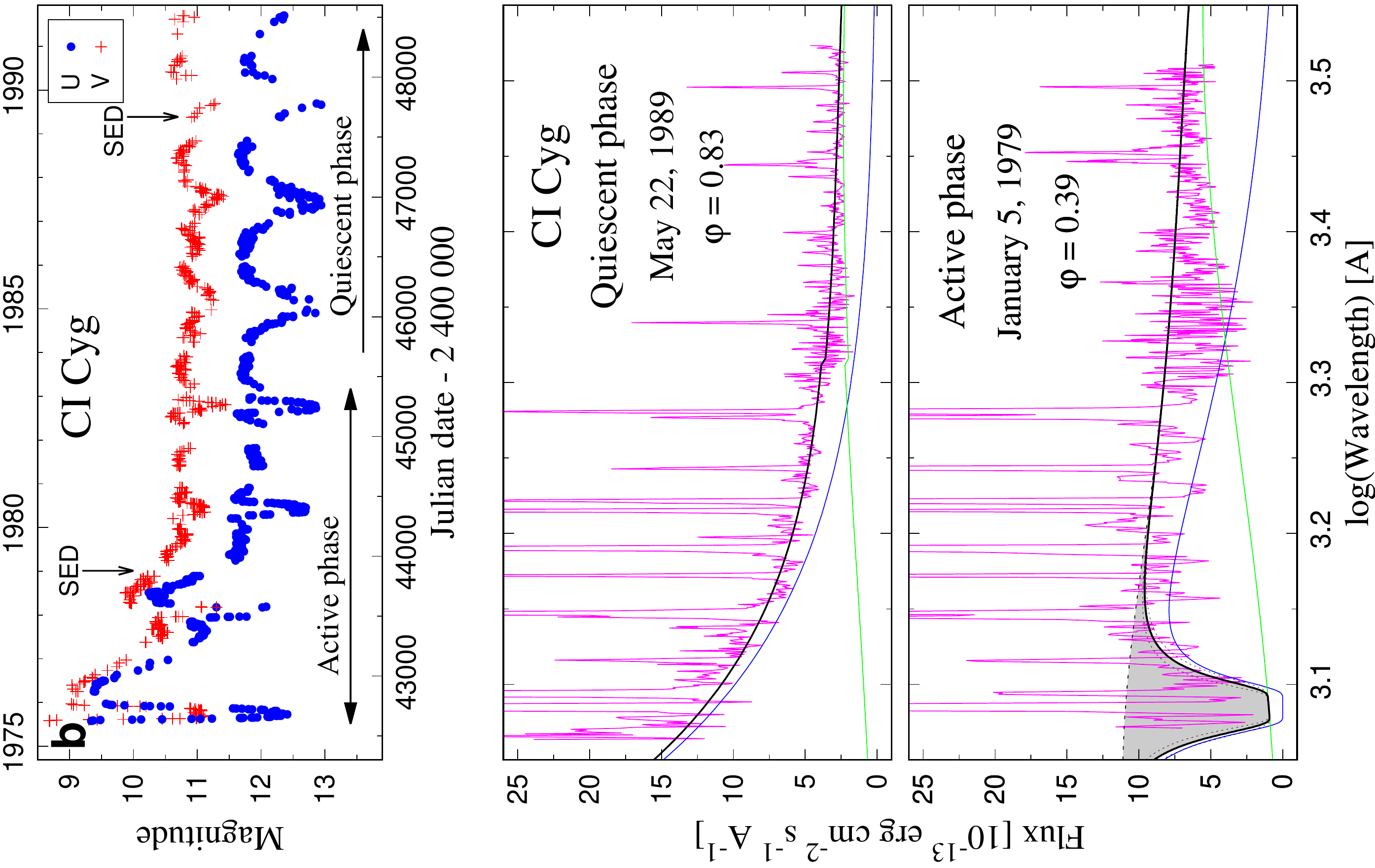}
           \includegraphics[angle=-90]{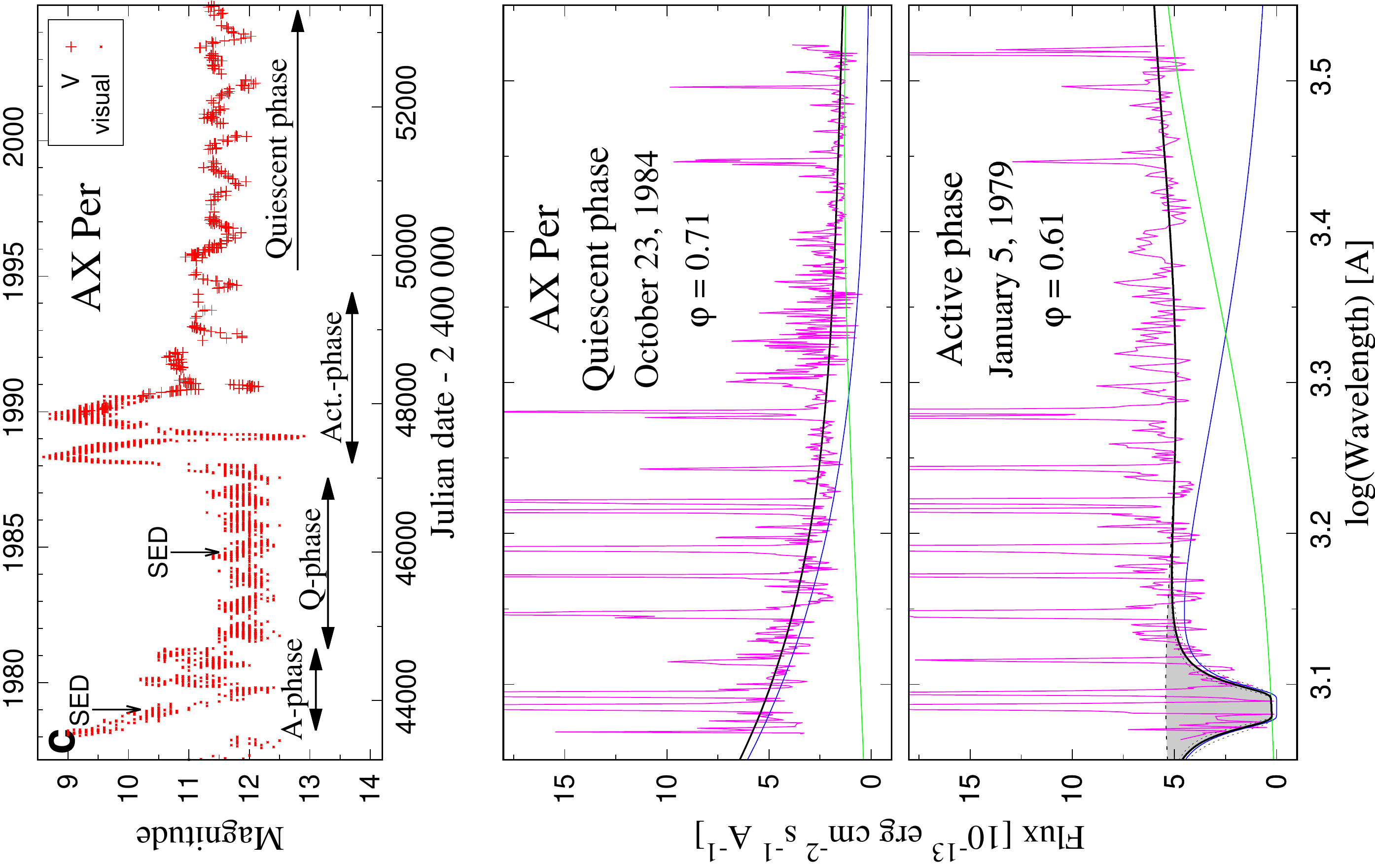}}

\vspace*{5mm}

\resizebox{\hsize}{!}
          {\includegraphics[angle=-90]{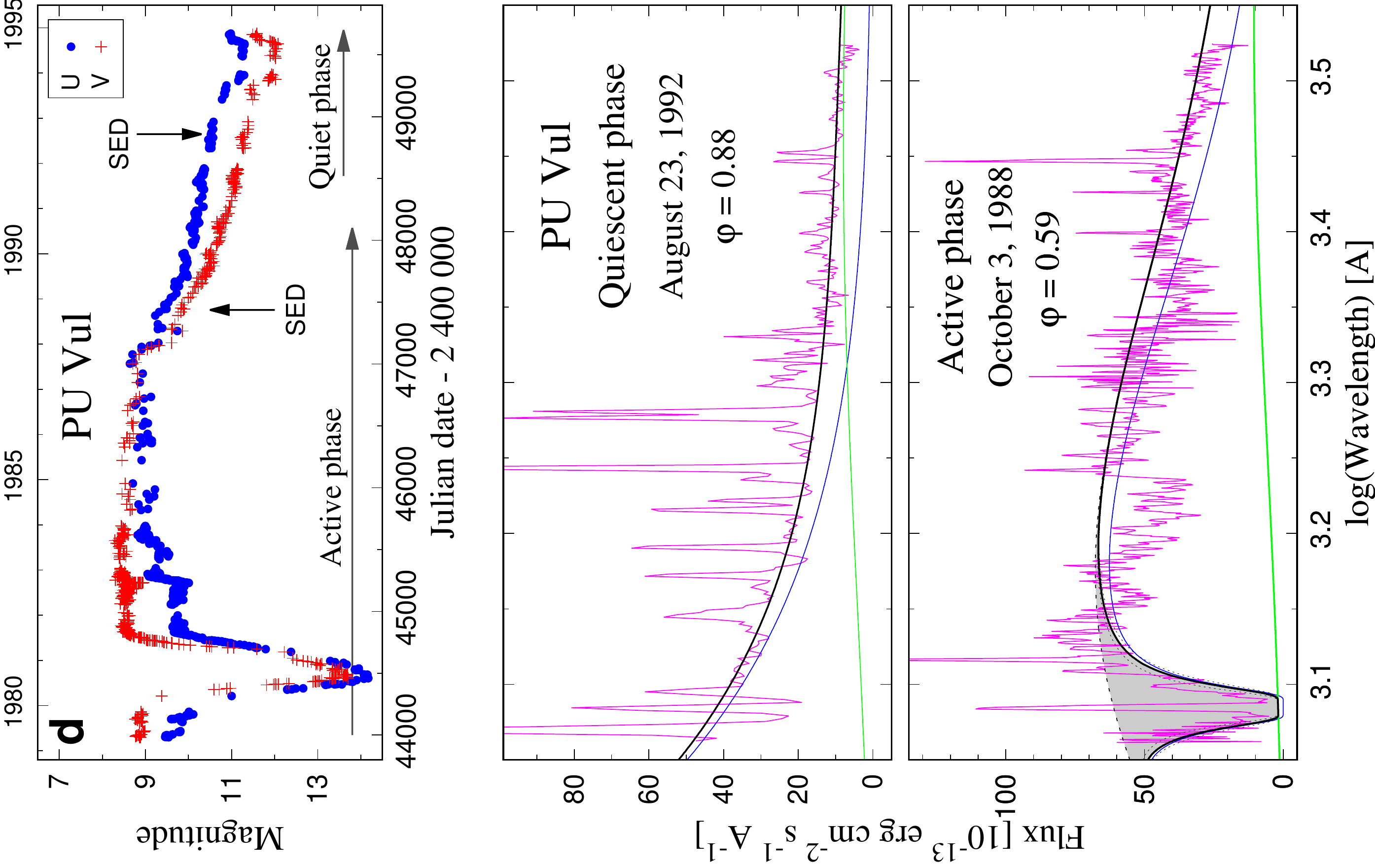}
           \includegraphics[angle=-90]{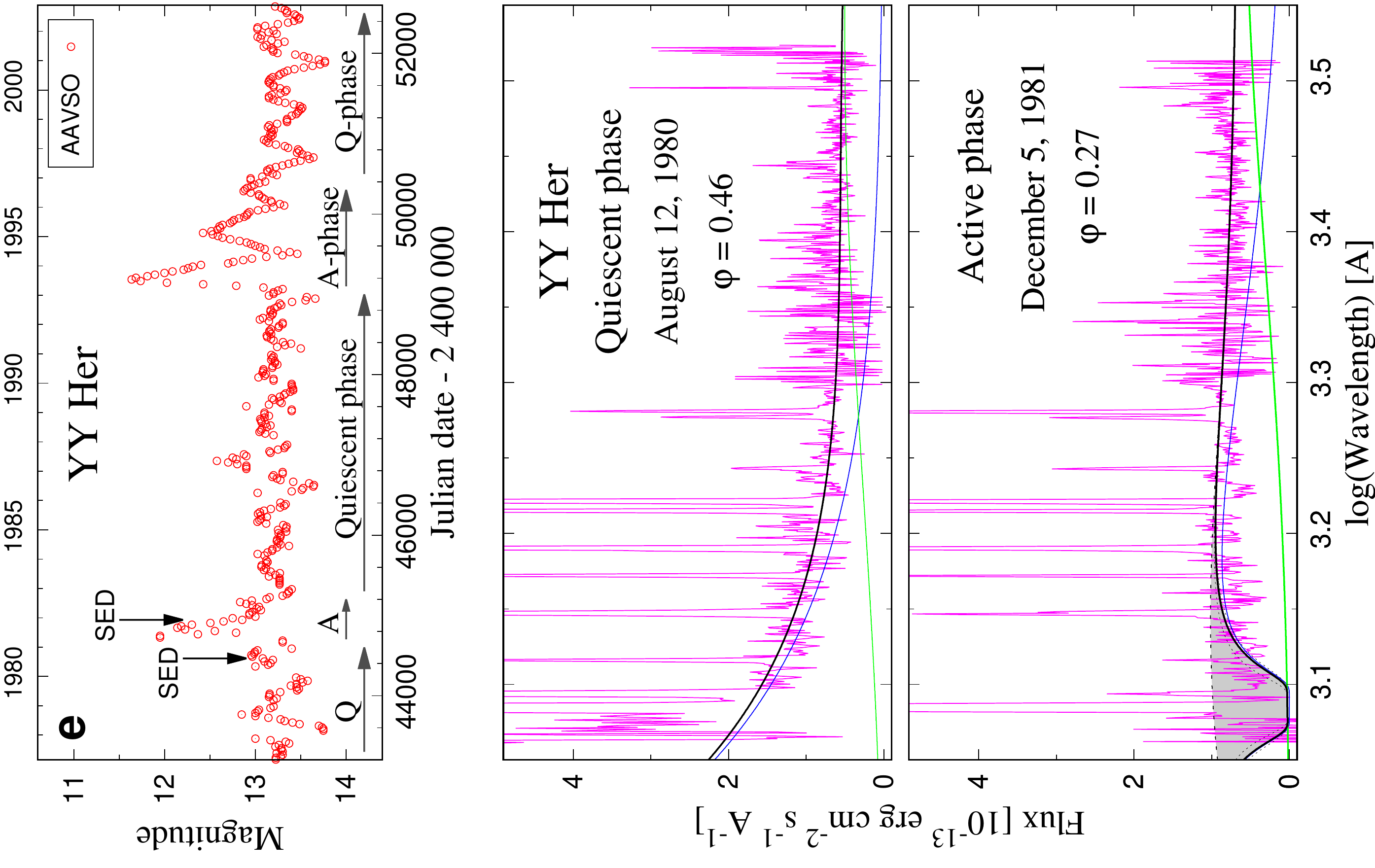}
           \includegraphics[angle=-90]{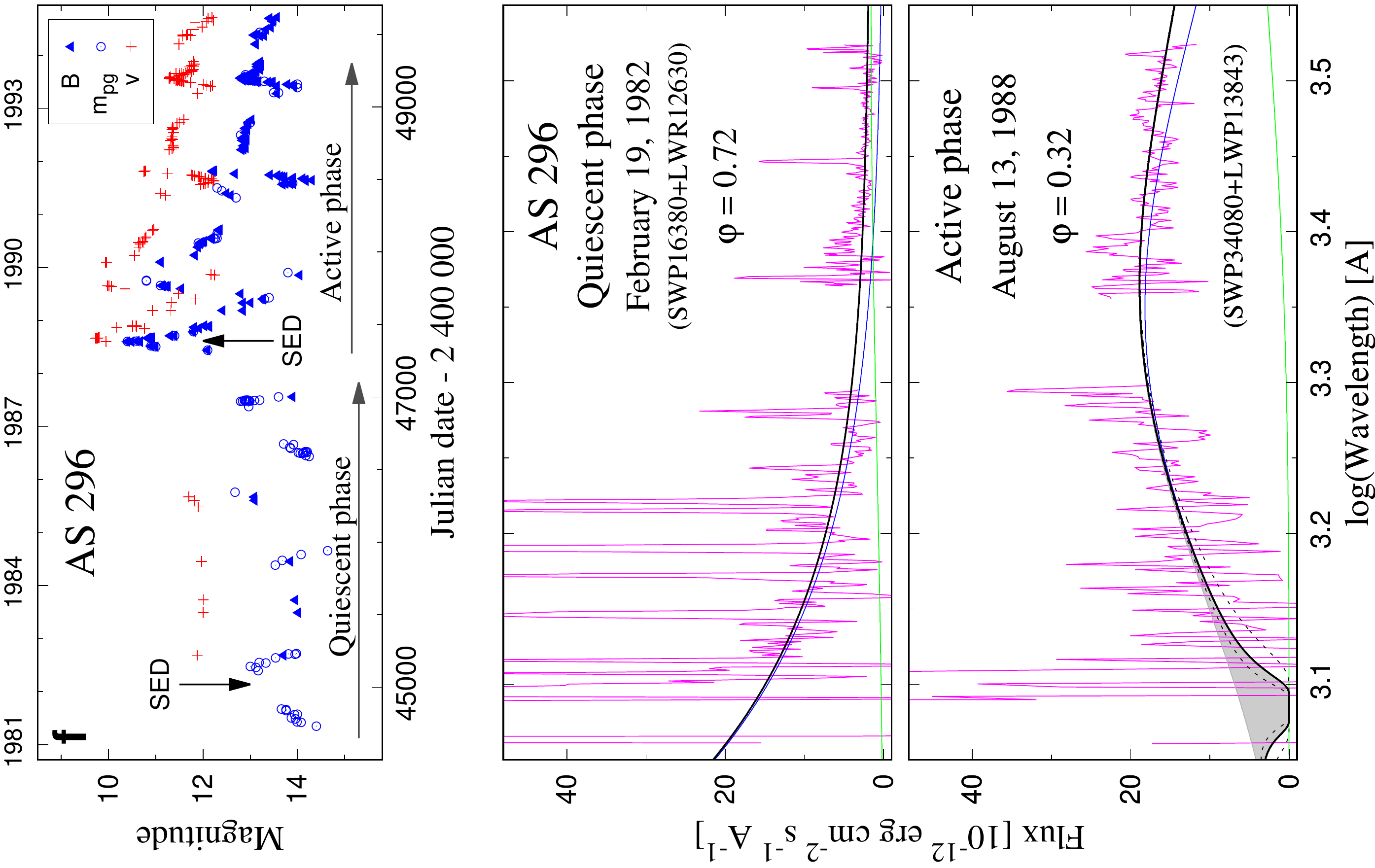}}
\end{center}
\caption{
Comparison of SEDs for our targets during quiescent and active 
phases. Meaning of lines (magenta, black, green, blue, and dotted) 
and denotation by i.c. as in Appendix~\ref{app:A}. Panels 
for BF~Cyg, CI~Cyg, AX~Per and YY~Her ({\bf a}, {\bf b}, {\bf c} 
and {\bf e}) are compiled according to \cite{2005A&A...440..995S}, 
and that for PU~Vul is adapted from \cite{2014ASPC..490..211S}. 
In panel {\bf f} we used photometry of \cite{
1989A&A...214L...5M,
1992AJ....104..262M,
1995AJ....109.1740M}. 
The dramatic change of the spectrum during the transition from 
quiescent to active phase is caused by the creation of 
a disk-like structure around the WD during outbursts, and 
subsequent formation of the neutral wind region in the orbital 
plane (Sect.~\ref{ss:res2}). 
Here, the former is evidenced by the emergence of deep eclipses 
in the light curves (top panels), while the latter is indicated 
by Rayleigh scattering on atomic hydrogen around the Ly$\alpha$ 
line (gray area). Note that AR~Pav was observed only during the 
active phase, so its UV spectrum from quiescence is not available. 
Instead, we show here an example of AS~296, although its far-UV 
spectrum is not well determined (see Appendix~\ref{app:targets}). 
          }
\label{fig:sedqa}
\end{figure*}

Figure~\ref{fig:sedqa} of this appendix compares the profile of 
the ultraviolet continuum for the investigated objects during 
the quiescent and active phases (Sect.~\ref{ss:res1}). 
%
%
\clearpage
%
%
\section{Discussion of the selection of the final set of targets}
\label{app:targets}
In this appendix, we explain why some of the systems suggested 
in Sect.~\ref{ss:targets} as candidates cannot be included in 
the resulting set of our targets. 
First, we verify if non-eclipsing systems from the second group 
(i.e., showing the far-UV continuum depression; Z~And, TX~CVn 
and CD-43$^{\circ}$14304) have also a high orbital inclination. 
Second, we comment on some eclipsing systems from the first set 
(i.e., showing the optical eclipses), whose far-UV spectra 
are underexposed (BX~Mon, AS~338, and AS~296), and disputed 
object CH~Cyg. The following comments are relevant: 

Z~And and TX~CVn sporadically show eclipses in their optical 
light curves during active phases, suggesting that their orbital 
inclination $i$ is quite high \citep[see][]{2004CoSka..34...45S}. 
A high $i$ for both systems is also constrained by 
the two-temperature type of their UV spectrum 
\citep[see Figs.~3 and 6 of][]{2005A&A...440..995S} -- a typical 
spectral feature of symbiotic binaries with a high $i$ 
that emerges during active phases (Sect.~\ref{s:intro}). 
In addition, an analysis of the 2003 eclipse of Z~And 
suggested $i > 76^{\circ}$ \citep[][]{2003A&A...401L..17S}. 
On the basis of polarimetric observations, 
\cite{2010AJ....140..235I} derived $i=74^{\circ}\pm 14^{\circ}$, 
and analyzing the Rayleigh scattering effect during quiescence, 
\cite{2012A&A...547A..45S} obtained 
$i=59^{\circ}-2^{\circ}/+3^{\circ}$, which was later refined 
to $63.6^{\circ}-12.2/+6.9^{\circ}$ by \cite{2017PASP..129f7001S}. 
A high value of $i$ for Z~And is independently supported by the 
obscuration of the O\,{\scriptsize I}]\,$\lambda$1641\,\AA\ 
emission line around the inferior conjunction of the giant 
\citep[][]{2010A&A...515A.108S,2012A&A...547A..45S}. 
As concerns to CD-43$^{\circ}$14304, there are no relevant 
observations to estimate directly its $i$. Here, a relatively 
high value of $i$ is suggested by the ionization structure 
during active phases that is constrained for CD-43$^{\circ}$14304 
by the two-temperature type of the UV spectrum shown by its model 
SED \citep[see Fig.~5 of][]{2005A&A...440..995S}. 
That these three systems have a high $i$, but they are not strictly 
eclipsing, explains why the corresponding $N_{\rm H}^{\rm obs}$ 
values are lower than those measured for eclipsing systems 
(see Fig.~\ref{fig:nhfi}), as expected according to the ionization 
structure during active phases. 

As concerns to BX~Mon, its very faint far-UV continuum does not 
make it possible to evaluate the effect of Rayleigh scattering. 
%
We excluded CH~Cyg, because its activity is accretion-powered, and 
thus does not show typical features of Z~And-type outbursts 
\citep[e.g.,][]{1988A&A...198..150M,2021AstL...47..235T}. 
In addition, its basic configuration (binary or triple system) and 
orbital inclination are uncertain -- possible eclipses observed 
during 1992 and 1994 \citep[][]{1996A&A...308L...9S} never repeated 
again \citep[see][ for a review]{2009ApJ...692.1360H}. 
That is probably why the Rayleigh scattering effect was observed 
only during a short period (from $\sim$1993 to $\sim$1995), although 
CH~Cyg is active from $\sim$1977 (see its light 
curve\footnote{e.g., https://www.aavso.org/LCGv2/}). 
%
Further, FN~Sgr was observed in the UV only during 
quiescence \citep[e.g.,][]{2005A&A...440..239B}. 
%

Finally, spectra of AS~338 and AS~296 are strongly underexposed 
in the far-UV, which precludes a correct determination of 
$N_{\rm H}^{\rm obs}$ by Rayleigh scattering 
(see Fig.~\ref{fig:sediue}, bottom row panels). 
This is because the light from both stars is strongly 
reddened with the color excess 
$E_{\rm B-V}$ = 0.5 \citep[][]{2000AstL...26..162E} and 
1.0 \citep[][]{1992AJ....104..262M} for AS~338 and AS~296, 
respectively. 
Therefore, if we attribute the total attenuation of the far-UV 
radiation to Rayleigh scattering, we obtain only the upper 
limits of $N_{\rm H}^{\rm obs}$, or very uncertain values 
(see Fig.~\ref{fig:nhfi}). 
\clearpage
%
%
\section{Characteristic parameters of targets}
\label{app:param}
In this appendix, we briefly summarize the basic characteristics 
of the selected targets that are suitable for the aim of this 
paper -- the eclipsing symbiotic binaries with a well-exposed 
far-UV spectrum acquired during the active phase 
(see Sect.~\ref{ss:targets}). 

Recent ephemerides of the eclipses and/or the inferior conjunction 
of the red giant, distances, and the $E_{\rm B-V}$ interstellar 
reddening parameter of the targets are found in Table~\ref{tab:nh}, 
while Table~\ref{tab:param} of this appendix lists the physical 
parameters of their main radiation sources. 
In addition, Table~\ref{tab:apar} (Appendix~\ref{app:par}) presents 
these parameters for the hot component and the nebula, we obtained 
by modeling their UV SEDs in this paper (Sect.~\ref{ss:ray}), while 
Table~\ref{tab:qpar} of Appendix~\ref{app:par} collects these 
parameters for the spectra from quiescent phases, plotted in 
Figs.~\ref{fig:qa} and \ref{fig:sedqa}. 
Below we briefly describe their photometric observations with 
regard to the appearance of eclipses in their light curves 
during active phases. 

BF~Cyg: 
In $\sim$1894 BF~Cyg experienced a nova-like explosion 
\citep[][]{1941BHarO.915...17J}, after which its brightness 
was gradually declining for about 70 years, occasionally 
showing brightenings lasting for several years (Z~And-type 
of outbursts), with maxima of around 1920, 1956, 1968-74, 
1990, and 2006 \citep[][]{2006MNRAS.366..675L,
                          1997MNRAS.292..703S,
                          2006CBET..596....1M,
                          2019CoSka..49...19S}.
During the 1989-94 active phase, \cite{1992IBVS.3780....1S} 
reported for the first time the appearance of a primary minimum 
in the $UBV$ light curves (eclipse), which was relatively narrow 
and rectangular in profile, located around the position of the 
inferior conjunction of the giant. Later, the eclipsing nature 
of BF~Cyg was also confirmed by observing the effect of Rayleigh 
scattering near the eclipse during the quiescent phase 
\citep[][]{1996AJ....111.1329P}. 
The following development of the wave-like orbital-related variation 
during 1995 -- 2005 reflects a quiescent phase of BF~Cyg 
(see Fig.~\ref{fig:sedqa}a of Appendix~\ref{app:B} here). 

CI~Cyg: 
The eclipsing nature of CI~Cyg is best documented by its major 
active phase that began in 1975 \citep[][]{1976IBVS.1169....1B}, 
when narrow minima, eclipses, developed in the light curve during 
the first four 855.25-d cycles from the maximum, with the subsequent 
change to a broad wave-like variation since $\sim$1984 
\citep[see Fig.~2 of][ and Fig.~\ref{fig:sedqa}b 
of Appendix~\ref{app:B} here]{1991SvA....35..154B}. This is 
a textbook example of the typical photometric features of 
the active and quiescent phases of SySts reflecting the very 
different ionization structure of the hot component (see 
Sects.~\ref{s:intro} and \ref{ss:iona}; Figs.~\ref{fig:qa}c 
and \ref{fig:qa}f for a summary). 

YY~Her: 
\cite{1997A&A...323..113M} identified four Z~And-type outbursts 
with additional six brightenings in the historical (1890--1996) 
light curve of YY~Her. The authors also found four deep and sharp 
minima in the light curve, located near the zero orbital phase 
(i.e., at the light minima in their quiescent light curve). 
However, they rejected their eclipsing nature because they did 
not find any reason for this interpretation compared to the 
eclipses observed in SySts. On the other hand, they found no 
other reasonable explanation for these minima. 
During the 1993-98 main active phase, the visual light curve 
of YY~Her showed a relatively narrow V-shaped minimum in 
1994.43, around the zero orbital phase, suggesting a high 
orbital inclination, but less than 90$^{\circ}$ 
\citep[see e.g., Fig.~16 of][]{2005A&A...440..995S}. 
The eclipsing nature of YY~Her was confirmed by measuring the 
eclipse effect in the $U$-light curve during 1997 
\citep[][]{2000ARep...44..190T}. 
The transition to the quiescent phase was similar to that observed 
for AX~Per or CI~Cyg \citep[see Fig.~1 of][]{2001AstL...27..703T}. 
After entering the quiescent phase, only a few brighter periods 
have been recorded 
\citep[][]{2013ATel.4996....1M,2021ATel14464....1M}. 

AR~Pav: 
The eclipsing nature of AR~Pav was already revealed by the first 
photographic observations made between 1889 and 1936.5 by 
\cite{1937AnHar.105..491M}, who classified it as an unusual 
eclipsing P~Cyg type binary with an orbital period of 605 days. 
Later, optical spectroscopy showed that AR~Pav is a symbiotic 
binary \citep[see][ and references therein]{1986syst.book.....K}. 
Observing changes in the eclipse profile and systematic changes 
in their timing from cycle to cycle, a strong out-of-eclipse 
variability \citep[e.g.,][]{1974MNRAS.167..635A,
                            1994A&A...287..829B,
                            2000MNRAS.311..225S,
                            2019CoSka..49...19S}, 
and also the spectroscopic variability in the UV/optical domain 
\citep[e.g.,][]{2001A&A...366..972S,2002A&A...387..139Q} reflect 
continuous activity of AR~Pav. 
The persistent presence of narrow deep minima -- eclipses -- in 
the light curve and current activity indicate that AR~Pav is in 
a long-lasting active phase. 

AX~Per: 
The historical light curve of AX~Per (since 1887) is characterized 
by long quiescent phases, occasionally interrupted by Z~And-type 
outbursts of typically several years, usually showing more than 
one brightening 
\citep[see e.g., Figs.~1 in][]{2011A&A...536A..27S,
                               2013AJ....146..117L}. 
The eclipsing nature of AX~Per was unambiguously revealed during 
the 1988-1994 active phase by the appearance of narrow minima 
around the inferior conjunction of the giant 
\citep[][]{1990CoSka..20...99S,1991IBVS.3603....1S}. 
Figure~\ref{fig:sedqa}c of Appendix~\ref{app:B} shows this part 
of the light curve together with the following development of 
the quiescent phase since $\sim$1995. 
The nature of low stages between outbursts as the quiescent phase 
of symbiotic star was proven by \cite{1982PASP...94..165K}, who 
revealed the wave-like variation in the light curve between the 
$\sim$1950 and $\sim$1978 outbursts with a period of 681.6 days, 
which he ascribed to the orbital period. 

PU~Vul 
is a symbiotic nova, discovered during its outburst in 1979 by 
\cite{1979IAUC.3344....1K} and \cite{1979IAUC.3348....2A}. 
From $\sim$1977 to $\sim$1979, PU~Vul 
brightened by $\sim$5 magnitudes in the $m_{\rm pg}/B$ band. 
Subsequently, from 1980.1 to 1981.4 the light curve of the 
nova showed a narrow (relative to the orbital period; 
see below) $\sim$5 mag deep minimum, after which continued at 
a bright stage (a supergiant phase) up to $\sim$1987.5 when 
PU~Vul began a gradual fading till $\sim$1999 (the nebular phase). 
Meanwhile, the light curve showed another relatively narrow but 
only $\sim$0.5 mag deep minimum around 1994.25 
\citep[see Fig.~1 of][ and references therein]{2018RAA....18...98T}. 
\cite{1992A&A...259..525V} interpreted the first deep minimum 
as an eclipse of the outbursting star by the M giant companion. 
The appearance of the second 1994 eclipse 
allowed \cite{1996A&A...307..470N} to estimate the orbital period 
of the binary to be 4900$\pm$100 days, later refined by 
\cite{2012BaltA..21..150S} to 4897$\pm$10 days using three 
primary minima. 
Since $\sim$2000, the development of a broad wave in the light 
curve with minima around the inferior conjunction of the giant 
\citep[][]{2018RAA....18...98T}\footnote{See also the current 
AAVSO light curve: \url{https://www.aavso.org/LCGv2/}.}
indicates that PU~Vul is currently in the quiescent phase. 
%
%
\begin{table*}
\caption{Characteristic parameters of the radiation sources of 
our targets. 
Red Giant: Spectral type $ST$, radius $R_{\rm RG}$ ($R_{\odot}$), 
           luminosity $L_{\rm RG}$ ($L_{\odot}$). 
Hot component: Effective radius $R_{\rm WD}^{\rm eff}$ ($R_{\odot}$), 
               temperature $T_{\rm BB}$(10$^4$\,K), luminosity 
               $L_{\rm WD}$ ($L_{\odot}$), during quiescent (Q) 
               and active (A) phases. 
Nebula: Emission measure $EM$ ($10^{60}\,{\rm cm^{-3}}$). 
}
\label{tab:param}
\begin{center}
\begin{tabular}{ccccccccc}
\hline
\hline
\noalign{\smallskip}
                                 &
\multicolumn{3}{c}{Red Giant}    &
\multicolumn{3}{c}{Hot Component}&
\multicolumn{1}{c}{Nebula$^{c}$} &
                                 \\
                                                     &
\multicolumn{3}{c}{---------------------------}      &
\multicolumn{3}{c}{----------------------------------------------------} &
                                                     &
                                                     \\
Object                          &
$ST$                            & 
$R_{\rm RG}$                    & 
$L_{\rm RG}$                    & 
$R_{\rm WD}^{\rm eff}$          & 
$T_{\rm BB}$                    & 
$L_{\rm WD}$                    &
$EM$                            & 
Reference$^{\dagger}$          \\
                           & 
                           & 
                           & 
                           & 
Q/A                      & 
Q/A                      & 
Q/A                      & 
Q/A                      & 
                           \\ 
\noalign{\smallskip}
\hline
\noalign{\smallskip}
BF~Cyg & M\,5 & 150 & 2700 
       & 0.79/7.1 & 5.5/2.15 & 5200/16800$^{a}$
       & 3.1/12 
       & 1, 2, 3, 11 \\
%
\noalign{\smallskip}
CI~Cyg & M\,5.5& 180 & 3400 
       & 0.11/0.67 &11.5/2.8 & 1700/1540$^{a}$  
       & 0.48/1.1  
       & 2, 3, 4 \\
%
\noalign{\smallskip}
YY~Her & M\,4& 110 & 1600 
       & $<$0.14/0.89 &$>$10.5/2.2 & $>$2100/3800$^{a}$  
       & 1.1/0.7-1.3  
       & 2, 3$^{e}$ \\
%
\noalign{\smallskip}
AR~Pav & M\,5& 139 & 2300 
       & --/1.9 & --/2.2 & --/6400$^{a}$  
       & --/4.6   
       & 2, 3, 5, 12 \\
%
\noalign{\smallskip}
AX~Per & M\,4.5    & 102    & 1200 
       & 0.07/0.42 & $\sim$10/0.6 & 400-900/$\sim$1200$^{a}$  
       & 0.25/0.30-0.71   
       & 2, 3 \\
%
 ---   & M\,5.6& 115 & 1500 
       & --/6.2-11 & --/0.53-0.63 & --/600-1400$^{a}$  
       & --/0.35-0.82$^{d}$   
       & 6 \\
%
\noalign{\smallskip}
PU~Vul & M\,6-7& 282 & 3820 
       & $\approx$0.03/8.2$^{b}$  
       & $\approx$20/2$^{b}$  
       & $\approx$1000/9700$^{b}$  
       & --/11$^{b}$   
       & 7, 8, 9, 10 \\
%
\noalign{\smallskip}
\hline
\end{tabular}
\end{center}
{\bf Notes:}\\
$^{a}$ the luminosity of the $10^5$\,K hot ionising source 
       that generates the observed $EM$ during active phases 
       \citep[][]{2017A&A...604A..48S}, \\
$^{b}$ as on October 3, 1988 (see Table~\ref{tab:apar}), \\
$^{c}$ symbiotic nebulae radiate at characteristic electron 
     temperature $T_{\rm e}$ = 13\,000 -- 19\,000\,K during 
     quiescent phases \citep[][]{1991A&A...248..458M,
     2005A&A...440..995S}, while during active phases, an additional 
     high-temperature nebula with $T_{\rm e}\gtrsim$30\,000\,K 
     is indicated \citep[see Figs.~3, 5, 8, 10, 16, 19 and 21 of][ and 
     Table~\ref{tab:apar} in Appendix~\ref{app:par} 
     here]{2005A&A...440..995S}, \\
$^{d}$ the out-of-eclipse values, \\
$^{e}$ as on December 5, 1981, for the active phase, \\
$^{\dagger}$ References: 
1 - \cite{1991A&A...248..458M},
2 - \cite{1999A&AS..137..473M},
3 - \cite{2005A&A...440..995S},
4 - \cite{2012AN....333..242S},
5 - \cite{2001A&A...366..972S},
6 - \cite{2011A&A...536A..27S},
7 - \cite{2012ApJ...750....5K},
8 - \cite{2012BaltA..21..150S},
9 - \cite{1998IBVS.4571....1C},
10 - \cite{2018MNRAS.479.2728C},
11 - \cite{1989AJ.....98.1427M},
12 - \cite{2002A&A...387..139Q}.
\normalsize
\end{table*}
%
%
%
\clearpage
\section{Physical parameters of the ultraviolet continuum}
\label{app:par}
%
%
Table~\ref{tab:apar} of this appendix presents other physical 
parameters of our targets given by the variables of 
Eq.~(\ref{eq:uvsed}) that determines the continuum models for 
active phases, while Table~\ref{tab:qpar} lists these parameters 
for the spectra from quiescent phases that are used in 
Figs.~\ref{fig:qa} and \ref{fig:sedqa}. 
The modeling is introduced in Sect.~\ref{ss:ray}. 
\begin{table*}
\caption{Parameters from model SEDs during active phases 
(Table~\ref{tab:nh}): $R_{\rm WD}^{\rm eff}$, $T_{\rm BB}$, 
$T_{\rm e}$, $EM$ for the given date of observation, and 
a minimum of the $\chi^2_{\rm red}$ function. 
}
\label{tab:apar}
\begin{center}
\begin{tabular}{ccccccc}
\hline
\hline
\noalign{\smallskip}
Object                          &
Date                            &
$R_{\rm WD}^{\rm eff}$          & 
$T_{\rm BB}$                    & 
$T_{\rm e}$                     & 
$EM$                            & 
$\chi^2_{\rm red}$ / d.o.f.     \\
                                &
yyyy-mm-dd                      &
($R_{\odot}$)                   & 
(K)                             & 
(K)                             & 
(10$^{60}$\,cm$^{-3}$)          & 
                               \\
%
\noalign{\smallskip}
\hline
\noalign{\smallskip}
BF~Cyg & 1987-08-11 & 1.7$\pm0.2$ & 25000$\pm2000$ & 30000$\pm5000$  & 
                      3.0$\pm0.5$ & $\dagger$ \\
       & 1987-11-16 & 2.0$\pm0.2$ & 25000$\pm2000$ & 20000$\pm3000$  & 
                      3.4$\pm0.5$ & $\dagger$ \\
       & 1988-03-19 & 1.1$\pm0.1$ & 45000$\pm5000$ & 18000$\pm2000$ &
                      2.6$\pm0.3$ & 1.4 / 21 \\
       & 1990-06-30 & 6.3$\pm0.8$ & 21500$\pm2000$ & 42000$^b\pm5000$ &
                      6.9$\pm0.7$ & $\dagger$ \\
       & 1990-07-06 & 6.5$\pm0.6$ & 21000$\pm2000$ & 30000$^b\pm5000$ &
                      8.4$\pm0.8$ & $\dagger$  \\
       & 1990-11-05 & 5.8$\pm0.5$ & 25000$\pm2000$ & 42000$^b\pm5000$ &
                      11$\pm 1$     & $\dagger$ \\
CI~Cyg & 1979-01-05 & 0.50$\pm0.05$ & 28000$\pm2000$ & 20000$^a$ &
                      0.64$\pm0.07$ & $\dagger$ \\
       & 1979-06-11 & 0.60$\pm0.1$   & 25000$\pm4000$ & $\sim$15000 &
                      $\sim$0.37    & $\star$ \\
       & 1979-06-29 & $\sim$0.54    & $\sim$25000 & $\sim$15000 &
                      $\sim$0.40    & $\star$ \\
       & 1980-01-30 & 0.52$\pm0.07$ & 25000$\pm2000$ & 20000$\pm2000$ &
                      0.43$\pm0.06$ & $\dagger$ \\
       & 1980-04-14 & $\sim$0.44    & $\sim$25000    & $\sim$40000$^b$ &
                      $>$0.26       & $\times$ \\
       & 1980-08-01 & 0.44$\pm0.07$ & 25000$\pm3000$ & 20000$\pm3000$ &
                      0.37$\pm0.06$ & $\dagger$ \\
       & 1980-08-18 & 0.49$\pm0.08$ & 25000$\pm3000$ & 20000$\pm3000$ &
                      0.40$\pm0.07$ & $\dagger$ \\
       & 1980-08-28 & 0.46$\pm0.05$ & 28000$\pm2000$ & 15000$^a$/40000$^b$ &
                      0.61$\pm0.05$ & $\dagger$ \\
       & 1980-11-14 & 0.38$\pm0.05$ & 28000$\pm3000$ & 25000$\pm3000$ &
                      0.52$\pm0.07$ & $\dagger$ \\
       & 1981-01-08 & 0.34$\pm0.04$ & 28000$\pm2000$ & 25000$\pm3000$ &
                      0.46$\pm0.05$ & $\dagger$ \\
       & 1981-08-14 & $\sim$0.36    & $\sim$28000    & $\sim$25000 &
                      $\sim$0.46 & $\star$ \\
       & 1981-12-11 & 0.51$\pm0.05$ & 25000$\pm2000$ & 20000$\pm2000$ &
                      0.25$\pm0.04$ & $\dagger$ \\
YY~Her & 1981-12-04 & 1.1$\pm0.2$   & 22000$\pm2000$ & 15000$\pm2000$ &
                      0.66$\pm0.2$  & 1.8 / 17 \\
AR~Pav & 1980-10-30 & 2.7$\pm0.3$   & 16000$\pm1500$ & 14000$^a$/40000$^b$ &
                      3.2$\pm0.4$   & 1.4 / 14 \\
       & 1980-11-01 & 2.7$\pm0.3$   & 16000$\pm1500$ & 15500$^a$/40000$^b$ &
                      3.3$\pm0.4$   & 1.4 / 15 \\
       & 1980-11-02 & 3.1$\pm0.3$   & 16000$\pm1500$ & 15500$^a$/40000$^b$ &
                      3.3$\pm0.4$   & 1.5 / 17 \\
       & 1980-11-03 & 3.0$\pm0.3$   & 16000$\pm1500$ & 15500$^a$/40000$^b$ &
                      4.0$\pm0.5$   & 1.4 / 17 \\
       & 1982-07-01 & 3.5$\pm0.4$   & 16000$\pm1500$ & 17000$^a$/40000$^b$ &
                      4.0$\pm0.5$   & 1.8 / 17 \\
       & 1984-03-07 & 4.4$\pm0.5$   & 16000$\pm1500$ & 17000$\pm2000$/40000$^b$ &
                      3.7$\pm0.5$   & 1.6 / 12 \\
       & 1992-07-08 & 3.9$\pm0.5$   & 16000$\pm1500$ & 17000$^a$/40000$^b$ &
                      2.6$\pm0.3$   & 1.2 / 11 \\
       & 1984-04-10 & 5.5$\pm0.5$   & 15500$\pm1500$ & 13500$^a$/40000$^b$ &
                      4.3$\pm0.5$   & 1.8 / 13 \\
       & 1979-07-17 & 3.5$\pm0.5$   & 19000$\pm2000$ & 15800$^a$/40000$^b$ &
                      3.9$\pm0.4$   & $\dagger$ \\
       & 1981-05-10 & 2.6$\pm0.3$   & 20000$\pm2000$ & 15500$\pm2000$/40000$^b$ &
                      4.0$\pm0.5$   & 0.90 / 18 \\
       & 1978-05-16 & $\sim$3.6     & $\sim$17000 & -- &
                       --           & $\times$  \\
       & 1978-08-08 & 2.9$\pm0.5$   & 17500$\pm2000$ & 15500$^a$/40000$^b$ &
                      3.5$\pm0.7$   & 2.0 / 15 \\
       & 1978-11-11 & 3.8$\pm0.5$   & 16000$\pm2000$ & 14000$^a$/40000$^b$ &
                      3.3$\pm0.4$   & $\dagger$ \\
       & 1980-07-30 & 4.6$\pm0.7$   & 16000$\pm2000$ & 15000$^a$/40000$^b$ &
                      5.3$\pm0.6$   & $\dagger$ \\
       & 1982-04-07 & 3.2$\pm0.4$   & 16000$\pm2000$ & 40000$^b$ &
                   $\sim$3.2        & 0.39 / 127 \\
AX~Per & 1978-12-31 & 0.90$\pm0.2$  & 20000$\pm2000$ & 15000$\pm2000$ &
                      0.41$\pm0.06$ & $\dagger$ \\
       & 1979-01-05 & 0.90$\pm0.1$  & 20000$\pm1000$ & 15000$\pm2000$ &
                      0.40$\pm0.05$ & 0.80 / 18 \\
PU~Vul & 1988-06-29 & 6.0$\pm0.7$ & 23000$\pm2000$ & 15000$\pm2000$ &
                      12$\pm2$    &  2.4 / 18 \\
       & 1988-07-16 & 9.8$\pm1.1$ & 19000$\pm1500$ & 14000$\pm2000$ &
                      1.3$\pm0.3$ & 1.5 / 21 \\
       & 1988-10-03 & 8.2$\pm1.0$ & 20000$\pm2000$ & 20500$\pm3000$ &
                      11$\pm 2$   & 3.9 / 212 \\
       & 1989-04-08 & 7.2$\pm0.8$ & 22000$\pm2000$ & 20000$^b$ &
                      5.3$\pm 1$  & 2.4 / 16 \\
       & 1989-05-19 & 5.5$\pm0.6$ & 24000$\pm2000$ & 31000$\pm3000$ &
                      9.9$\pm 2$  & 1.5 / 15 \\
       & 1989-09-25 & 4.2$\pm0.5$ & 27000$\pm3000$ & 24000$\pm3000$ &
                      5.4$\pm 0.7$  & 0.88 / 15 \\
\noalign{\smallskip}
\hline 
\end{tabular}
\end{center}
\end{table*}
\addtocounter{table}{-1}
\begin{table*}[!ht]
\begin{center}
\caption{continued}
\begin{tabular}{ccccccc}
\hline
\noalign{\smallskip}
AS~338 & 1989-10-09 & $\approx$2.9& $\approx$15000 & $\approx$15000 &
                      $\approx$0.4  & $\star$ \\
AS~296 & 1988-08-13 & $\approx$8.7& $\approx$13000 & $\approx$10000 &
                      $\approx$0.1  & $\star$ \\
       & 1988-11-11 & $\approx$4.9  & $\approx$15000 & $\approx$15000 &
                      $\approx$0.1  & $\star$ \\
       & 1990-03-13 & $\approx$4.7  & $\approx$16000 & $\approx$20000 &
                      $\approx$0.1  & $\star$ \\
\noalign{\smallskip}
\hline
\hline
\end{tabular}
\end{center}
{\bf Notes:}\\
$^a$ -- fixed value, \\
$^b$ -- high-temperature nebula ($T_{\rm e}$ fixed) 
        filling in the residual light around Ly$\alpha$, \\
$\dagger$ -- from a comparison with a set of synthetic models 
             (Sect.~\ref{ss:ray}), \\
$\star$ -- underexposed spectrum, \\
$\times$ -- only SWP spectrum available
%
\end{table*}
%
\begin{table*}
\caption{As in Table~\ref{tab:apar}, but for the spectra from 
quiescent phases that are plotted in Figs.~\ref{fig:qa} and 
\ref{fig:sedqa}. 
}
\label{tab:qpar}
\begin{center}
\begin{tabular}{ccccccccc}
\hline
\hline
\noalign{\smallskip}
Object                          &
Spectrum                        &
Date                            &
Julian Date                     &
$R_{\rm WD}^{\rm eff}$          & 
$T_{\rm BB}$                    & 
$T_{\rm e}$                     & 
$EM$                            & 
Note                           \\
                                & 
                                & 
yyyy-mm-dd                      &
JD~2\,44...                     &
($R_{\odot}$)                   & 
(K)                             & 
(K)                             & 
(10$^{60}$\,cm$^{-3}$)          & 
                               \\
%
\noalign{\smallskip}
\hline
\noalign{\smallskip}
BF~Cyg & SWP13477+LWR10132 & 1981-03-13 & 44676.84$^a$ 
       & 0.37  & 100000$^b$ & $\sim$5000 & $\sim$0.22 & $\dagger$ \\
       & SWP28734+LWP08681 & 1986-07-22 & 46633.58 
       & 0.48  & 100000$^b$ & $\sim$22000& $\sim$3.3  & $\dagger$ \\
       & SWP58385+LWP32695 & 1996-09-26 & 50352.80 
       & 0.44  &  95000 &       22800& 2.5            & Sk05$^c$  \\
CI~Cyg & SWP36321+LWP15571 & 1989-05-22 & 47669.04 
       & 0.088 & 115000 &       24000& 0.31           & Sk05    \\
YY~Her & SWP09773+LWR08493 & 1980-08-12 & 44464.18 
       &$<$0.14&$>$105000&      21500& 1.1            & Sk05    \\
AX~Per & SWP24278+LWP04619 & 1984-10-23 & 45997.21 
       &$<$0.10&$>$66000&       25000& 0.25           & Sk05    \\
PU~Vul & SWP45415+LWP23757 & 1992-08-23 & 48858.25 
       &$<$0.64&$>$79000& $\sim$25000&$\sim$9.6       & Sk14    \\
AS~296 & SWP16380+LWR12630 & 1982-02-19 & 45020.11
       & 0.22  & 100000$^b$ &$\sim$20000& $\sim$0.72  & $\dagger$/$\star$ \\
\noalign{\smallskip}
\hline
\hline
\end{tabular}
\end{center}
{\bf Notes:}\\
$^a$ -- orbital phase $\sim$0.1; 
        $N_{\rm H}^{\rm obs}\sim 2\times 10^{23}$\cmd, \\
$^b$ -- fixed value, \\
$^c$ -- for the distance $d=3.4$\,kpc (Table~\ref{tab:nh}), \\
$\dagger$ -- from a comparison with a set of synthetic models 
             (Sect.~\ref{ss:ray}), \\
Sk05 -- according to \cite{2005A&A...440..995S}, \\
Sk14 -- according to \cite{2014ASPC..490..211S}, \\
$\star$ -- underexposed spectrum
%
\end{table*}
%
\clearpage
%
%
%
%
\section{Ly$\alpha$ line profiles from spherically symmetric 
         expanding shell}
\label{app:lya}
In this appendix, we test whether the Ly$\alpha$ absorption 
line profile that can be formed in the expanding wind from 
the WD during the active phases could be as broad as the 
attenuation due to Rayleigh scattering on H$^0$ atoms in 
the wind from the red giant. 

For this purpose, we used the code {\small SHELLSPEC} 
\cite[][]{2004CoSka..34..167B} that solves a simple radiative 
transfer along the line of sight in the moving circumstellar 
medium (CM) assuming local thermodynamic equilibrium (LTE) and 
single scattering approximations. 
The LTE model determines the basic CM structure, temperature, 
mass density, and electron density as functions of depth. 

Here, the CM has the form of spherically symmetric wind from 
the WD, whose optically thick/thin interface represents the 
WD's pseudophotosphere \citep[originally suggested for classical 
novae outbursts by][]{1994ApJ...437..802K}. 
Synthetic Ly$\alpha$ line profiles are calculated using an object 
{\small SHELL} in the {\small SHELLSPEC} code. The radius of the 
inner, $R_{\rm in}$, and outer, $R_{\rm out}$, sphere is 
set to 1 and 5\ro, respectively. 
We assume that the line profile is formed in the wind from the WD 
that conforms to wind properties derived from observations during 
active phases: The mass-loss rate, $\dot{M}$ of 
a few times $10^{-6}$\myr\ \citep[e.g.,][]{2006A&A...457.1003S}, 
and the pseudophotosphere expansion velocity of $\approx 200$\kms\ 
\cite[e.g.,][]{2011PASP..123.1062M} are considered. 
The latter is given by radial velocities of absorption 
components in P~Cygni profiles of hydrogen Balmer lines measured 
mainly at the beginning of outbursts. 
For simplicity, its value was assumed to be constant, and for 
turbulent motions, we adopted a constant velocity of 10\kms. 

Next, we choose the initial temperature and density at the inner 
radius of the shell, $T_{\rm in}$ and $\rho_{\rm in}$, deep 
inside the pseudophotosphere (i.e., where the optical depth 
in the continuum is $\gg$1), and find 
what temperature $T_1$ and density $\rho_1$ correspond to the 
wind medium with an optical thickness of 1, assuming it to 
represent the WD's pseudophotosphere. Its radius, $R_1$ 
is given by the mass continuity equation that determines 
the densities of the expanding wind. 
The input parameters, $T_{\rm in}$ and $\rho_{\rm in}$ were chosen 
so that the resulting parameters, $T_1$, $\rho_1$, $\dot{M}$, 
and $R_1$, were comparable to those determined from 
observations (Tables~\ref{tab:apar} and \ref{tab:lya}). 

\begin{figure*}
\begin{center}
\resizebox{\hsize}{!}
          {\includegraphics[angle=-90]{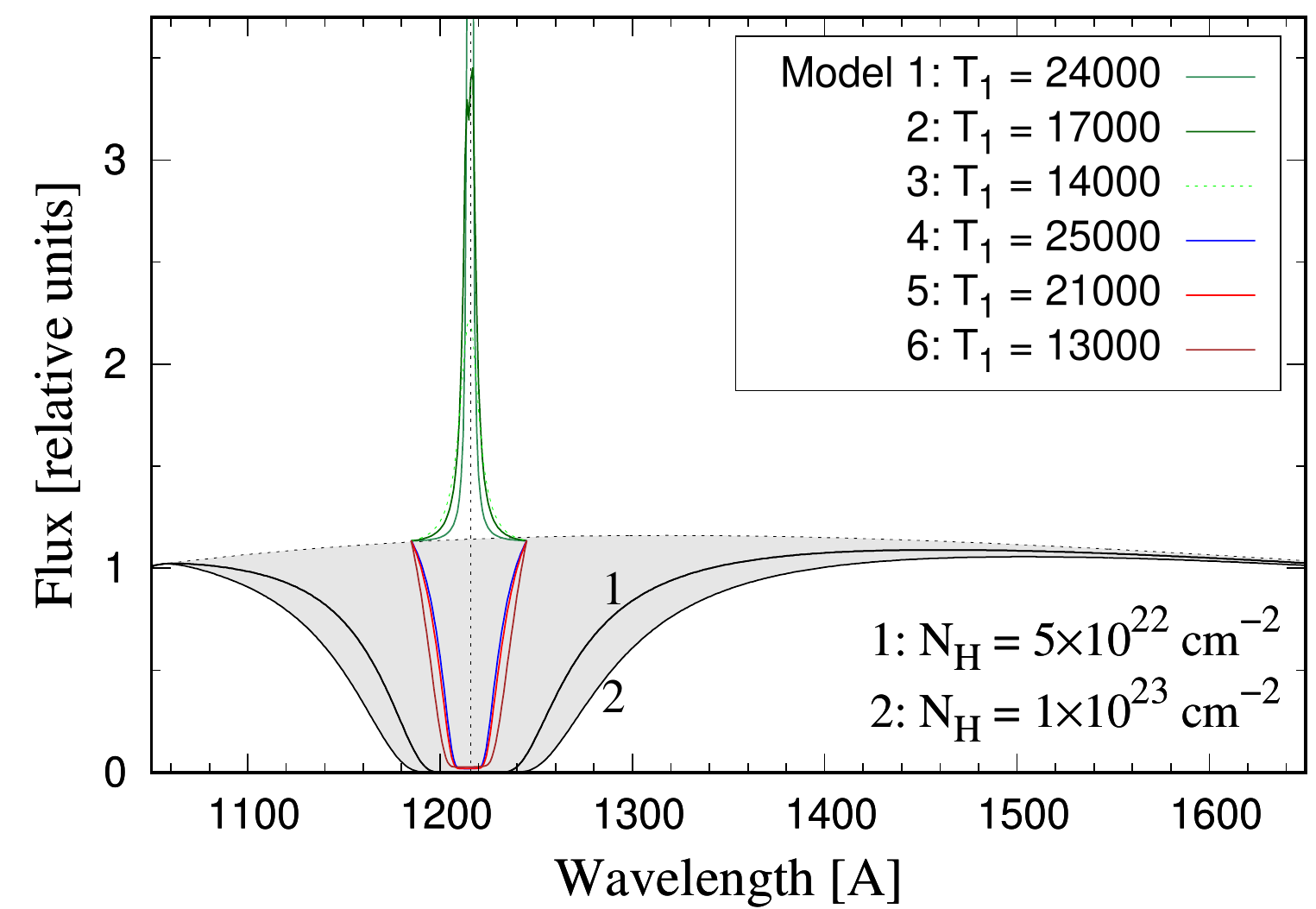}
           \includegraphics[angle=-90]{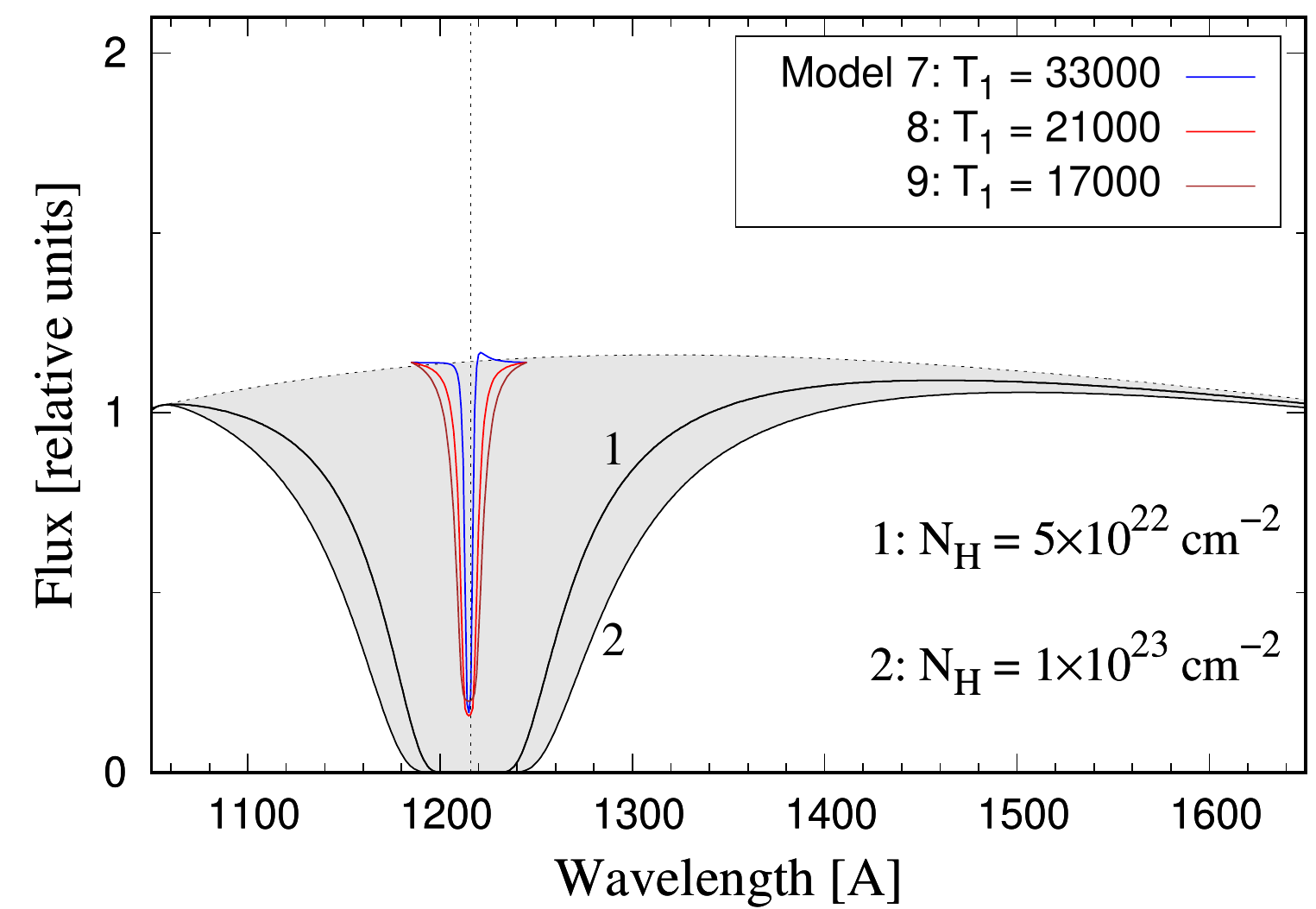}
           \includegraphics[angle=-90]{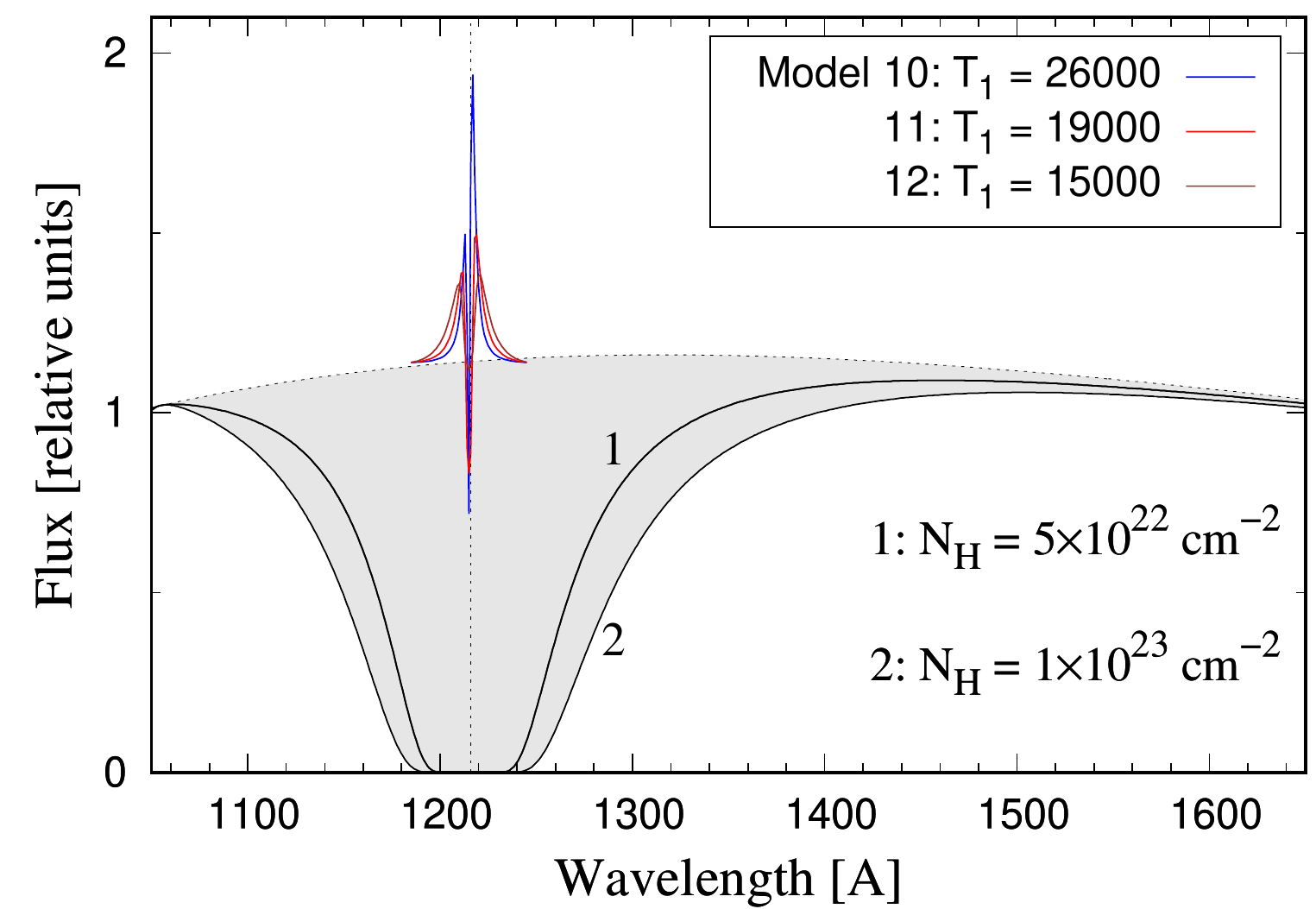}}
\end{center}
\caption{
Comparison of the UV continuum attenuation due to Rayleigh 
scattering on H$^0$ atoms for two values of $N_{\rm H}$ 
(grey area) with the Ly$\alpha$ line profile that may form 
in the expanding wind from the WD. 
The modeling of the Ly$\alpha$ line is described in 
Appendix~\ref{app:lya}. The model parameters are summarized in 
Table~\ref{tab:lya}. 
The dotted line represents the unattenuated continuum of the 
pseudophotosphere with a temperature $T_{\rm BB} = 22000$\,K. 
          }
\label{fig:appE}
\end{figure*}

As the emitted flux at the radial distance, $r$ from the WD could 
basically satisfy the proportionality, $\sigma T^4 \propto r^{-2}$, 
we first provided calculations using the temperature gradient, 
$T(r)/T_{\rm in} = (r/R_{\rm in})^{-1/2}$. 
For suitable parameters, $T_{\rm in} = 30000 - 40000$\,K and 
$\rho_{\rm in}$ = a few times $10^{-10} - 10^{-9}$\,g\,cm$^{-3}$, 
all the corresponding profiles have a dominant emission component 
(see Fig.~\ref{fig:appE}\,a, green profiles), $R_1$ is of a few 
solar radii, $T_1$ between 14000 and 24000\,K, and $\dot{M}$ 
of a few times $10^{-6} - 10^{-5}$\myr\ (Table~\ref{tab:lya}). 
%

In order to obtain the absorption component in the Ly$\alpha$ 
profile, we assumed a steeper temperature gradient, 
$T(r)/T_{\rm in} = (r/R_{\rm in})^{-1}$. 
The corresponding profiles, calculated for reasonable initial 
values of $T_{\rm in}$ and $\rho_{\rm in}$, are shown in 
Fig.~\ref{fig:appE}, where they are compared 
with the attenuation of the UV continuum around the Ly$\alpha$ 
line due to Rayleigh scattering on $5\times 10^{22}$ and 
$1\times 10^{23}$ hydrogen atoms per cm$^{2}$. Despite the fact 
that we chose rather lower values of $N_{\rm H}$ 
(see Fig.~\ref{fig:nhfi}), the attenuation effect by Rayleigh 
scattering is clearly dominant. 
The specific very broad wings from Rayleigh scattering 
are significantly broader than the `atmospheric' profile of 
the Ly$\alpha$ line\footnote{Note that the gas pressure at 
the expanding pseudophotosphere is zero, so the broadening 
of spectral lines by the Stark effect does not apply.}. 
%

Finally, it is important to note that if the width of the 
`atmospheric' Ly$\alpha$ absorption profile was dominant, 
it would not be possible to measure the orbital phase 
dependence of $N_{\rm H}^{\rm obs}$. 
Instead, the $N_{\rm H}^{\rm obs}$ values would be scattered 
along the orbital phase, as they would be dependent on 
the properties of the WD's pseudophotosphere, i.e., on the 
object and its current activity. This is not observed. 
In our case, the optically thick pseudophotosphere plays only 
the role of blocking ionization photons from the central WD, 
which has nothing to do with its properties. 
%
%
\begin{table*}
\caption{Input parameters, $T_{\rm in}$, $\rho_{\rm in}$, and 
resulting parameters, $T_1$, $\rho_1$, $\dot{M}$, $R_1$, for 
modeling the Ly$\alpha$ profile formed in the wind from the WD 
above its pseudophotosphere using the code {\small SHELLSPEC}. 
Parameters and modeling are described in Appendix~\ref{app:lya}. 
Corresponding profiles are shown in Fig.~\ref{fig:appE}. 
}
\label{tab:lya}
\begin{center}
\begin{tabular}{ccccccc}
\hline
\hline
\noalign{\smallskip}
Model                           &
$T_{\rm in}$                    &
$\rho_{\rm in}$                 &
$T_1$                           &
$\rho_1$                        &
$\dot{M}$                       &
$R_1$                           \\
No.                             &
(K)                             &
(${\rm g\,cm^{-3}}$)            &
(K)                             &
(${\rm g\,cm^{-3}}$)            &
($M_{\odot}\,{\rm yr}^{-1}$)    &
($R_{\odot}$)                   \\
%
%
\noalign{\smallskip}
\hline
\noalign{\smallskip}
1$^a$ & 30000 & $1\times 10^{-10}$  & 24000 & $4.4\times 10^{-11}$ 
               & $1.9\times 10^{-6}$ & 1.5 \\
2$^a$ & 30000 & $4\times 10^{-10}$  & 17000 & $3.9\times 10^{-11}$ 
               & $7.7\times 10^{-6}$ & 3.2 \\
3$^a$ & 30000 & $2\times 10^{-9}$   & 14000 & $9.5\times 10^{-11}$ 
               & $3.9\times 10^{-5}$ & 4.6 \\
4$^b$ & 40000 & $9\times 10^{-11}$  & 25000 & $3.5\times 10^{-11}$ 
               & $1.7\times 10^{-6}$ & 1.6 \\
5$^b$ & 40000 & $1\times 10^{-10}$  & 21000 & $2.8\times 10^{-11}$ 
               & $1.9\times 10^{-6}$ & 1.9 \\
6$^b$ & 40000 & $1.5\times 10^{-10}$ & 13000 & $1.7\times 10^{-11}$ 
               & $2.9\times 10^{-6}$ & 3.0 \\
7$^b$ & 50000 & $1\times 10^{-10}$  & 33000 & $4.4\times 10^{-11}$ 
               & $1.9\times 10^{-6}$ & 1.5 \\
8$^b$ & 50000 & $2\times 10^{-10}$  & 21000 & $3.5\times 10^{-11}$ 
               & $3.9\times 10^{-6}$ & 2.4 \\
9$^b$ & 50000 & $3\times 10^{-10}$  & 17000 & $3.6\times 10^{-11}$ 
               & $5.8\times 10^{-6}$ & 2.9 \\
10$^b$& 60000 & $2\times 10^{-10}$  & 26000 & $3.8\times 10^{-11}$ 
               & $3.9\times 10^{-6}$ & 2.3 \\
11$^b$& 60000 & $4\times 10^{-10}$  & 19000 & $3.9\times 10^{-11}$ 
               & $7.7\times 10^{-6}$ & 3.2 \\
12$^b$& 60000 & $8\times 10^{-10}$  & 15000 & $5.0\times 10^{-11}$ 
               & $1.5\times 10^{-5}$ & 4.0 \\
\noalign{\smallskip}
\hline
\hline
\end{tabular}
\end{center}
{\bf Notes:}\\
$^a$ -- models for temperature gradient, 
        $T(r)/T_{\rm in} = (r/R_{\rm in})^{-1/2}$, \\
$^b$ -- $T(r)/T_{\rm in} = (r/R_{\rm in})^{-1}$. 
\end{table*}
%
%
%
%
%
%
%
\section{On the effect of interstellar extinction on 
         modeling the UV continuum}
\label{app:ebv}
The strong attenuation of radiation by interstellar dust particles 
around the wavelength of 2175\,\AA\ is often used to determining 
the characteristic interstellar absorption from the UV spectrum. 
   Since the profile of the true continuum can be determined with 
an accuracy of about 10\%, this uncertainty also limits 
the value of the color excess so that the corrected profile of 
the continuum around 2175\,\AA\ does not show a local extreme. 
%

\begin{figure}
 \begin{center}
\resizebox{12cm}{!}
          {\includegraphics[angle=-90]{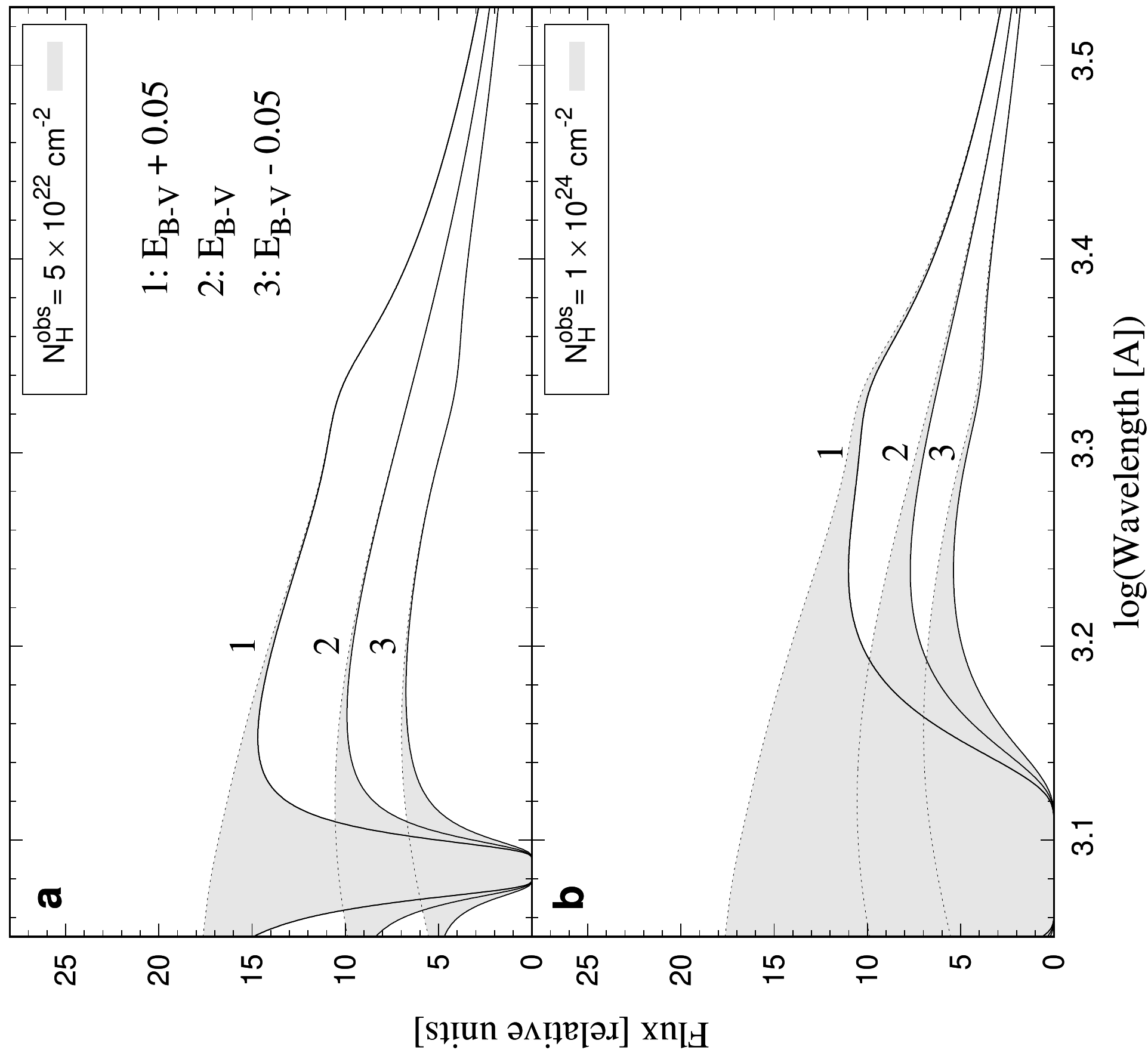}}
 \end{center}
\caption{
Effect of interstellar extinction in the range of IUE spectra 
shown for the Planck function ($T_{\rm BB}=22000$\,K) attenuated 
with $N_{\rm H}^{\rm obs}=5\times 10^{22}$ ({\bf a}) and 
$1\times 10^{24}$\cmd\ ({\bf b}), and dereddened with 
$E_{\rm B-V}\pm 0.05$\,mag (see text). 
Meaning of lines and the gray area as in Fig.~\ref{fig:sediue}. 
}
\label{fig:ebv1}
\end{figure}

Figure~\ref{fig:ebv1} shows what the effect of interstellar 
reddening looks like in the range of IUE spectra. 
It plots the Planck function for $T_{\rm BB}=22000$\,K, 
attenuated by Rayleigh scattering on $5\times 10^{22}$ 
(panel {\bf a}) and $1\times 10^{24}$ ({\bf b}) hydrogen atoms, 
both dereddened with $E_{\rm B-V}\pm 0.05$\,mag, where 
$E_{\rm B-V}$ is the correct value of the color excess. 
The value of $N_{\rm H}^{\rm obs}$ is given by the width of 
the absorption core and the extension of its wings due to 
Rayleigh scattering (Sect.~\ref{ss:ray}). Since the extinction 
curve is wavelength-dependent, increasing toward shorter 
wavelengths, one can expect that $N_{\rm H}^{\rm obs}$ values 
determined from the (broad) continuum depression around the 
Ly$\alpha$ line can be affected also by the uncertainty in 
$E_{\rm B-V}$. Moreover, the determination of lower values of 
$N_{\rm H}^{\rm obs}$ that attenuate a narrower range of the 
spectrum will be probably less affected by a change in color 
excess than high $N_{\rm H}^{\rm obs}$ values that attenuate 
a wider range of the spectrum. 

\begin{figure*}
 \begin{center}
\resizebox{\hsize}{!}
          {\includegraphics[angle=-90]{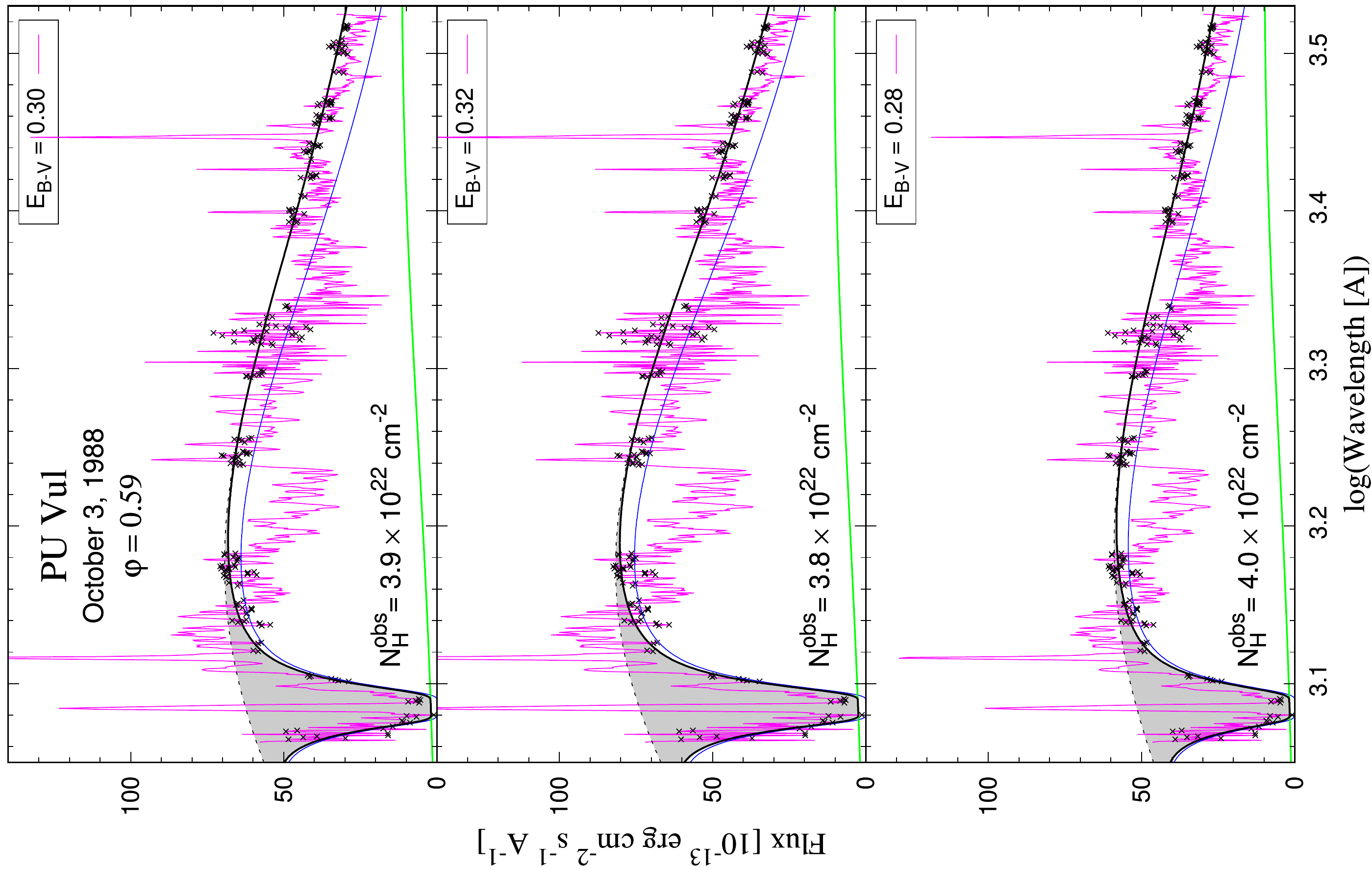}
           \includegraphics[angle=-90]{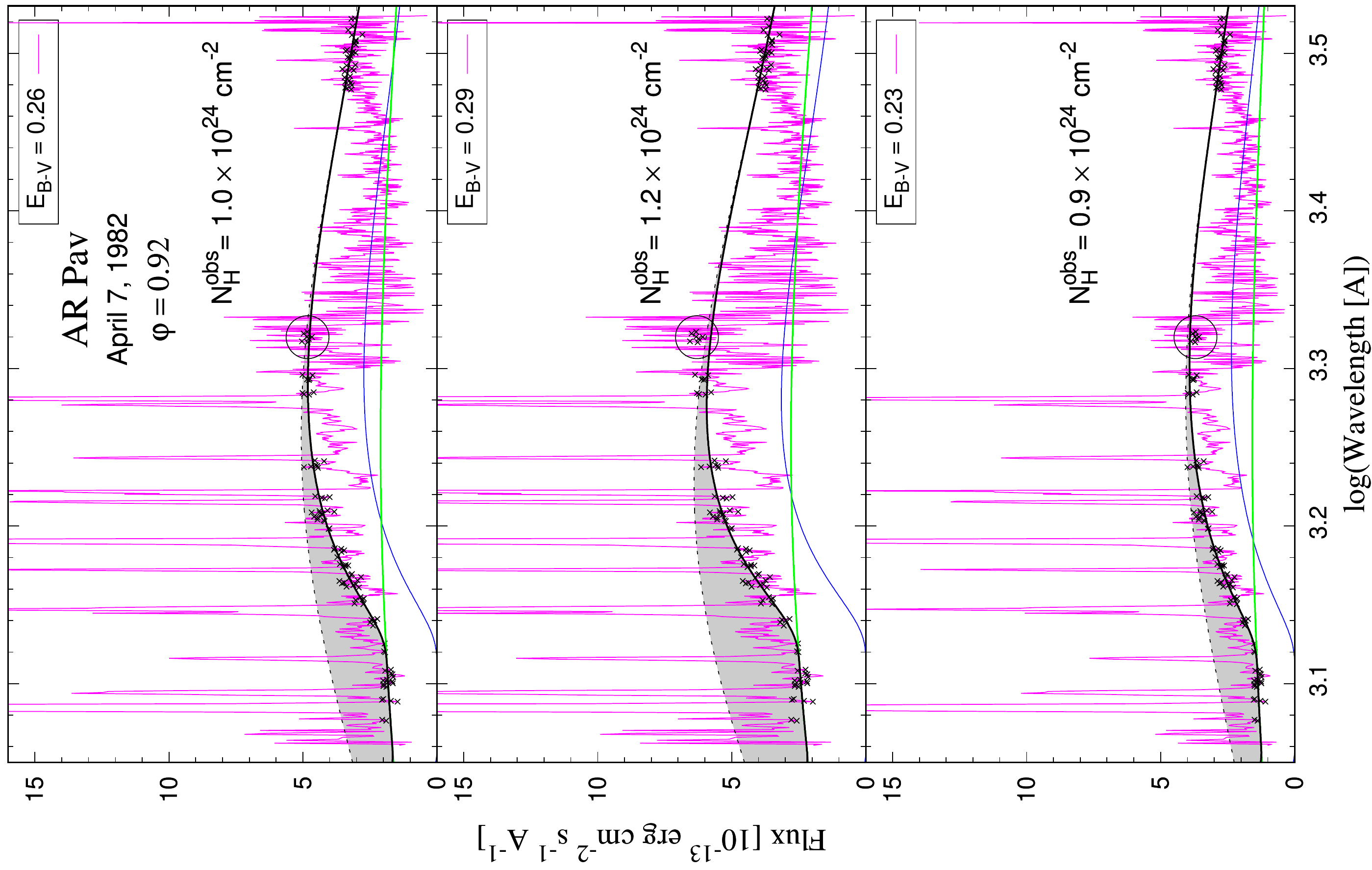}}
 \end{center}
\caption{
A possible effect of interstellar extinction on the determination 
of $N_{\rm H}^{\rm obs}$ illustrated for PU~Vul (left) and AR~Pav 
(right). The former and the latter represent the cases with 
relatively small and high values of $N_{\rm H}^{\rm obs}$ 
measured around the superior and the inferior conjunction of 
the red giant, respectively. 
The crosses represent the flux points that we modeled. 
The large spread of selected points near to $\lambda$2175\,\AA\ 
for PU~Vul is responsible for the high $\chi^2_{\rm red}$ value, 
while the small spread of these points for AR~Pav (circled) 
demonstrates the sensitivity of the spectrum in this region 
to interstellar extinction. 
Meaning of lines and the gray area as in Fig.~\ref{fig:sediue}. 
Corresponding parameters are summarized in Table~\ref{tab:ebv}. 
}
\label{fig:ebv2}
\end{figure*}

Therefore, in order to test how the uncertainties of the color 
excess affect the determination of $N_{\rm H}^{\rm obs}$, we 
selected one spectrum with a narrow absorption and another 
one with very broad absorption, which are typical for 
the superior and inferior conjunction of the red giant, 
respectively. 
Specifically, we modeled the spectrum of PU~Vul exposed on 
October 3, 1988 ($\varphi = 0.59$, 
$N_{\rm H}^{\rm obs} = 3.9\times 10^{22}$\cmd) and the spectrum of 
AR~Pav taken on April 7, 1982 ($\varphi = 0.92$, 
$N_{\rm H}^{\rm obs} = 1.0\times 10^{24}$\cmd) dereddened with 
different values of $E_{\rm B-V}$. For PU~Vul we used 
$E_{\rm B-V}=0.30\pm 0.04$, although \cite{2012ApJ...750....5K} 
give an uncertainty of only $\pm$0.02, and for AR~Pav we adopted 
$E_{\rm B-V}=0.26\pm 0.03$ according to models of 
\cite{1984ApJ...279..252K}. 
The corresponding SED models are shown in Fig.~\ref{fig:ebv2}, 
and their parameters are introduced in Table~\ref{tab:ebv}. 

In the case of PU~Vul (a narrower absorption), $N_{\rm H}^{\rm obs}$ 
decreases with increasing $E_{\rm B-V}$ for acceptable 
$\chi^2_{\rm red}$ (Table~\ref{tab:ebv}). 
This is because dereddening the spectrum with a larger value of 
$E_{\rm B-V}$ will increase the height of the absorption 
core relative to its long-wavelength wing, leading to 
a narrowing of the absorption core and thus a decrease of 
the corresponding $N_{\rm H}^{\rm obs}$. 
According to our models of PU~Vul, $E_{\rm B-V}$ uncertainties 
of $\sim$13\% correspond to errors in $N_{\rm H}^{\rm obs}$ of 
less than 10\% (see Table~\ref{tab:ebv}). 

In the case of AR~Pav, the SED modeling revealed the opposite 
behavior. This is due to the fact that a strong component 
of the hot nebular continuum contributes to the far-UV, which 
shifts the absorption core well above the zero value 
(see Fig.~\ref{fig:ebv2}, right). 
As a result, dereddening the spectrum with a higher value of 
$E_{\rm B-V}$ will increase the level of the far-UV 
spectrum more than its mid-UV part, which leads to relative 
smoothing of the absorption core that corresponds to 
a higher value of $N_{\rm H}^{\rm obs}$. 
Here, the uncertainties in $E_{\rm B-V}$ of $\pm$0.03\,mag 
($\sim$12\%) cause errors in $N_{\rm H}^{\rm obs}$ of 
10 -- 20\%. However, the uncertainly of 20\%, corresponding to 
the model for $E_{\rm B-V}=0.029$, is overestimated because 
of the worse model -- the flux-points near $\lambda$2175\,\AA\ 
are above the model which increases the value of $\chi^2_{\rm red}$ 
(see Fig.~\ref{fig:ebv2} and Table~\ref{tab:ebv}). 
Finally, we also tested a model with $E_{\rm B-V}=0.35$ that 
can be found in the literature (i.e., the uncertainty of 
$\sim$35\%). However, the points near $\lambda$2175\,\AA\ 
were $\sim$25\%\ above the model, and $\chi^2_{\rm red}$ 
increased by a factor of $\sim$2 (Table~\ref{tab:ebv}). 

According to these examples, we assume that interstellar 
reddening uncertainties may cause additional errors in 
the $N_{\rm H}^{\rm obs}$ determination of about 10\%. 
%
%
%
%
%
\begin{table*}
\caption{The best-fit parameters corresponding to SED models 
of PU~Vul and AR~Pav plotted in Fig.~\ref{fig:ebv2}. Models are 
calculated for different values of the color excess $E_{\rm B-V}$. 
Denotation and meaning as in Tables~\ref{tab:nh} and \ref{tab:apar}. 
}
\label{tab:ebv}
\begin{center}
\begin{tabular}{cccccccc}
\hline
\hline
\noalign{\smallskip}
Object                          &
$E_{\rm B-V}$                   &
$N_{\rm H}^{\rm obs}$           &
$R_{\rm WD}^{\rm eff}$          & 
$T_{\rm BB}$                    & 
$T_{\rm e}$                     & 
$EM$                            & 
$\chi^2_{\rm red}$ / d.o.f.     \\
                                &
(mag)                           &
(cm$^{-2}$)                     &
($R_{\odot}$)                   & 
(K)                             & 
(K)                             & 
(10$^{60}$\,cm$^{-3}$)          & 
                               \\
%
\noalign{\smallskip}
\hline
\noalign{\smallskip}
PU~Vul & 0.26 & $4.2\times 10^{22}$ & 7.5 & 19500 & 19000 
              & 9.3                 & 4.4 / 212        \\
       & 0.28 & $4.0\times 10^{22}$ & 8.1 & 19500 & 20000 
              & 9.7                 & 4.1 / 212        \\
       & 0.30 & $3.9\times 10^{22}$ & 8.2 & 20000 & 20000 
              & 11                  & 3.9 / 212        \\
       & 0.32 & $3.8\times 10^{22}$ & 9.0 & 20000 & 22000 
              & 11                  & 3.8 / 212        \\
       & 0.34 & $3.5\times 10^{22}$ & 9.8 & 20000 & 22000 
              & 12                  & 3.8 / 212        \\
AR~Pav & 0.23 & $0.9\times 10^{24}$ & 3.5 & 15000 & 40000$^a$
              & 2.4                 & 0.40 / 127        \\
       & 0.26 & $1.0\times 10^{24}$ & 3.2 & 16000 & 40000$^a$
              & 3.2                 & 0.39 / 127        \\
       & 0.29 & $1.2\times 10^{24}$ & 2.8 & 17500 & 40000$^a$
              & 4.2                 & 0.46 / 127        \\
       & 0.35 & $1.2\times 10^{24}$ & 2.7 & 19000 & 40000$^a$
              & 7.0                 & 0.99 / 127        \\
\noalign{\smallskip}
\hline
\hline
\end{tabular}
\end{center}
{\bf Notes:}\\
$^a$ -- fixed value
\end{table*}
\clearpage
\bibliography{windd.bib}{}
\bibliographystyle{aasjournal}
\end{document}